\documentclass[12pt]{article}
\usepackage{amsmath}
\usepackage{graphicx,psfrag,epsf}
\usepackage{enumerate}
\usepackage{natbib}
\usepackage{url} 

\newcommand{\blind}{0}

\usepackage{amssymb}
\usepackage{color, colortbl, xcolor} 
\usepackage{bm}
\usepackage{setspace}
\usepackage{caption}
\usepackage{subcaption}
\usepackage{array}
\usepackage{booktabs}
\usepackage[polish,english]{babel}

\newtheorem{theorem}{Theorem}

\begin{document}
	\def\spacingset#1{\renewcommand{\baselinestretch}%
		{#1}\small\normalsize} \spacingset{1}

	
	\if0\blind
	{
		\title{\bf Identification of structural shocks in Bayesian VEC models with two-state Markov-switching heteroskedasticity}
		\author{Justyna Wróblewska\thanks{
				This work was financially supported through a grant from the National Science Center (NCN, Poland) under decision no. UMO-2018/31/B/HS4/00730.}\hspace{.2cm}\\
			Department of Econometrics and Operations Research,\\ Krakow University of Economics\\
			and \\
			Łukasz Kwiatkowski\\
			Department of Econometrics and Operations Research,\\ Krakow University of Economics}
		\maketitle
	} \fi
	
	\if1\blind
	{
		\bigskip
		\bigskip
		\bigskip
		\begin{center}
			{\LARGE\bf Identification of structural shocks in Bayesian VEC models with two-state Markov-switching heteroskedasticity}
		\end{center}
		\medskip
	} \fi
	
	\bigskip
	\begin{abstract}
		We develop a Bayesian framework for cointegrated structural VAR models identified by two-state Markovian breaks in conditional covariances. The resulting structural VEC specification with Markov-switching heteroskedasticity (SVEC-MSH) is formulated in the so-called B-parameterization, in which the prior distribution is specified directly for the matrix of the instantaneous reactions of the endogenous variables to structural innovations. We discuss some caveats pertaining to the identification conditions presented earlier in the literature on stationary structural VAR-MSH models, and revise the restrictions to actually ensure the unique global identification through the two-state heteroskedasticity. To enable the posterior inference in the proposed model, we design an MCMC procedure, combining the Gibbs sampler and the Metropolis-Hastings algorithm. The methodology is illustrated both with a simulated as well as real-world data examples.
	\end{abstract}
	
	\noindent%
	\vfill
	
	Declarations of interest: none
	
	\newpage
	\spacingset{1.8} 
	
	\section{Introduction}
	\label{sec:Introduction}

	Since the seminal work by \citet{sims1980macroeconomics}, both structural and reduced-form Vector Autoregressive (VAR) models have been used extensively in empirical macroeconomic studies. A reduced-form VAR provides a potent statistical tool for capturing the dynamics of multivariate macroeconomic time series, with the model's parameters being uniquely identified solely by data. However, when structural analyses are of interest (such as forecast error variance decomposition, impulse response functions, historical decompositions, and forecast scenarios), then a VAR model in its structural form is required. In a structural VAR (SVAR), it is identification of orthogonal (or, structural) shocks that is of vital importance, with the task hopefully delivering  results that are not only statistically valid, but also meaningful from an economic perspective. 
	
	The identification of structural shocks boils down to identification of the SVAR model's parameters, a problem that a reduced-form VAR does not suffer from. Traditional approaches to the task include linear (typically, zero) or sign restrictions, which can be imposed on the system variables' reactions at given horizons (see, e.g., \citet{arias2018inference}, \citet{rubio2010structural}, but also  \citet{rubio2005markov} in the context of SVAR models identified through heteroskedasticity). Most often, such a number of the restrictions is introduced that ensures the so-called just identification (as opposed to over-identification), which, however, precludes testing thereof. On the other hand, the over-identified models do admit testing the over-identifying restrictions, but the approach proves viable only under the validity of the remaining constraints.
	
	
	The above-listed problems can be circumvented by extending a standard homoskedastic and Gaussian VAR model to account for additional features of the shocks, such as heteroskedasticity (modeled by various approaches) and non-Gaussianity (see, e.g., \citet{kilian2017structural}, \citet{lutkepohl2017structural} for reviews). Such generalizations may (although do not have to) successfully aid the identification, so that the traditional restrictions (imposed a priori) become over-identifying, and thus testable. Nonetheless, only statistical identification is directly facilitated then, with no guarantee of economically sound interpretations whatsoever. Therefore, figuring out the economic meaning of obtained results is an additional, follow-up issue, which even under statistically identified shocks can still pose quite a challenge.
	
	The line of research on identification through regime-switching heteroskedasticity (although not in VAR, but simultaneous equations models) started with the works by \citet{sentana2001identification}, \citet{rigobon2003identification}, and \citet{rigobon2003measuring}. However, in the latter, the Authors notice that the very idea of linking the errors' variances with estimation and identification of the model parameters traces back to \citet{wright1928tariff}.
	
	In this paper, we adopt the idea of structural shocks being identified through heteroskedasticity for identification of Bayesian cointegrated (as opposed to stationary) VAR systems with volatility breaks governed by a two-state Markov chain. As such, our study contributes to the area of research devoted to VAR models with Markov-switching heteroskedasticity (VAR-MSH), started by \citet{lanne2010structural}, and further developed by \citet{netsunajev2013reaction}, \citet{velinov2013can}, \citet{herwartz2014structural}, \citet{lutkepohl2014disentangling}, \citet{lutkepohl2017structural}, \citet{velinov2015stock}, \citet{lanne2016data}, \citet{lutkepohl2016structural}, \citet{lutkepohl2018relation}, \citet{velinov2018importance}, \citet{lutkepohl2021testing} (see also \citet{lutkepohl2020bayesian} for a more comprehensive literature review). The cited works utilized the frequentist statistical framework for estimation and inference, whereas \citet{kulikov2013identifying}, \citet{wozniak2015assessing}, \citet{lanne2016data} \citet{lutkepohl2020bayesian} resort to the Bayesian approach. 
	To the best of our knowledge, out of all of the papers cited above, only the four: \citet{velinov2013can}, \citet{herwartz2014structural}, \citet{lutkepohl2016structural}, and \citet{velinov2018importance} consider cointegrated (as opposed to stationary) VARs with the structural shocks identified through regime-switching volatility, with the methodology developed within the frequentist estimation setting. Naturally, stationary VAR systems are methodologically more straightforward to handle (which obviously, does not preclude them from still proving empirically valid in many cases). However, taking long-run relationships into account can not only enhance the model's capability of capturing data dynamics, but may also provide additional restrictions aiding the identification by enabling one to isolate permanent shocks from ones that exert only transient impact on modeled variables (see, e.g., \citet{king1991stochastic}, \citet{GonzaloNg2001}, \citet{lutkepohl2005new}).
	
	As can be inferred from the above, the current state of pertinent research misses out Bayesian (as opposed to frequentist) structural VAR models featuring both cointegration and Markov-switching volatility. Hence, filling this gap is the main aim of this paper\footnote{Note that Bayesian Markov-switching VAR models with (possibly regime-changing) cointegration have already been considered in the literature, although only in their reduced form, and thus without the structural context; see \citet{cui2012bayesian}, \citet{jochmann2015regime}, \citet{sugita2016bayesian}, and \citet{hauzenberger2021stochastic}.}.
	
	Another aim of our present work is to revisit the conditions under which a VAR (whether stationary or cointegrated) model with Markovian volatility breaks is uniquely and globally identified. The discussion builds on Theorem 1 stated in \citet{lutkepohl2020bayesian}, for which we deliver a counterexample to show that yet another constraint (beyond the ones required by the theorem) is actually required to ensure the global identification of a two-state VAR-MSH system, and thus the theorem warrants only a local identification (up to the ordering of structural innovations). Then, we formalize our finding through a relevant theorem (with the proof presented in Appendix~\ref{sec:appendixProof}).
	
	The remainder of the paper is organized as follows. The next section presents a two-state SVEC-MSH model underlying our methodological framework, and revisits a central issue of structural shocks identification therein. In Section 3, we develop a Bayesian approach to estimation of the model in question, further illustrated by an empirical analysis of a dividend discount model for the U.S. economy (Section 4), previously considered by \citet{binswanger2004stock}, \citet{louis2010stock}, \citet{velinov2013can}, and \citet{lutkepohl2016structural}. Section 5 concludes. Finally, our work is accompanied by a supplementary material presenting i) details on the prior structure, ii) a Markov Chain Monte Carlo algorithm for posterior sampling, iii) a simulated data study showcasing the methodology developed in the paper, and iv) additional empirical results.
	
	\section{Model framework}
	\label{sec:Model}
	\subsection{The sampling model}
	\label{subsec:Sampling_model}
	The basic structure underlying our modeling framework is a linear $n$-variate vector autoregressive process $y_t=(y_{t1}, ..., y_{tn})'$ of order $p$ ($y_t \sim$VAR(\textit{p})), in the structural vector error correction (SVEC) form, with deterministic terms and a regime-switching conditional covariance matrix:
	\begin{equation}
		\Delta y_t = \alpha\beta'\tilde{y}_{t-1}+\sum_{i=1}^{p-1}\Gamma_i\Delta y_{t-i}+\Phi D_t+B\varepsilon_t, \quad t=1, 2, ..., T,
		\label{eqn:SVEC_MSH_model}
	\end{equation}  
	\begin{equation}
		\varepsilon_t|S_t, \lambda_{S_t}\sim N^{(n)} \left(0, \Lambda_{S_t}\right),
		\label{eqn:Cond_cov_matrix}
	\end{equation}
	where $\tilde{y}_{t-1}' = (y_{t-1}',d_{t-1}')$, $d_{t-1}$ denotes deterministic components restricted to cointegrating relations, $\alpha$ in an $n\times r$ matrix of adjustment coefficients, $\beta' = (\tilde{\beta}',\phi')$ is an $r\times \tilde{n}$ cointegration matrix $(\tilde{n}\ge n)$ of a full rank $r$, $r$ is the number of cointegration relationships ($0<r<n$, if they exist), or $r=0$ (for a non-stationary but non-cointegrated system), or $r=n$ (for a stationary VAR), and $D_t$ comprises deterministic variables (e.g., an unrestricted constant, inducing linear trends in $y_t$, and possibly seasonal dummies; see \citet{juselius2006cointegrated}). Contemporaneously and serially uncorrelated error terms (i.e. structural shocks) are arrayed in $\varepsilon_t=(\varepsilon_{t1}, ..., \varepsilon_{tn})'$, with their regime-dependent, diagonal covariance matrix $\Lambda_{S_t}=diag(\lambda_{S_t})$, where $\lambda_{S_t}=(\lambda_{S_t,1}, ..., \lambda_{S_t,n})'$ denotes an $n$-variate vector of the conditional variances associated with volatility state $S_t$, i.e. $\lambda_{S_t,i}=Var(\varepsilon_{ti}|S_t,\lambda_{S_t,i})$ (note that, from a Bayesian perspective, conditioning on $\lambda_{S_t,i}$ as well as on $\lambda_{S_t}$ in Eq.~(\ref{eqn:Cond_cov_matrix}) is needed). The state-dependent covariance matrix of the reduced-form errors, $u_t=B\varepsilon_t$, is then given as $\Sigma_{S_t}=V(u_t|S_t, \lambda_{S_t})=B\Lambda_{S_t}B'$. Finally, $B=[b_{ij}]_{i,j=1,...,n}$, termed as the structural matrix, comprises the instantaneous responses $b_{ij}$ of the $i^{th}$ variable, $y_{ti}$, on the $j^{th}$ structural shock, $\varepsilon_{tj}$. For the sake of normalization, we set $b_{ii}=1$ for $i=1, 2, ..., n$, so that the signs of the shocks are fixed, leaving their identification hinged on heteroskedasticity and possibly also additional (linear) restrictions (see \citet{lutkepohl2020bayesian})\footnote{Note that \citet{lutkepohl2020bayesian} assume that the normalization restriction, $b_{ii}=1$ for $i=1, 2, ..., n$, leaves the identification only to the changes of the relative variances, thereby resolving also the issue of the ordering of the shocks. However, as demonstrated in this paper, it is not the case.}. Simultaneously, $B^{-1}$ reflects the contemporaneous relationships between the endogenous variables.
	
	Similarly to \citet{lanne2010structural} and \citet{lutkepohl2020bayesian} (among others), we assume that the variables $S_1, S_2, ..., S_T$ form a two-state ergodic and homogeneous Markov chain, $S_t\in{\{1,2\}}$, that governs the regime changes of structural shocks' variances\footnote{Notice that we restrict our considerations to a two-state model only. Generalization of the theoretical results presented in this paper, for systems with more regimes is deferred for our future work.}. The properties of the chain are determined by the transition matrix $\mathbf{P}=[p_{ij}]_{i,j=1,2}$, where $p_{ij}=\Pr( S_t=j|S_{t-1}=i)$, for $i,j\in\{1,2\}$, is the transition probability from state $i$ at time $t-1$ to state $j$ at time $t$. Thus, the elements of each row $p_i=(p_{i1},\ p_{i2})$, $i=1,2$, sum up to 1.
	
	Note that, according to Eq.~(\ref{eqn:SVEC_MSH_model}), we decide to work directly with a structural VEC (SVEC) model instead of embarking on with a VAR structure and then recovering its VEC form parameters. This allows us to apply the B-model parameterization to the SVEC form, which is a common choice for a structural VEC framework (see \citet{lutkepohl2005new}, p. 369). For the model at hand, the Beveridge-Nelson MA representation of $y_t$ takes the form:
	\begin{equation}
		y_t=\Xi B \sum_{i=1}^{t}{\varepsilon_i}+B\sum_{j=0}^{\infty}{\Xi^*_j \varepsilon_{t-j}}+\Xi\Phi\sum_{i=1}^t D_i+\sum_{j=0}^{\infty}{\Xi^*_j (\Phi D_{t-j}+\alpha\phi'd_{t-1-j})}+y_0^{*},\nonumber
		\label{eqn:Beveridge_Nelson}
	\end{equation}
	where $y_0^*$ depends on the initial values, matrices $\Xi_j^*$, $j=0, 1, ...$, are some functions of the model parameters and are absolutely summable, thus forming such a convergent sequence that $\sum_{j=0}^{\infty}{\Xi^*_j \varepsilon_{t-j}}$ is an I(0) process. Finally, $\Xi=\beta_{\bot}(\alpha_{\bot}'\Gamma\beta_{\bot})^{-1}\alpha_{\bot}'$, with $\Gamma = I_n-\sum_{i=1}^{k-1}\Gamma_i$, is the matrix of the long-run multipliers so that $\Xi B$ contains the long-run effects of the structural innovations\footnote{Matrices $\alpha_{\bot}$ and $\beta_{\bot}$ span the orthogonal complements of $sp(\alpha)$ and $sp(\tilde{\beta})$, respectively, so they are $n\times(n-r)$ full column rank matrices, such that $\alpha'\alpha_{\bot} = 0$, $rank(\alpha, \alpha_{\bot}) = n$, $\tilde{\beta}'\beta_{\bot} = 0$, and $rank(\tilde{\beta}, \beta_{\bot}) = n$.}. 
	
	Notice that the rank of $\Xi$ equals $n-r$, which is also the rank of $\Xi B$, since $B$ is non-singular (see, e.g., \citet{johansen1996likelihood}, and \citet{lutkepohl2005new} for details). The property can by utilized to identify the shocks. Since $rank(\Xi B)= n-r$, this matrix can contain $r$ zero columns at most, allowing for maximally $r$ shocks to be only of a transient nature. To identify $B$, as many as $n(n-1)/2$ restrictions are required. Assuming that exactly $r$ shocks are transient, we impose $n(n-r)$ restrictions. Remaining restrictions that are needed to identify the shocks in both of the groups, need to be imposed on: $B$ (for the transient shocks; note that they correspond to the zero columns in $\Xi B$), and the non-zero columns of $\Xi B$ (for the permanent shocks); see, e.g., \cite{king1991stochastic}, \cite{GonzaloNg2001}, \cite{lutkepohl2005new}.

	\subsection{Identification of the SVEC-MSH model}
	\label{subsec:Identification}
	
	In this research, we identify structural shocks via Markovian volatility breaks, which contributes to the line of research initiated in a VAR framework by \citet{lanne2010structural}, and later developed by, among others, \citet{lutkepohl2016structural}, \citet{lutkepohl2020bayesian}. In the first of the cited works, a stationary SVAR-MSH is considered within a frequentist estimation setting. On the other hand, \citet{lutkepohl2016structural} shift the focus on cointegrated systems in a SVEC-MSH form (again, within the frequentist framework). Finally, \citet{lutkepohl2020bayesian} develop a Bayesian approach to estimation and inference of a SVAR-MSH model, limiting their attention, however, only to the stationary case.
	
	The point of departure for our considerations presented below, is Theorem 1 formulated in \citet{lutkepohl2020bayesian}, providing conditions for ensuring the identification of a stationary structural VAR system with Markov-switching heteroskedasticity. Of note, in the cited paper, the so-called A-parameterization of a VAR system is employed (so that the instantaneous dependencies between the endogenous variables are explicitly modeled), as opposed to the B-parameterization followed in our present work  (see Eq.~(\ref{eqn:SVEC_MSH_model})), so that the immediate responses of the endogenous variables to the shocks are explicitly modeled. Such a parameterization is a typical choice for cointegrated VARs in the VEC form, for being conducive to disentangling the permanent from only transient shocks (see \citet{lutkepohl2005new}).
	
	Finally, before we venture into the key matters underlying this section, let us invoke a remark often raised in the literature on various approaches to structural VAR identification, namely that distilling structural shocks through regime-switching heteroskedasticity (among other approaches, which use for example non-Gaussianity of the errors or more elaborate volatility structures such as GARCH; see \citet{kilian2017structural}) is an attempt of merely a statistical identification of the shocks, and thus do not necessarily provide results that are of valid economic meaning, which remains to be figure out following up the statistical procedure (see, e.g., \citet{lutkepohl2016structural}, \citet{lutkepohl2020bayesian}). Nevertheless, a statistically identified VAR/VEC system enables testing traditionally imposed identifying restrictions (see, e.g., \citet{lanne2010structural}, \citet{lutkepohl2016structural}, \citet{kilian2017structural}, \citet{lutkepohl2020bayesian}).
	
	To identify a two-state SVEC-MSH model, one could think of applying Theorem 1 formulated by \citet{lutkepohl2020bayesian}, although formulated therein for a stationary (rather than cointegrated) structural VAR model with finite-state Markov-switching heteroskedasticity (see also \citet{lanne2010structural}). Upon adaptation to our present notation, and assuming only a two-state system, we quote the theorem below.
	
	\begin{theorem}
		Let $\Sigma_m$, $m=1,2$, be positive-definite symmetric $n\times n$ matrices and $\Lambda_m=diag(\lambda_{m,1}, ..., \lambda_{m,n})$, $m=1,2$, be $n\times n$ diagonal matrices with positive diagonals. Suppose there exists a non-singular $n\times n$ matrix $B$ with unit main diagonal such that $\Sigma_m = B\Lambda_m B'$, $m=1,2$. Let $\omega_{2,i}=\lambda_{2,i}/\lambda_{1,i}$, $i=1,...,n$, be the i\textsuperscript{th} structural shock's (state 2) variance relative to state 1. Then, the $k^{th}$ row of $B$ is unique if $\forall_{i\in\{1,...,n\}\setminus \{k\}}\ \omega_{2,k} \ne \omega_{2,i}$.
		\label{Theorem_Lutke_Wozniak}
	\end{theorem}
	
	Note that if the conditions stipulated in the above theorem hold for each of the rows of $B$, then the entire matrix $B$ is identified. 
	
	Since under the model parameterization considered in our paper, matrix $B$ features ones on its main diagonal (thereby restricting the signs of the shocks), Theorem~\ref{Theorem_Lutke_Wozniak} should ensure the global identification of the system. However, consider the following (counter)example with $n=2$ variables only: $\lambda_1=\left(\begin{array}{cc}1 & 0.7\end{array} \right)$, $\lambda_2=\left(\begin{array}{cc}0.2 & 0.1\end{array} \right)$, and $B=\left(\begin{array}{cc}
		1 & -0.2  \\
		0.5 & 1 
	\end{array} \right)$, so that the following reduced-form state-specific covariance matrices are obtained:
	$\Sigma_1=\left(\begin{array}{cc}
		1.028 & 0.36  \\
		0.36 & 0.95 
	\end{array} \right)$ and
	$\Sigma_2=\left(\begin{array}{cc}
		0.204 & 0.08  \\
		0.08 & 0.15 
	\end{array} \right)$. The resulting relative (to state 1) structural variances are as follows: $\omega_{2,1}= \lambda_{2,1}/\lambda_{1,1}=\frac{1}{5}$ for the first variable, and $\omega_{2,2}=
	\lambda_{2,2}/\lambda_{1,2}=\frac{1}{7}$, respectively, so that $\omega_2=\left(\begin{array}{cc} \frac{1}{5} & \frac{1}{7} \end{array} \right)'$. The identification condition stated in Theorem~\ref{Theorem_Lutke_Wozniak} would simply require that the difference (contrast) between these two ratios be non-zero, which is the case here, indeed: $\omega_{2,2}-\omega_{2,1}=-\frac{2}{35}\ne 0$. However, and to one's surprise, simple calculations show that the very same reduced-form covariance matrices, $\Sigma_1$ and $\Sigma_2$, can be obtained also for a different set of $\lambda_1$, $\lambda_2$ and $B$, namely: $\tilde{\lambda}_1=\left(\begin{array}{cc}0.028 & 0.25\end{array} \right)$, $\tilde{\lambda}_2=\left(\begin{array}{cc}0.004 & 0.05\end{array} \right)$ and $\tilde{B}=\left(\begin{array}{cc}
		1 & 2  \\
		-5 & 1 
	\end{array} \right)$. The corresponding vector of the relative variances is now 
	$\tilde{\omega}_2=\left(\begin{array}{cc}
		\tilde{\lambda}_{2,1}/\tilde{\lambda}_{1,1} & \tilde{\lambda}_{2,2}/\tilde{\lambda}_{1,2}
	\end{array} \right)' = \left(\begin{array}{cc}
		\frac{1}{7} & \frac{1}{5}
	\end{array} \right)'$, which indicates that $\tilde{\omega}_{2,2}-\tilde{\omega}_{2,1}=\frac{2}{35}\ne 0$, so the condition underlying Theorem~\ref{Theorem_Lutke_Wozniak} is also met here. Notice that $\tilde{\omega}_{2,2}-\tilde{\omega}_{2,1}=-(\omega_{2,2}-\omega_{2,1})$, which actually arises from nothing else but $\tilde{\omega}_2$ being equal to the reordered $\omega_2$. Indeed, $\tilde{\omega}_{2,1}=\omega_{2,2}$ and $\tilde{\omega}_{2,2}=\omega_{2,1}$.
	
	It follows from the example above that: i) the theorem provided by \citet{lutkepohl2020bayesian} actually does not ensure the \textit{global} identification, enabling one only to identify the shocks locally up to their order; ii) apart from the condition already stipulated in Theorem~\ref{Theorem_Lutke_Wozniak}, the global identification requires restricting also the order of the relative changes of the state-dependent structural variances, $\omega_{2,i}$, $i=1, ..., n$. The latter remark leads to formulation of the following theorem (which can be viewed as a 'revised' version of the one formulated by \citet{lutkepohl2020bayesian}) of the global identification of the structural two-state VAR/VEC-MSH model (we provide the proof in Appendix~\ref{sec:appendixProof}).
	
	\begin{theorem}
		Let $\Sigma_m$, $m=1,2$, be positive-definite symmetric $n\times n$ matrices and $\Lambda_m=diag(\lambda_{m,1}, ..., \lambda_{m,n})$, $m=1,2$, be $n\times n$ diagonal matrices with positive diagonals. Suppose there exists a non-singular $n\times n$ matrix $B$ with unit main diagonal such that $\Sigma_m = B\Lambda_m B'$, $m=1,2$. Let $\omega_{2,i}=\lambda_{2,i}/\lambda_{1,i}$, $i=1,...,n$, be the i\textsuperscript{th} structural shock's (state 2) variance relative to state 1. Then, under a fixed order of the elements of the vector $\omega_2=(\omega_{2,1},\omega_{2,2},\dots,\omega_{2,n})'$, the $k^{th}$ row of $B$ is unique if  $\forall_{i\in\{1,...,n\}\setminus \{k\}}\ \omega_{2,k} \ne \omega_{2,i}$.
		\label{Theorem_ours}
	\end{theorem}
	
	The only difference between the two theorems is this additional fixing of the order of $\omega_{2,i}$, $i=1, ..., n$. As we point it out in the proof of Theorem~\ref{Theorem_ours} (see Appendix~\ref{sec:appendixProof}), the reason for which Theorem~\ref{Theorem_Lutke_Wozniak} actually fails to guarantee the global identification is that in its proof it has been assumed that the state-dependent variances, $\lambda_{m,i}$ ($m=1,2$ and $i=1,\ldots,n$) are unique. Indeed, under such an assumption, it can be shown that there exists also a unique matrix $B$. However, the restriction of unique structural regime-specific variances does not appear justifiable (contrary to the uniqueness of the reduced-form regime-specific covariance matrices, obviously). And so, as the proof in Appendix~\ref{sec:appendixProof} presents, under free $\lambda_{m,i}$s, the global identification of $B$ (and $\lambda_{m,i}$s) requires fixing the order of the relative (to state 1) structural variances.
	
	In view of the above, one can ponder on possible severity with which the failure to restrict the order of $\omega_{2,i}$s may affect estimation results. Obviously, the issue remains of a rather empirical nature, with particular outcomes hinging conceivably on assumed prior distributions, data set at hand, and even the properties of designed posterior distribution sampler. We illustrate some aspects of this otherwise general problem in Section 3 of the Supplementary material, through simulated-data-based studies (also using the case delivered above as the counterexample for Theorem~\ref{Theorem_Lutke_Wozniak}).
	
	\section{Bayesian two-state SVEC-MSH model}
	\label{sec:Bayesian_SVEC_MSH}  
	
	This section is devoted to specification of a Bayesian two-state SVEC-MSH model. In Section~\ref{subsec:Bayesian_SVEC_MSH_likelihood}, we write down the likelihood function, preceded by introduction of a suitable matrix notation. Section~\ref{subsec:Bayesian_SVEC_MSH_prior} displays the prior structure of our choice. A detailed presentation of the MCMC posterior sampling routine (combining the Gibbs and Metropolis-Hastings algorithms), along with the underlying full conditional posteriors, is deferred to the Supplementary material (see Section 2 therein).

	\subsection{The likelihood function}
	\label{subsec:Bayesian_SVEC_MSH_likelihood}
	
	Writing down the likelihood function for the model at hand requires introduction of some notation conducive thereto. In that regard, our framework draws on \citet{jochmann2015regime}, who developed a Bayesian Markov-switching VEC model in the reduced (instead of structural) form. Also note that their approach enables regime changes of all the model parameters (i.e. including the ones of the conditional mean), whereas in the structural VEC-MSH model considered here, only switches of the conditional covariance matrix are allowed, with the remaining parameters (and thus, the shocks transmission mechanism, in particular) held constant throughout (see also \citet{sugita2016bayesian}). Such a limitation, however, is typically entertained in the literature on structural shocks identification through heteroskedasticity (see, e.g., \citet{lutkepohl2020bayesian} in the context of Bayesian, although stationary VAR-MSH models; and \citet{lutkepohl2016structural} in the context of cointegrated, although frequentist VAR-MSH framework).
	
	For each of the two states ($m=1, 2$), define the matrix of the observations allocated to state $m$:
	
	\begin{equation}
		Z_0^{(m)} = Z_1^{(m)}\beta\alpha'+ Z_2^{(m)}\Gamma + E^{(m)}B',
		\label{eqn:SVEC-MSH_matrix}
	\end{equation}
	along with its vectorization
	\begin{equation}
		vec(Z_0^{(m)}) = \left(I_n\otimes Z_1^{(m)}\right)vec(\beta\alpha') + \left(I_n\otimes Z_2^{(m)}\right)vec(\Gamma) + \left(B\otimes I_{T_m}\right)vec(E^{(m)}),
		\label{eqn:SVEC-MSH_vectorization}
	\end{equation}
	where $T_m$ is the number of the observations assigned to state $m$. All observations are then collected in
	\begin{equation}
		\tilde z_0 = \tilde z_1vec(\beta\alpha') + \tilde z_2\gamma + \tilde{u},\quad \tilde u\vert S \sim N(0,\tilde\Sigma),
		\label{eqn:SVEC-MSH_final}
	\end{equation}
	where $\tilde z_0 = \left(\begin{array}{c}vec(Z_0^{(1)})\\vec(Z_0^{(2)})\end{array}\right)$, $\tilde z_1 = \left(\begin{array}{c}I_n\otimes Z_1^{(1)}\\I_n\otimes Z_1^{(2)}\end{array}\right)$, $\tilde z_2 = \left(\begin{array}{c}I_n\otimes Z_2^{(1)}\\I_n\otimes Z_2^{(2)}\end{array}\right)$,\\ $\gamma=vec(\Gamma)$, $\tilde u = \left(\begin{array}{c}u^{(1)}\\u^{(2)}\end{array}\right)$, $u^{(m)} = vec\left(U^{(m)}\right) = \left(B\otimes I_{T_m}\right)vec(E^{(m)})$, for $m=1,2$, and $S=(S_1, S_2, \dots, S_T)$, denotes the vector of the Markov chain states $S_t\in\{1, 2\}$ at $t=1, 2, \dots, T$. By $\tilde\Sigma$ we denote the (block-diagonal) conditional covariance matrix of $\tilde u$ (given the state vector, $S$), i.e. $\tilde\Sigma = diag\left(B\Lambda^{(1)} B'\otimes I_{T_1},\, B\Lambda^{(2)} B'\otimes I_{T_2}\right)$. 
	
	Using the above notation, we can write down the likelihood function (conditional on the vector of states):\footnote{Initial conditions: $y_{-p+1},\dots,y_{-1}, y_0$ (fixed at pre-sample values) and $S_0$ (a random variable), are suppressed from the notation.}
	\begin{equation}
		\begin{aligned}
			& p(y|S, \theta)   = (2\pi)^{-\frac{nT}{2}}\left|\tilde{\Sigma}\right|^{-\frac{1}{2}}\exp\left\{-\frac{1}{2}\tilde{u}'\tilde{\Sigma}^{-1}\tilde{u}\right\}=\\
			& =(2\pi)^{-\frac{nT}{2}}|B|^{-T}\left(\prod_{m=1}^2|\Lambda_m|^{-\frac{T_m}{2}}\right)\exp\left\{-\frac{1}{2}tr\left[\sum_{m=1}^2\left((B\Lambda_mB')^{-1}E^{(m)'}E^{(m)}\right)\right]\right\}=\\
			& =(2\pi)^{-\frac{nT}{2}}|B|^{-T}\left(\prod_{m=1}^2\prod_{i=1}^n\lambda_{m,i}^{-\frac{T_m}{2}}\right)\exp\left\{-\frac{1}{2}tr\left[\sum_{m=1}^2\left(\Lambda_m^{-1}B^{-1}E^{(m)'}E^{(m)}(B')^{-1}\right)\right]\right\},
		\end{aligned}
		\label{eqn:likelihood}
	\end{equation}
	with $\theta$ collecting all model parameters, i.e. $\theta = (\alpha, \beta, \Gamma, b, \lambda_1, \lambda_2, p_{11}, p_{22})$, where $b$ represents only the free entries of matrix $B$ (so that $vec(B) = Qb +q$, where $Q$ is an $n^2\times d_b$ known matrix, and $q$ is an $n^2\times 1$ known vector), and $\lambda_m=diag(\Lambda_m)$, $m=1, 2$. Notice that $\theta$ involves directly the structural variances for both states ($\lambda_1$ and $\lambda_2$) rather than only the first-state variances ($\lambda_1$) along with their relative changes in the second state ($\omega_2$). Although the latter parameterization was of choice in \citet{lutkepohl2020bayesian}, where it proved conducive to a testing setup developed therein, we decide to work here directly with structural variances in both states instead, as we prefer to introduce prior information directly on the regime-changing parameters. That way, we can ensure that the joint prior distribution for these parameters is symmetric (with respect to relabeling the states), and thereby ensure the shocks identification to arise only from an order restriction imposed explicitly on $\frac{\lambda_{2,i}}{\lambda_{1,i}}$, and not from asymmetric prior information for $\lambda_1$ and $\lambda_2$ (introduced inadvertently when parameterizing the model through $\lambda_1$ and $\omega_2$, instead).
	
	Obviously, under no further constraints on the structural variances, the likelihood function given by Eq.~(\ref{eqn:likelihood}) suffers from a perfect symmetry with respect to relabeling of the states. To ensure a unique identification of the regimes, we impose relevant restrictions on the prior distribution, as discussed in the following section.

	\subsection{The prior distribution}
	\label{subsec:Bayesian_SVEC_MSH_prior}
	
	Specification of our Bayesian two-state SVEC-MSH statistical model is completed with a prior distribution of the following structure (with specific densities of our choice provided later on):
	
	\begin{equation}
		p(\theta) \propto p(\alpha_*,\beta_*,\Gamma)p(b\vert \nu_b)p(\nu_b)p(\lambda_1, \lambda_2 \vert s^\lambda_1,s^\lambda_2)p(s^\lambda_1)p(s^\lambda_2)p(p_{11})p(p_{22}),    
		\label{eqn:prior}
	\end{equation}
	where the priors for $\alpha_*$ (i.e. the non-normalized matrix of adjustment coefficients), $\beta_*$ (the non-normalized matrix of cointegrating vectors) and $\Gamma$ are independent, but only up to the restriction on the spectral radius $\rho(\mathbf{A})$ of the companion matrix $\mathbf{A}=\mathbf{A}(\alpha_*,\beta_*,\Gamma)$ to be less or equal one, thereby preventing the explosiveness of the underlying VAR process (though still allowing for its non-stationarity):
	$$p(\alpha_*,\beta_*,\Gamma)\propto p(\alpha_*)p(\beta_*)p(\Gamma)\mathbb{I}(\rho(\mathbf{A})\leq 1),$$
	with $\mathbb{I}(\cdot)$ denoting the indicator function.
	
	Notice that, according to Eq.~(\ref{eqn:prior}), for the three: $b$, $\lambda_1$ and $\lambda_2$, which are the key parameters to model identification, we specify hierarchical priors, with the approach commonly recognized for its general merits in terms of facilitation of Bayesian estimation. In particular, the joint prior distribution of the structural shocks' state-specific variances takes the form:
	\begin{equation}
		p(\lambda_1, \lambda_2\vert s^\lambda_1, s^\lambda_2) \propto \left(\prod_{m=1}^{2} \prod_{i=1}^{n} p(\lambda_{m,i}\vert s_{m,i}^{\lambda})p(s_{m,i}^{\lambda}) \right)
		\mathbb{I}(\lambda_{1,l}>\lambda_{2,l}) 
		\mathbb{I}(\omega_2\in \mathcal{U}^n),
		\label{eqn:prior_lambda}
	\end{equation}
	where $s^\lambda_m=(s^\lambda_{m,1},s^\lambda_{m,2},\dots,s^\lambda_{m,n})$ collects the hyperparameters for state $m\in\{1, 2\}$, and $\mathcal{U}^n=\left\{(x_1, x_2, ..., x_n)'\in \mathbb{R}^{n}: x_1<x_2<\dots<x_n \right\}$ so that the expression $\mathbb{I}( \omega_2\in \mathcal{U}^n)$ imposes a uniqueness restriction of the form: $\omega_{2,1}<\omega_{2,2}<\dots,<\omega_{2,n}$. Note, however, that from a formal perspective any other order defining the $\mathcal{U}^n$ set (and thus, the form of the inequality between $\omega_{2,1}, ..., \omega_{2,n}$) is equally acceptable, leaving any specific choice of the ordering to be resolved empirically in a 'data-driven' way. The other indicator function term in Eq.~(\ref{eqn:prior_lambda}), i.e. $\mathbb{I}\left(\lambda_{1,l}>\lambda_{2,l} \right)$ for a given $l\in\{1, 2, ..., n\}$ (a particular choice of which remains of empirical nature), intends to ensure the regime identification, thereby preventing the label switching of structural variances. Finally, notice that both restrictions introduce some dependence between the otherwise conditionally (given the hyperparameters) a priori independent variances.
	
	With details deferred to Section 1 of the Supplementary material, we only notice here that specific prior densities of our choice are the ones typically employed in VAR/VEC-MSH modeling, with their conditional conjugacy (except for $b$) enabling a use of a Gibbs sampler for posterior simulation (see Section 2 of the Supplementary material for a detailed presentation of the MCMC algorithm combining the Gibbs sampler with a Metropolis-Hastings step for sampling $b$). Specifically, the priors for the cointegration parameters, $\alpha_*$ and $\beta_*$, are standard in Bayesian cointegration literature (see \citet{koop2009efficient}), and have earlier been used also by \citet{jochmann2015regime} in reduced-form VEC-MSH models. A matrix normal prior distribution for the remaining VAR parameters, collected in $\Gamma$, is also a typical choice, with $\Omega_\Gamma$ admitting an optional introduction of some form of shrinkage to facilitate parsimony of the VAR lags structure (we apply the approach in Section \ref{sec:Empirical} of this paper as well as in Section 3 of the Supplementary material; see also \citet{jochmann2015regime}, \citet{koop2009efficient}, \citet{lutkepohl2020bayesian}).
	
	\section{Empirical illustration}
	\label{sec:Empirical}
	
	\subsection{Data and model setting}
	\label{sec:data_model_setting}
	
	For an empirical illustration of our methodology, we use one of the models analyzed previously by \citet{velinov2013can} (inspired by the works of \citet{binswanger2004stock} and \citet{louis2010stock}) and later by \citet{lutkepohl2016structural}, who focus on SVAR-MSH and SVEC-MSH specifications derived for a dividend discount model (DDM) for US data, relating the expected future discounted payoffs (dividends) to the real economic activity quantities such as the real GDP, industrial production, company earnings etc., thus contributing to the line of research initiated by \citet{ball1968empirical}. The model of our choice is the one denoted as Model VI in \citet{binswanger2004stock}, \citet{louis2010stock}, Model IV in \citet{velinov2013can}, and Model II in \citet{lutkepohl2016structural}, and consists of $n=3$ variables: real earnings ($E_t$), real interest rates ($r_t$), and real stock prices ($s_t$), so that $y_t=(E_t, r_t, s_t)'$. Comprising only three variables, the model may arguably appear rather simple (if not simplistic). Nevertheless, it is still quite popular in asset pricing, and as such it would serve well for our illustrative purposes.
	
	Following \citet{velinov2013can} and \citet{lutkepohl2016structural}, we use data from Robert Schiller's website\footnote{http://www.econ.yale.edu/shiller/data.htm} for earnings, but also for stock prices proxied by S\&P500 quotations (note that S\&P500 is also used in the two cited works, but with the data drawn from the Federal Reserve Economic Database, FRED). Both series are available already in real terms on the page, deflated by the CPI inflation rate. For the interest rate, we take the effective federal funds rate (\% p.a.), with the data obtained from FRED\footnote{Board of Governors of the Federal Reserve System (US), Federal Funds Effective Rate [FEDFUNDS], retrieved from FRED, Federal Reserve Bank of St. Louis; https://fred.stlouisfed.org/series/FEDFUNDS, July 25, 2022.}, deflated by using the CPI inflation rate. The data is quarterly (last business day of quarter), seasonally unadjusted, and in logs (except for the real interest rate), ranging from 1960:I to 2021:II, thus covering a series of the US economy recessions, including the global financial crisis of 2007-2009 and the outbreak of the Covid-19 pandemic. Note that the data range here differs markedly from the one in \citet{velinov2013can} and \citet{lutkepohl2016structural}, who use series from 1947:1 to 2012:I. Finally, to facilitate accounting for any potential seasonality not captured by seasonal dummies, we arbitrarily set $p=5$ lags in the underlying VAR model (thus, four lags in its VEC form), which necessitates sparing the first five data points available (i.e. from 1960:I to 1961:I) for the sake of initial conditions, which yields the actual sample's size of $T=241$. The specified number of lags is supported by a visual inspection of the ACF and PACF plots for the data (particularly the interest rate), and auxiliary examination of the residuals from non-Bayesian, homoskedastic VAR models of orders 1-6, with the results left unreported here for the sake of brevity.
	
	\subsection{General estimation results}
	\label{sec:general_estim_results}
	
	All results presented below are based on $500\,000$ MCMC draws, preceded by as many \textit{burn-in} iterations. The hyperparameters of the prior distributions are set the same as in the simulation study presented in the Supplementary material (see Section 3 therein). The states are identified through a restriction of the form:  $\lambda_{11}>\lambda_{21}$, i.e. the variance of the first structural shock in the first state is higher than the shock's variance in the second state: $Var(\varepsilon_{t1}|S_t=1, \theta)>Var(\varepsilon_{t1}|S_t=2, \theta)$. For unique identification of the model, the relative (to state 1) structural variances are subject to the constraint: $\omega_{2,1}<\omega_{2,2}<\omega_{2,3}$, with the choice of the ordering based on an ancillary examination of the posterior results obtained for a model without the constraint (yet retaining the state-identification restriction), strongly suggesting that particular order. Therefore, it appeared the most natural choice, remaining in tune with the data.
	
	For choosing the number of cointegration relationships, we resort to the Savage-Dickey density ratio (\textit{SDDR}; see, e.g., \citet{verdinelli1995computing}, \citet{mulder2022generalization}). The results (deferred to the Supplementary material; see Section 4.1 therein) indicate a huge superiority of the cointegrated models, with the specification featuring only one cointegration relation outperforming the others (although $r=2$ is a very close second), and therefore being selected here for further analysis.
	
	Figure~\ref{fig:data} presents the modeled data (with the initial conditions trimmed off) along with the posterior probabilities of the system residing in the first state (featured by a higher volatility of structural shocks), i.e. $Pr(S_t=1\vert y)$, $t=1,...,T$, and additionally, with indicated NBER-dated recessions of the US economy\footnote{https://www.nber.org/research/data/us-business-cycle-expansions-and-contractions}. The identification of the regimes is rather clear-cut, with apparent domination of the lower-volatility state, which also translates into its higher persistence ($Me(p_{22}\vert y)\approx 0.948$). Not surprisingly, the higher-volatility regime is of a more transient nature ($Me(p_{11}\vert y)\approx 0.784$, with a more diffuse posterior distribution, see panels~(a,b) in Figure~\ref{fig:lambdy_real}), largely corresponding to the NBER-dated contractions. Apparent additional switches to the first state, indicating a rise in the structural shocks' volatility, took place in 1987:I-IV and 2002:IV-2004:I. The former can be attributed to a relatively strong drop in stock prices accompanied by a relatively large increase in earnings (although both changes can only vaguely be discerned in Figure~\ref{fig:data}), while the latter -- to some abrupt changes in earnings. As can be inferred from the posterior distributions of $\lambda_{m,i}$, $m=1,\,2$, $i=1,\, 2,\, 3$ (see panels (c-h) in Figure~\ref{fig:lambdy_real}), all the three structural shocks are conditionally heteroskedastic, exhibiting markedly higher volatilities in the first regime, with the posterior medians of $\lambda_{1,i}$, $i=1,\, 2,\, 3$, exceeding the ones of $\lambda_{2,i}$ approximately by one to two orders of magnitude. This, in turn, leads to pronounced relative differences between these sets of parameters, reflected in the posterior probability of $\omega_{2,i}$, $i=1,\,2,\,3$, amassed entirely far from (and less than) unit. As noticed in Figure~\ref{fig:omegi_real}, the posterior distributions of the contrasts: $\omega_{2,2}-\omega_{2,1}$ and $\omega_{2,3}-\omega_{2,1}$, display a definitive separation from zero, thereby implying a very clear distinction between the first and the second shock. The results obtained for the last contrast, $\omega_{2,3}-\omega_{2,2}$, may appear somewhat less convincing due to a part of the posterior adhering to zero, the value of which, however, still falls beyond the 95\% highest posterior density interval. Moreover, the p-value of a Lindley-type test for $\omega_{2,3}-\omega_{2,2}=0$ equals 0.046 (with the test statistics at 3.986). Therefore, with a high posterior probability one may still assume that the second and the third shock are disentangled.
	
	With respect to the variables' instantaneous reactions to the structural shocks (see Figure 4 in the Supplementary material), it appears that only the stock prices' reaction to the first shock is significant (and positive), which will emerge highly intuitive in view of the economic identification of the shocks attempted below. On the other hand, the attention should be paid not only to the $B$ matrix itself, but also on the long-run matrix, $\Xi B$, which is key in this empirical example (see the following subsection).
	
	\subsection{Structural analyses and economic interpretation of the shocks}
	\label{sec:structural_analyses}
	
	Typically, in the model at hand, the identifying restrictions imposed on the long-run matrix assume the form (see, e.g., \citet{lutkepohl2016structural}):
	
	\begin{equation}
		\Xi B = \left(\begin{array}{ccc}\star&0&0\\\star&\star&0\\\star&\star&0\end{array}\right), 
		\label{eqn:restrictions}
	\end{equation}
	with the first two shocks identified as fundamental and related to the real earnings and the real interest rate, respectively (therefore, it is assumed that the second shock leaves the earnings unaffected in the long term). On the other hand, the third one is identified as non-fundamental (speculative), thereby of only a transient impact on the system. Below, we attempt to name the shocks identified through our statistical model, by means of typical structural analysis tools, and to assess the consistence of the results with the above-mentioned traditional long-run restrictions.
	
	Interestingly enough, pairing (\ref{eqn:restrictions}) with the impulse response functions (IRFs; see Figure~\ref{fig:IRF} as well as Figures 5-6 in Section 4.2 of the Supplementary material) indicates that the traditionally postulated restrictions are rejected by our empirical results, and are therefore of little use for the purpose of figuring out the economic meaning of the shocks\footnote{Notice, however, that the results seen in these figures do not need to be ordered column-wise in strict adherence with $\Xi B$ given by (\ref{eqn:restrictions}).}. Arguably, it could already put into question the very possibility of providing here sound economic interpretations for the otherwise statistically identified shocks (see, e.g., \citet{lutkepohl2020bayesian}). However, other structural analysis tools, including  forecast error variance decomposition (FEVD) and the point estimates of structural shocks, may still be of assistance.
	
	According to the FEVD results in Figure~\ref{fig:FEVD}, the first shock's contribution to  the earnings' forecast error variance is overwhelming both in the first and second state, which implies a fundamental role of the shock. Further, regarding the stock prices forecast error, it is interesting to observe a huge contribution of the third shock in the second (i.e. lower-volatility) regime, which indicates a speculative nature of the shock (see, e.g., \citeauthor{binswanger2004stock} (\citeyear{binswanger2004stock}, \citeyear{binswanger2004important}, \citeyear{binswanger2004G7}), \citet{louis2010stock}). Such a classification is further corroborated by a decreasing impact of the shock for forecasts of longer horizons (see, e.g., \citet{lee1998permanent}).
	
	Based on the above, we can identify the first two shocks as fundamental (to the real earnings and the real interest rate, respectively), while the third one -- as  speculative. This line of reasoning seems supported also by visual inspection of the shocks' point estimates, presented in Figure~\ref{fig:shocks}. The path for the second shock reveals the strongest swings at the turn of the 1970s and 1980s, which relates 'neatly' to the extremely sharp movements in the interest rate featuring that turbulent times in the US economy. This second type of fundamental impulses does not seem to contribute meaningfully to the forecast error variance of the stock prices (see Figure~\ref{fig:FEVD}), which is consistent with earlier findings by \citeauthor{lee1995response} (\citeyear{lee1995fundamentals}, \citeyear{lee1995response}, \citeyear{lee1998permanent}), and \citet{binswanger2004stock} (notice, however, that \citet{louis2010stock} point the opposite).
	
	As seen in Figure~\ref{fig:shocks}, the first shock manifests the most prominent values over the years 2008-2009, indicating their apparent association with the real earnings that suffered severe drops back then, followed soon by a swift rebound (see Figure~\ref{fig:data}). However, these strong movements in the earnings were accompanied by highly volatile, both downward and upward shifts in the interest rates (see Figure~\ref{fig:data}). In particular, a strong negative impulse of 2010, seen for the first shock in Figure~\ref{fig:shocks}, may be related to a sudden drop in the interest rate (see Figure~\ref{fig:data}). These findings effectively prevent us from a clear-cut classification of the first shock as one that strictly relates to the earnings only. Instead, it appears more fair to conclude that these two economic types of shocks: to the real earnings and the interest rate, are mixed up to some degree in this otherwise statistically identified shock.
	
	Finally, a rather irregular path of the third shock's estimates (see Figure~\ref{fig:shocks}) seems to resonate with its only speculative nature.
	
	Recall here that the traditional long-run zero restrictions are not met in our model. As argued by \citet{lutkepohl2016structural}, this may hinder valid IRF interpretations, thus adding further trouble in attempts to deliver sound and unique economic 'identities' of the shocks. The 68\% posterior quantile bands around the posterior medians of impulse response functions are displayed in Figure~\ref{fig:IRF}, with the blue and yellow bands corresponding to the first and second state, respectively (notice that the first- and second-state IRFs actually coincide up to a scaling, which is an artifact of only time-invariant autoregressive parameters, and thus the transmission mechanism in our model's specification). It appears that the first shock, identified earlier as fundamental and earnings-related, exerts a positive and lasting influence on both the earnings and stock prices, while remaining immaterial to the interest rate. Obviously, the result is intuitive (see also, e.g., \citet{lee1998permanent}).
	
	The other of the two fundamental shocks, related to the interest rate, positively affects the rate permanently and throughout the horizons, while remaining neutral to the earnings for about 2.5 years, yet causing them decline slowly afterwards. The result seems economically valid and to generally agree with one's expectations.
	
	On the whole, despite the results breaching the traditional restrictions to some extent, the first two shocks can still be regarded as fundamental. The third one, on the other hand, may raise some doubts. As implied by Figure~\ref{fig:IRF}, a positive impulse exerts a sustained, flat-lining throughout the horizons, and positive influence on the stock market, possibly indicating a permanent discrepancy (bubble) between the stock prices and fundamentals (although such a hypothesis should be formally investigated and tested within a model incorporating also dividends; see, e.g., \citet{chung1998fundamental}). Concerns are raised also by the earnings' response to the shock, with no significant reaction for about 1.5 year, followed by a systematic increase, perpetuating also in the long run (see Figures 5-6 in Section 4.2 of the Supplementary material), which simply questions only a speculative nature of the shock. Therefore, it may be the case that the shock still includes some 'admixture' of the fundamental impulses. On that note, \citet{binswanger2004stock} points out that the contribution of fundamental shocks to forecast variance is underestimated in a model featuring earnings, and thus, including in the model some real activity variable(s) instead, such as GDP or industrial production, could deliver more conclusive results.

	\section{Conclusions}
	\label{sec:Conclusion}
	
	In the paper, we developed a Bayesian framework for estimation and inference in a two-state structural VEC model with Markov-switching heteroskedasticity. In the process, we revisited the identification conditions stated previously in Theorem 1 by \citet{lutkepohl2020bayesian} in the context of (Bayesian) stationary VARs, to find them warranting only a local identification. We pinpointed the lacking restriction and showed that for the global identification of structural impulses, the ordering of the relative variances across the system's variables needs to be fixed. 
	
	Results of the simulated-data-based study (see Section 3 of the Supplement material) indicate that disregarding the additional restraint may manifest in more or less pronounced irregularities in the marginal posteriors of key model parameters, with all of the former being removed once a relevant ordering restriction is imposed on $\omega_{2,1}, \omega_{2,2}, \ldots, \omega_{2,n}$. On the other hand, the empirical illustration (Section~\ref{sec:Empirical}) shows that introducing all relevant restrictions on the prior distribution does not necessarily translates into fully convincing and economically sound identification of shocks that are statistically identified nonetheless. Thereby, combining our current setup with conventional approaches to identification, much in the spirit of \citet{herwartz2014structural}, could prove remedial here.
	
	\appendix
	\section{Proof of Theorem~\ref{Theorem_ours}}
	\label{sec:appendixProof}
	
	The proof presented here is a modification of the ones delivered in  \citet{lanne2010structural} and \citet{lutkepohl2020bayesian}.
	
	Consider two symmetric and positive-definite $n\times n$ matrices, $\Sigma_1$ and $\Sigma_2$, and assume that there exist two observationally equivalent decompositions thereof:
	\begin{enumerate}
		\item $\Sigma_1 = B\Lambda_1B'$ i $\Sigma_2 = B\Lambda_2B'$,
		\item $\Sigma_1 = \tilde{B}\tilde{\Lambda}_1\tilde{B}'$ i $\Sigma_2 = \tilde{B}\tilde{\Lambda}_2\tilde{B}'$,
	\end{enumerate}
	where $B$ and $\tilde{B}$ are two non-singular $n\times n$ matrices with ones on their main diagonals, $diag(B)=diag(\tilde{B})=(1, 1,\ldots, 1)$, whereas $\Lambda_{1}$, $\Lambda_{2}$, $\tilde{\Lambda}_{1}$ and $\tilde{\Lambda}_{2}$ are some diagonal matrices with positive diagonal entries, $diag(\Lambda_{i})=(\lambda_{i,1}, \lambda_{i,2},\ldots, \lambda_{i,n})$ and $diag(\tilde{\Lambda}_{i})=(\tilde{\lambda}_{i,1}, \tilde{\lambda}_{i,2},\ldots, \tilde{\lambda}_{i,n})$, $i=1, 2$.
	
	Now, define matrix $C=B\Lambda_1^{1/2}$. Then, for the first decomposition, we have $\Sigma_1 = CC'$ and $\Sigma_2 = C\Lambda_1^{-1/2}\Lambda_2\Lambda_1^{-1/2}C' = C\Omega_2C'$, where $\Omega_2 = \Lambda_2\Lambda_1^{-1} = diag(\lambda_{2,1}/\lambda_{1,1}, \lambda_{2,2}/\lambda_{1,2}, \dots, \lambda_{2,n}/\lambda_{1,n})$, so that $diag(\Omega_2) = \omega_2$. Similarly, for the alternative representation of $\Sigma_1$ and $\Sigma_2$, we define $\tilde{C}=\tilde{B}\tilde{\Lambda}_1^{1/2}$, and obtain $\Sigma_1 = \tilde{C}\tilde{C}'$ and $\Sigma_2 = \tilde{C}\tilde{\Omega}_2\tilde{C}'$, where $\tilde{\Omega}_2 = \tilde{\Lambda}_2\tilde{\Lambda}_1^{-1}$, so that $diag({\tilde\Omega}_2) = \tilde{\omega}_2$.
	
	There exists a non-singular $n\times n$ matrix $Q$ such that $\tilde{C} = CQ$. Pairing the two decompositions for $\Sigma_1$, we get $CC' = \tilde{C}\tilde{C}' = CQQ'C'$, so that $QQ' = I_n$, which indicates that $Q$ is orthonormal.
	
	From the decompositions of $\Sigma_2$, we obtain the identity: $C\Omega_2 C' = \tilde{C}\tilde{\Omega}_2\tilde{C}' = CQ\tilde{\Omega}_2Q'C'$, resulting in $\Omega_2 = Q\tilde{\Omega}_2'Q'$, and so $\Omega_2^{-1}Q\tilde{\Omega}_2 = Q$. The equality of the two matrices follows therefrom:
	$$\left(\begin{array}{cccc}
		q_{11}\frac{\tilde{\omega}_{2,1}}{\omega_{2,1}}&q_{12}\frac{\tilde{\omega}_{2,2}}{\omega_{2,1}}&\cdots&q_{1n}\frac{\tilde{\omega}_{2,n}}{\omega_{2,1}}\\
		q_{21}\frac{\tilde{\omega}_{2,1}}{\omega_{2,2}}&q_{22}\frac{\tilde{\omega}_{2,2}}{\omega_{2,2}}&\cdots&q_{1n}\frac{\tilde{\omega}_{2,n}}{\omega_{2,2}}\\
		\vdots&\vdots&\ddots&\vdots\\
		q_{n1}\frac{\tilde{\omega}_{2,1}}{\omega_{2,n}}&q_{n2}\frac{\tilde{\omega}_{2,2}}{\omega_{2,n}}&\cdots&q_{nn}\frac{\tilde{\omega}_{2,n}}{\omega_{2,n}}\\
	\end{array}\right) = 
	\left(\begin{array}{cccc}q_{11}&q_{12}&\cdots&q_{1n}\\q_{21}&q_{22}&\cdots&q_{2n}\\
		\vdots&\vdots&\ddots&\vdots\\q_{n1}&q_{n2}&\cdots&q_{nn}\end{array}\right),$$
	which gives us the system of equations ($i=1, 2,\ldots,n$):\\
	$\left\{\begin{array}{l}
		q_{1i}\frac{\tilde{\omega}_{2,i}}{\omega_{2,1}} = q_{1i}\\
		q_{2i}\frac{\tilde{\omega}_{2,i}}{\omega_{2,2}} = q_{2i}\\
		\vdots\\
		q_{ni}\frac{\tilde{\omega}_{2,i}}{\omega_{2,n}} = q_{ni}
	\end{array}\right. \Leftrightarrow
	\left\{\begin{array}{l}
		q_{1i}(\tilde{\omega}_{2,i}-\omega_{2,1}) = 0\\
		q_{2i}(\tilde{\omega}_{2,i}-\omega_{2,2}) = 0\\
		\vdots\\
		q_{ni}(\tilde{\omega}_{2,i}-\omega_{2,n}) = 0
	\end{array}\right. \Leftrightarrow\left\{\begin{array}{l}
		q_{1i} = 0\ \vee \ \tilde{\omega}_{2,i} = \omega_{2,1}, i = 1,2,\dots,n\\
		q_{2i} = 0\ \vee \ \tilde{\omega}_{2,i} = \omega_{2,2}, i = 1,2,\dots,n\\
		\vdots\\
		q_{ni} = 0\ \vee \ \tilde{\omega}_{2,i} = \omega_{2,n}, i = 1,2,\dots,n
	\end{array}\right.$\\

	Consider the $k^\text{th}$ row of the system, $k\in\{1,2,\dots,n\}$. Since $Q$ is non-singular, none of its rows can be a vector of zeros. Furthermore, it follows from the assumptions underlying Theorem~\ref{Theorem_ours} that the elements of $\tilde{\omega}_2$ are pairwise different. It follows therefrom, that $\exists! j\in\{1,2,\dots,n\}: \tilde{\omega}_{2,j} = \omega_{2,k}$ and $\forall i\in\{1,2,\dots,n\}\setminus\{j\}: q_{ki} = 0$. The orthonormality of $Q$ implies $\sum_{i=1}^nq_{ki}^2 = 1$. Therefore, $q_{kj}^2 = 1$, and thus, $q_{kj} = \pm 1$.
	
	According to the above, in each column and each row of $Q$ (which is non-singular and therefore, of pairwise different rows and columns), there is only one element equal either 1 or -1, with zeros elsewhere. The signs of these non-zero entries are determined in such a way as to ensure all the diagonal elements of $\tilde{C}$ to be, since $diag(\tilde{C}) = \left(\sqrt{\tilde{\lambda}_{1,1}},\sqrt{\tilde{\lambda}_{1,2}},\dots,\sqrt{\tilde{\lambda}_{1,n}}\right)$.
	
	The above reasoning leads to the conclusion that $Q$ shifts the order of the ratios of the shocks' first- and second-state variances. Thus, if the ordering of the ratios is fixed, then the only matrix meeting the conditions considered above is the identity matrix, $Q=I_n$, which ends the proof.

	\bibliographystyle{elsarticle-harv} 
	\bibliography{cas-refs}

\clearpage
	
\begin{figure}[h]
	\centering
	\includegraphics[width = \textwidth]{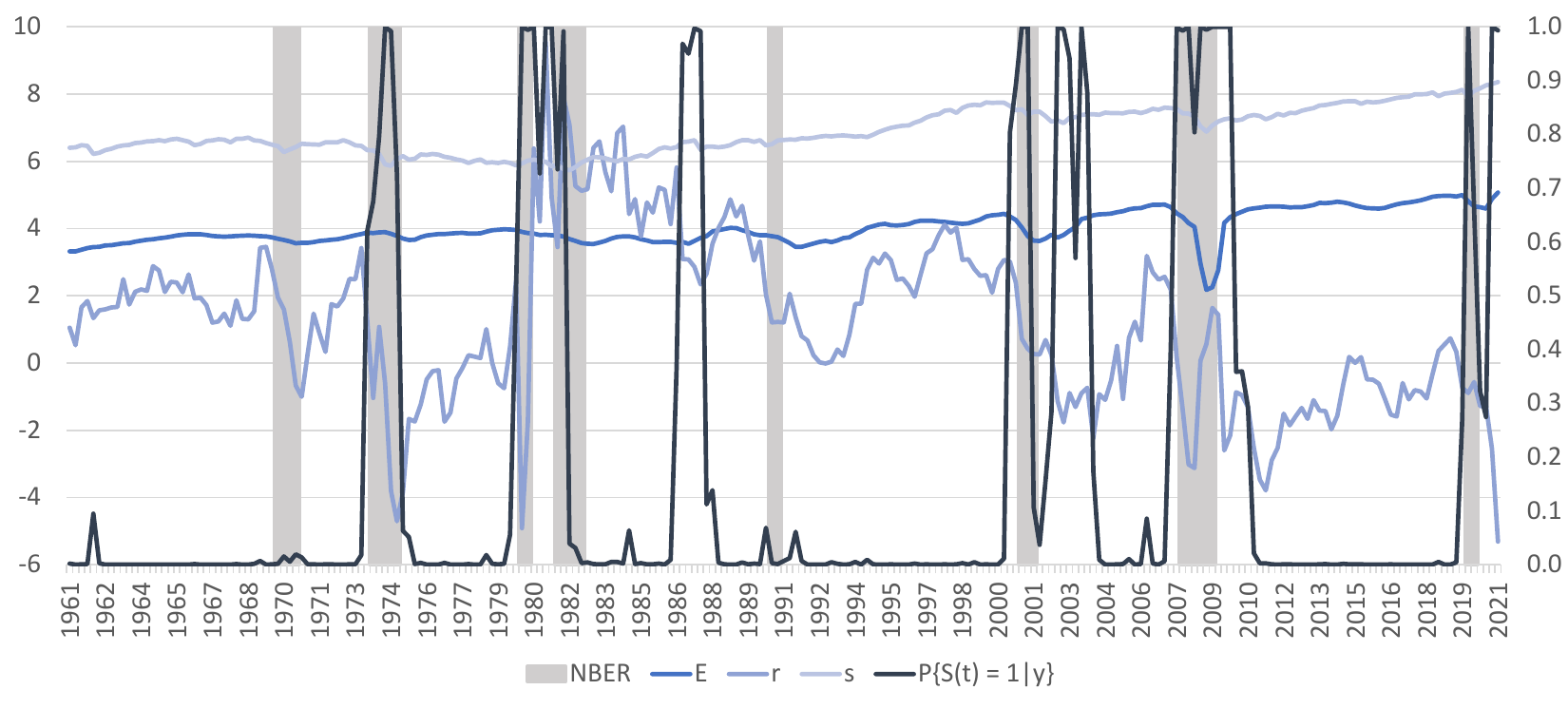}
	\caption{Modeled data, along with the posterior probabilities of the first state (the RHS axis) and NBER-dated recessions (grey shading).}
	\label{fig:data}
\end{figure}

\begin{figure}[h]
	\centering
	\begin{subfigure}[b]{0.4\textwidth}
		\centering
		\includegraphics[width=\textwidth]{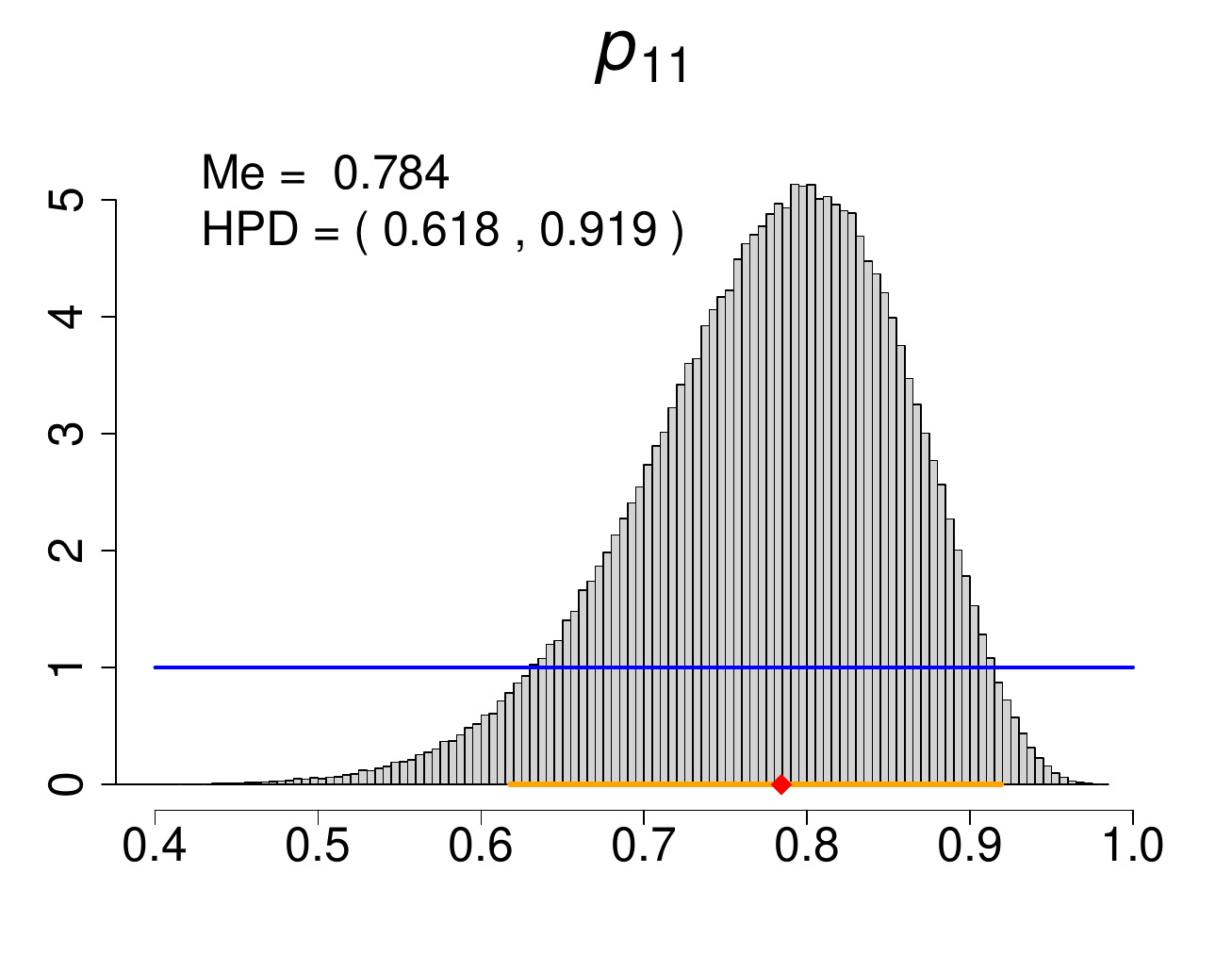}
		\caption{}
		\label{fig:real_p11}
	\end{subfigure}
	\hfill
	\begin{subfigure}[b]{0.4\textwidth}
		\centering
		\includegraphics[width=\textwidth]{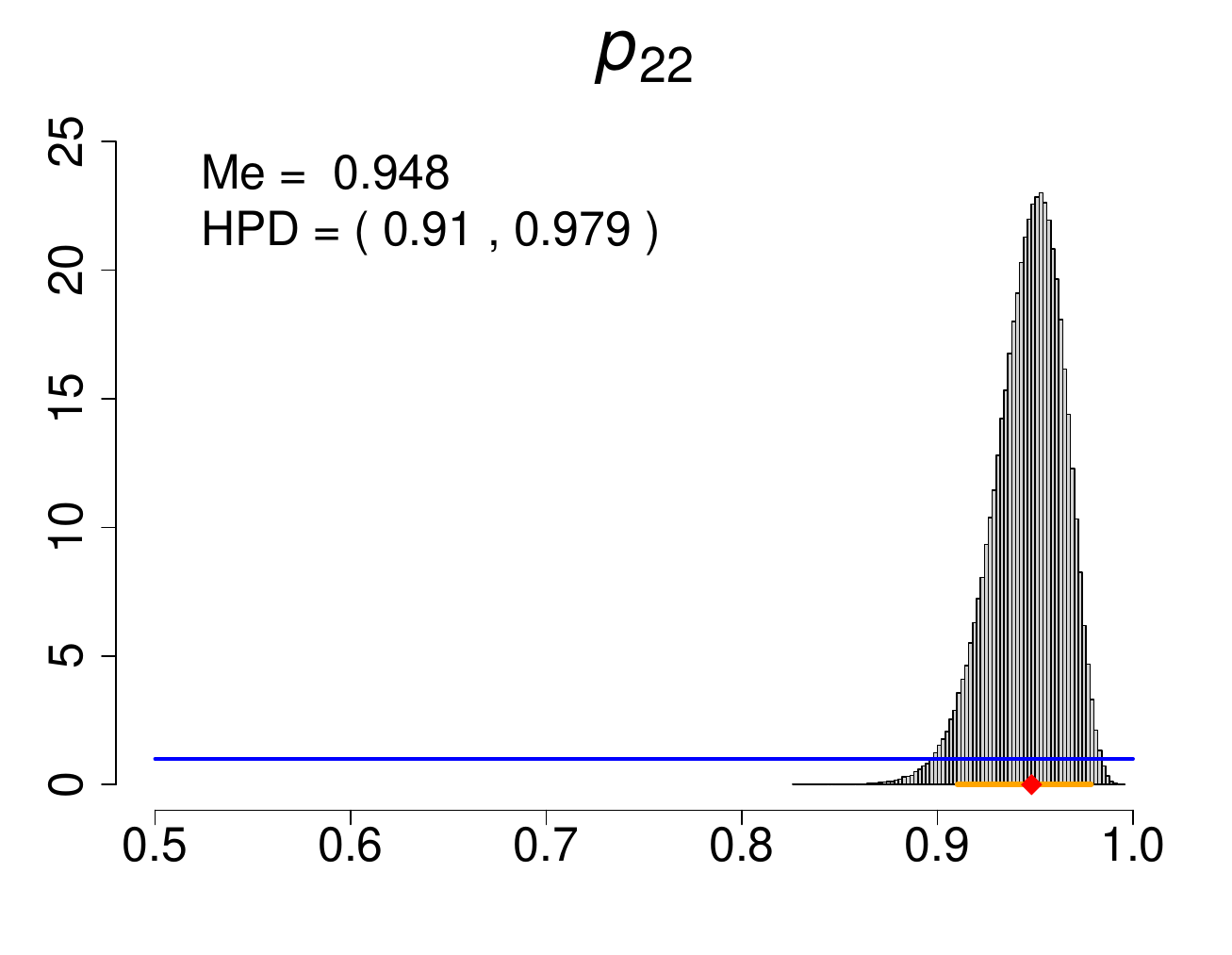}
		\caption{}
		\label{fig:real_p22}
	\end{subfigure}
	\vfill
	\begin{subfigure}[b]{0.4\textwidth}
		\centering
		\includegraphics[width=\textwidth]{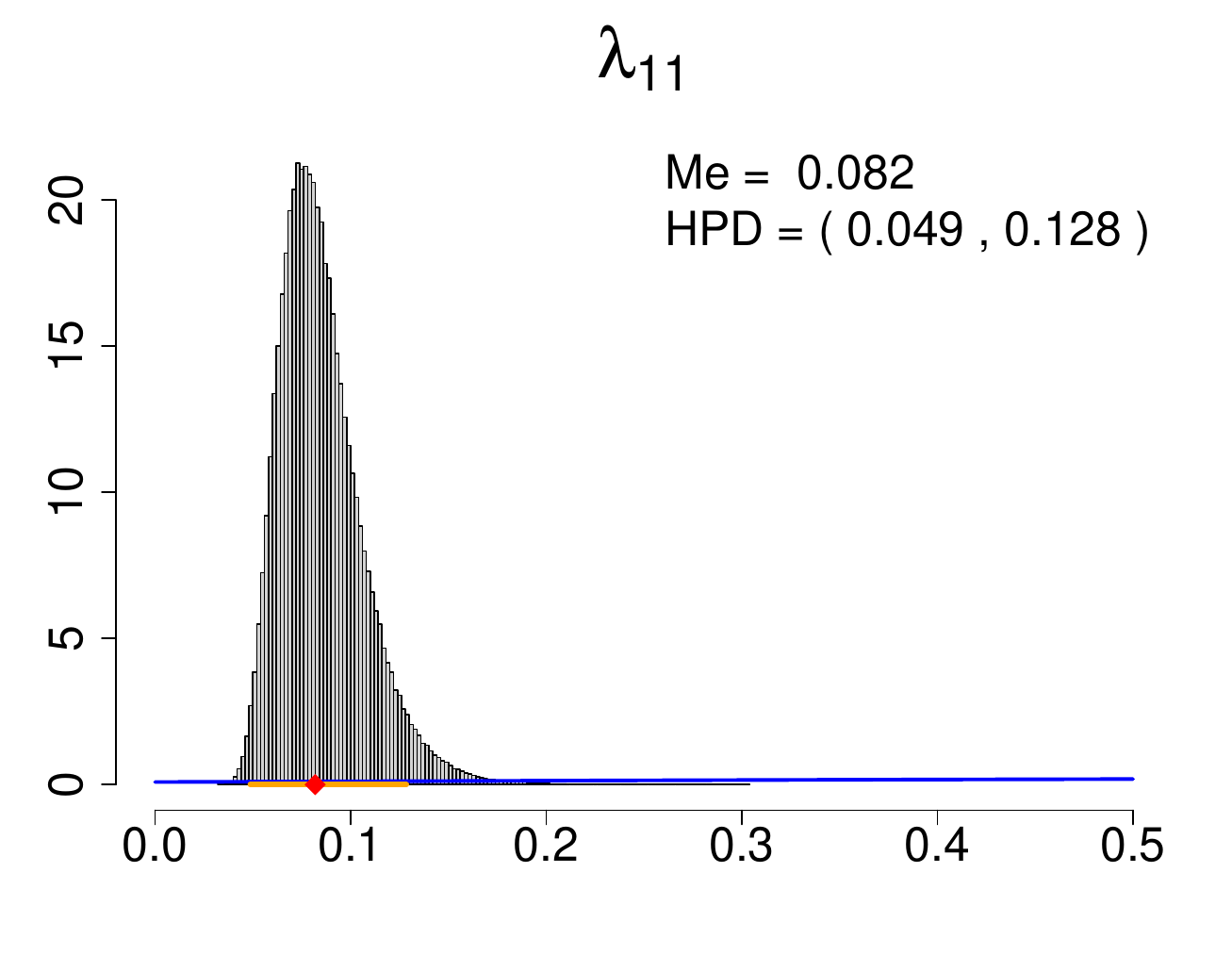}
		\caption{}
		\label{fig:real_l11}
	\end{subfigure}
	\hfill
	\begin{subfigure}[b]{0.4\textwidth}
		\centering
		\includegraphics[width=\textwidth]{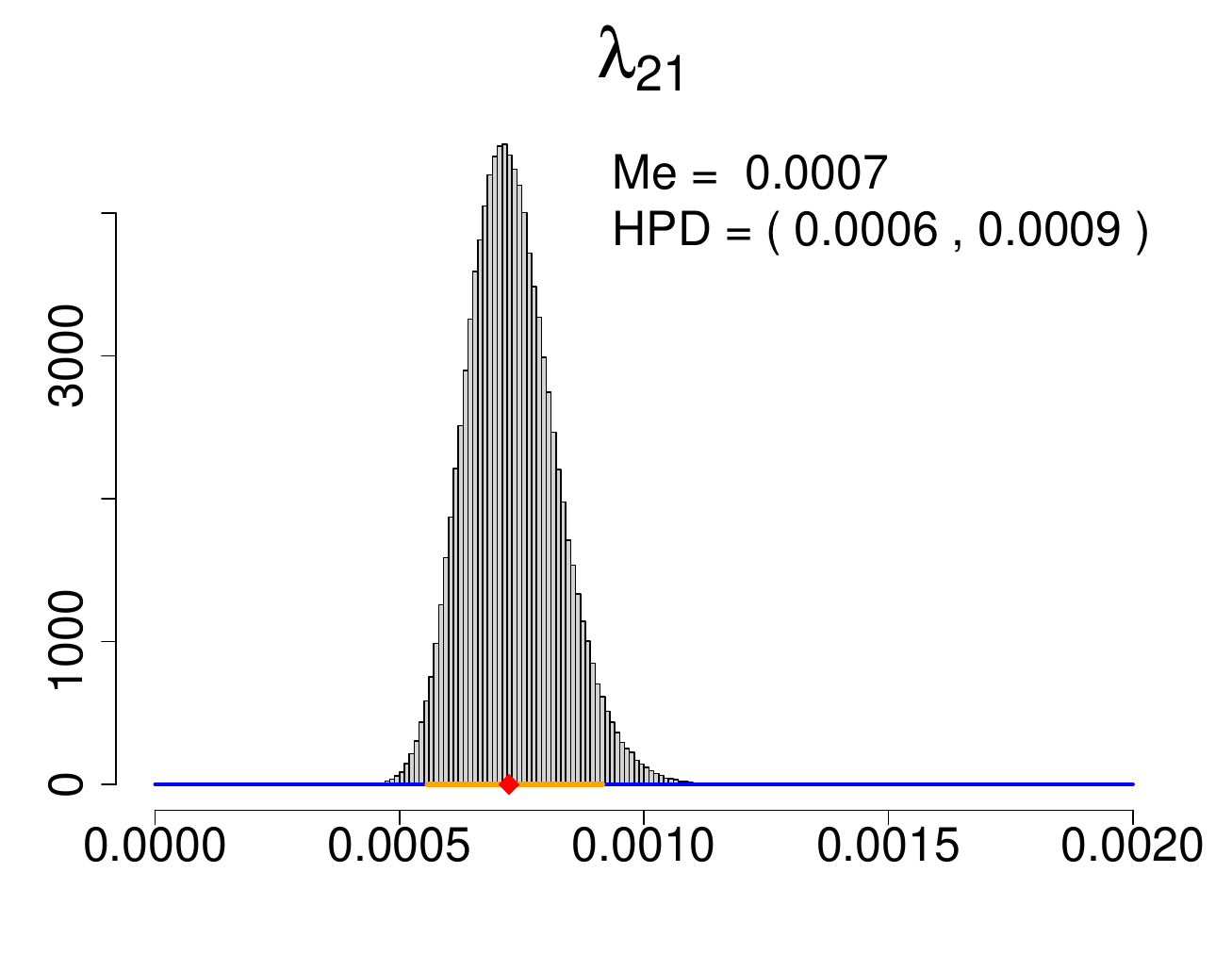}
		\caption{}
		\label{fig:real_l21}
	\end{subfigure}
	\vfill
	\begin{subfigure}[b]{0.4\textwidth}
		\centering
		\includegraphics[width=\textwidth]{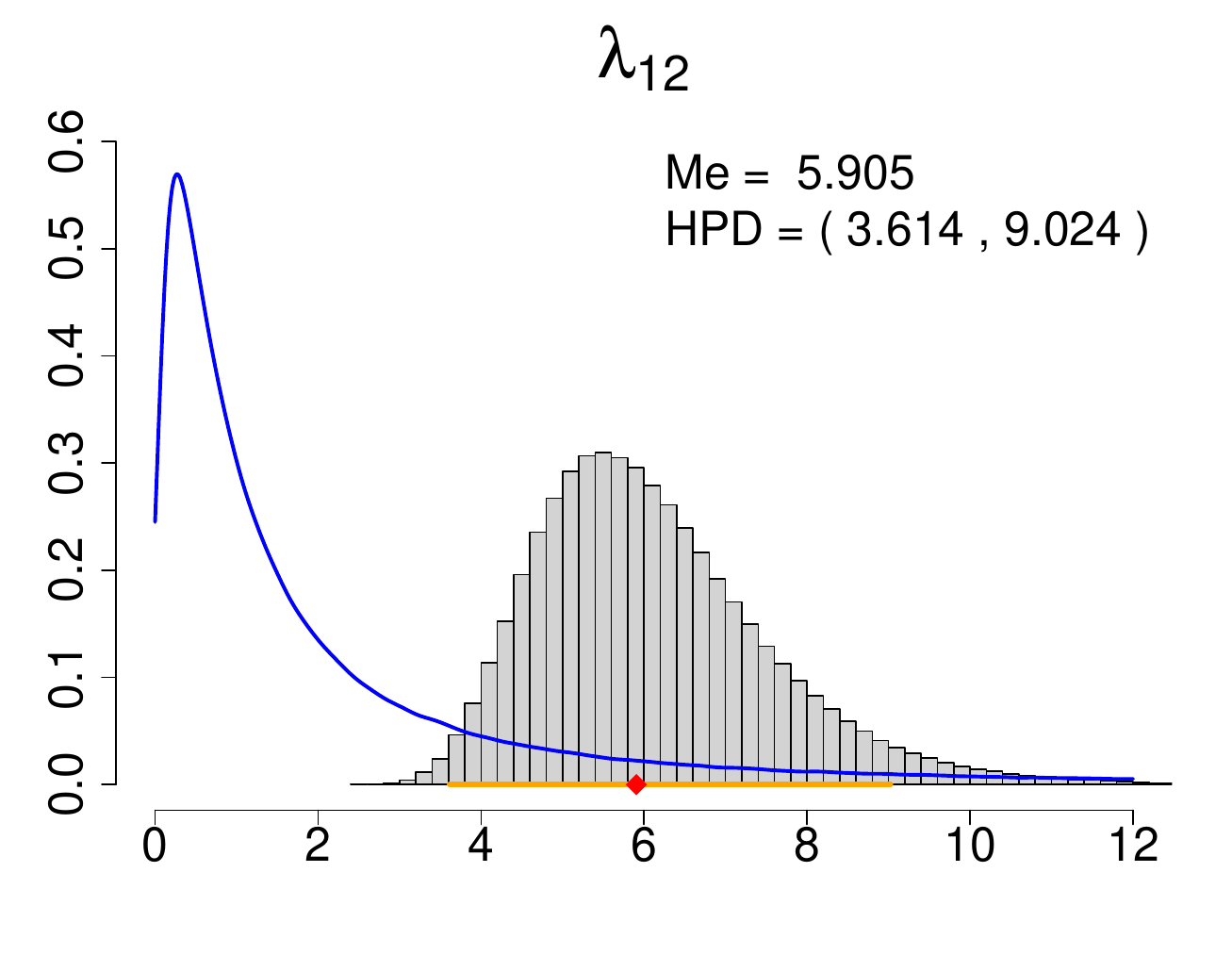}
		\caption{\tiny $HPD = \left(3.614, 9.024\right),\ Me=5.905$}
		\label{fig:real_l12}
	\end{subfigure}
	\hfill
	\begin{subfigure}[b]{0.4\textwidth}
		\centering
		\includegraphics[width=\textwidth]{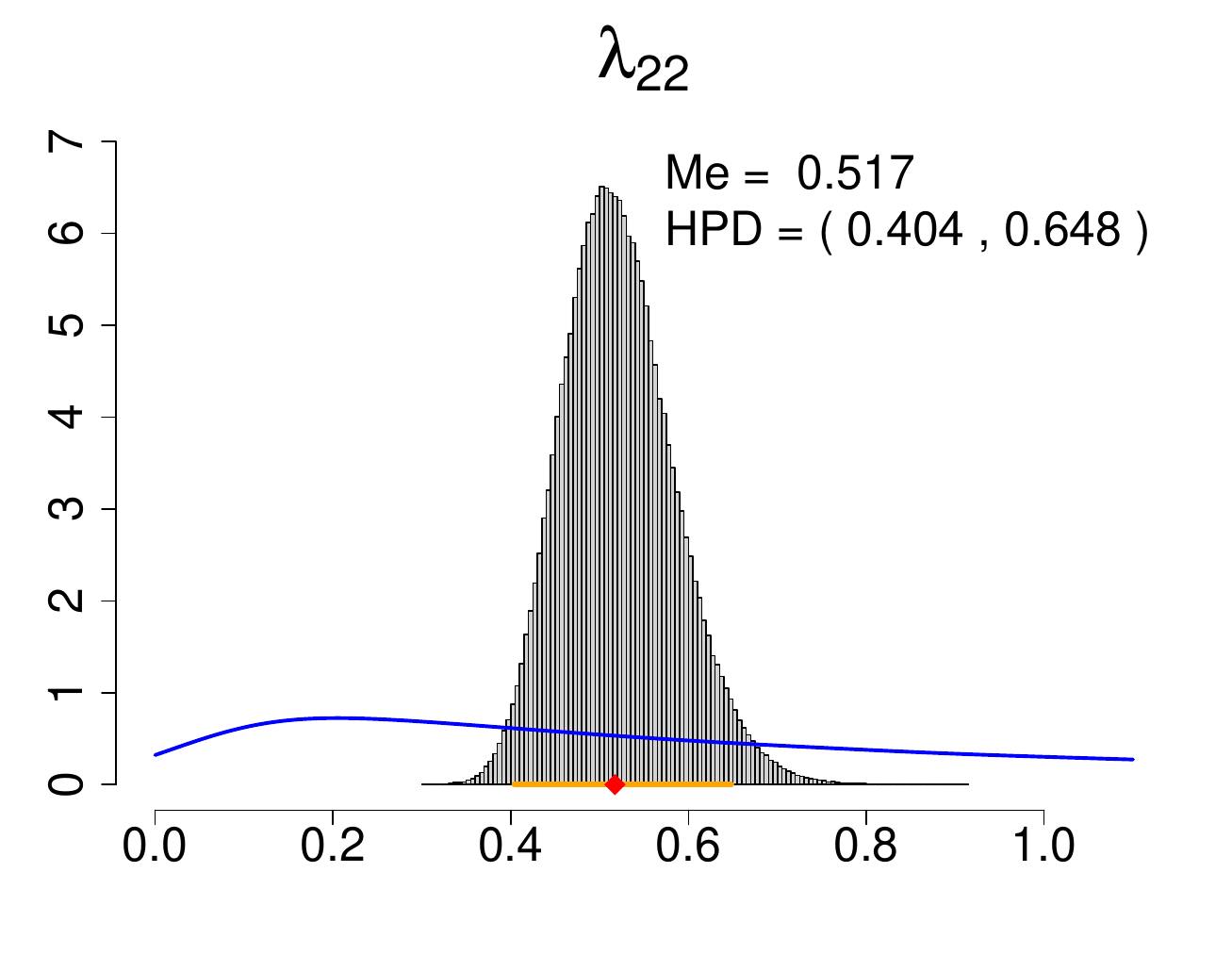}
		\caption{}
		\label{fig:real_l22}
	\end{subfigure}
	\vfill
	\begin{subfigure}[b]{0.4\textwidth}
		\centering
		\includegraphics[width=\textwidth]{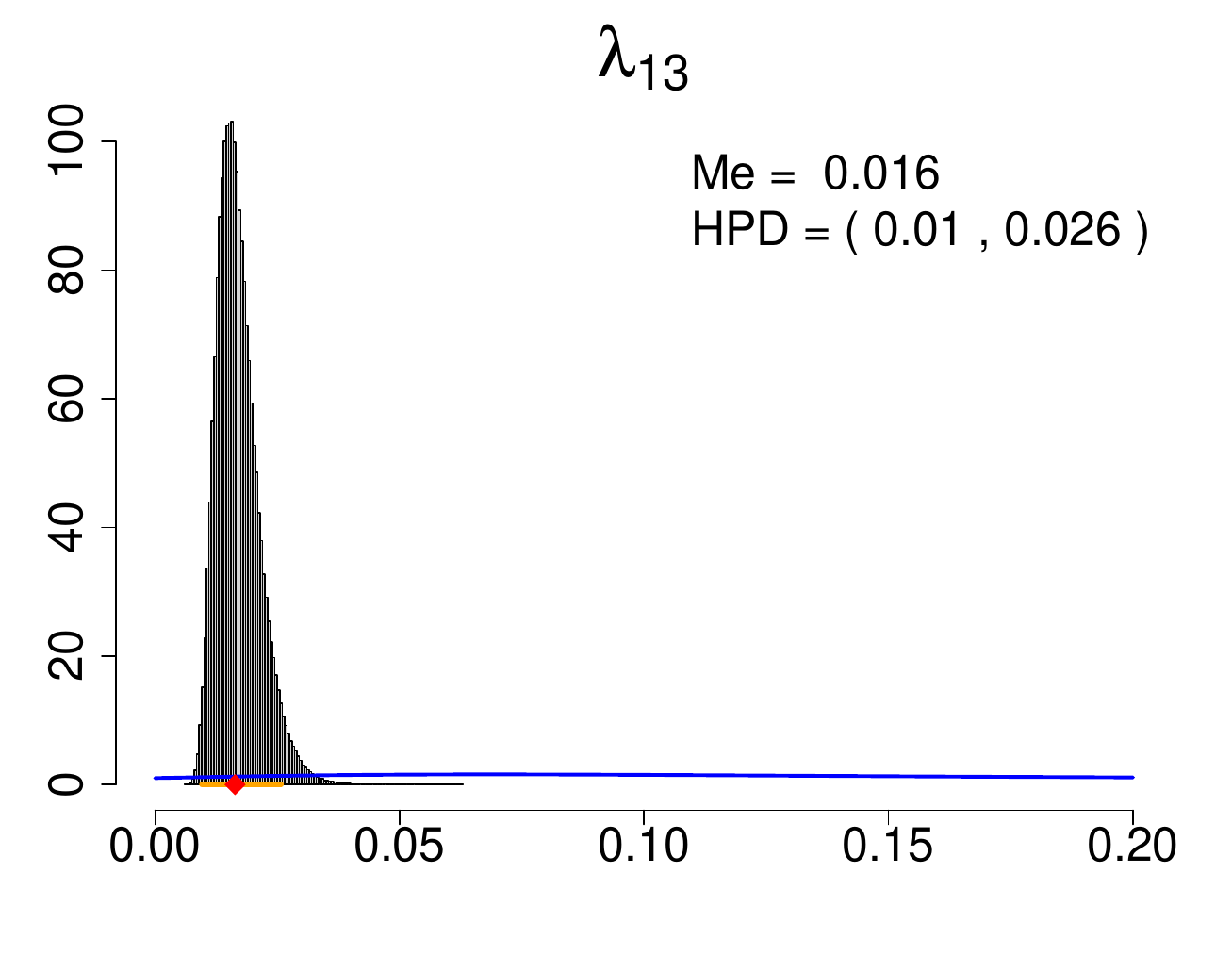}
		\caption{}
		\label{fig:real_l13}
	\end{subfigure}
	\hfill
	\begin{subfigure}[b]{0.4\textwidth}
		\centering
		\includegraphics[width=\textwidth]{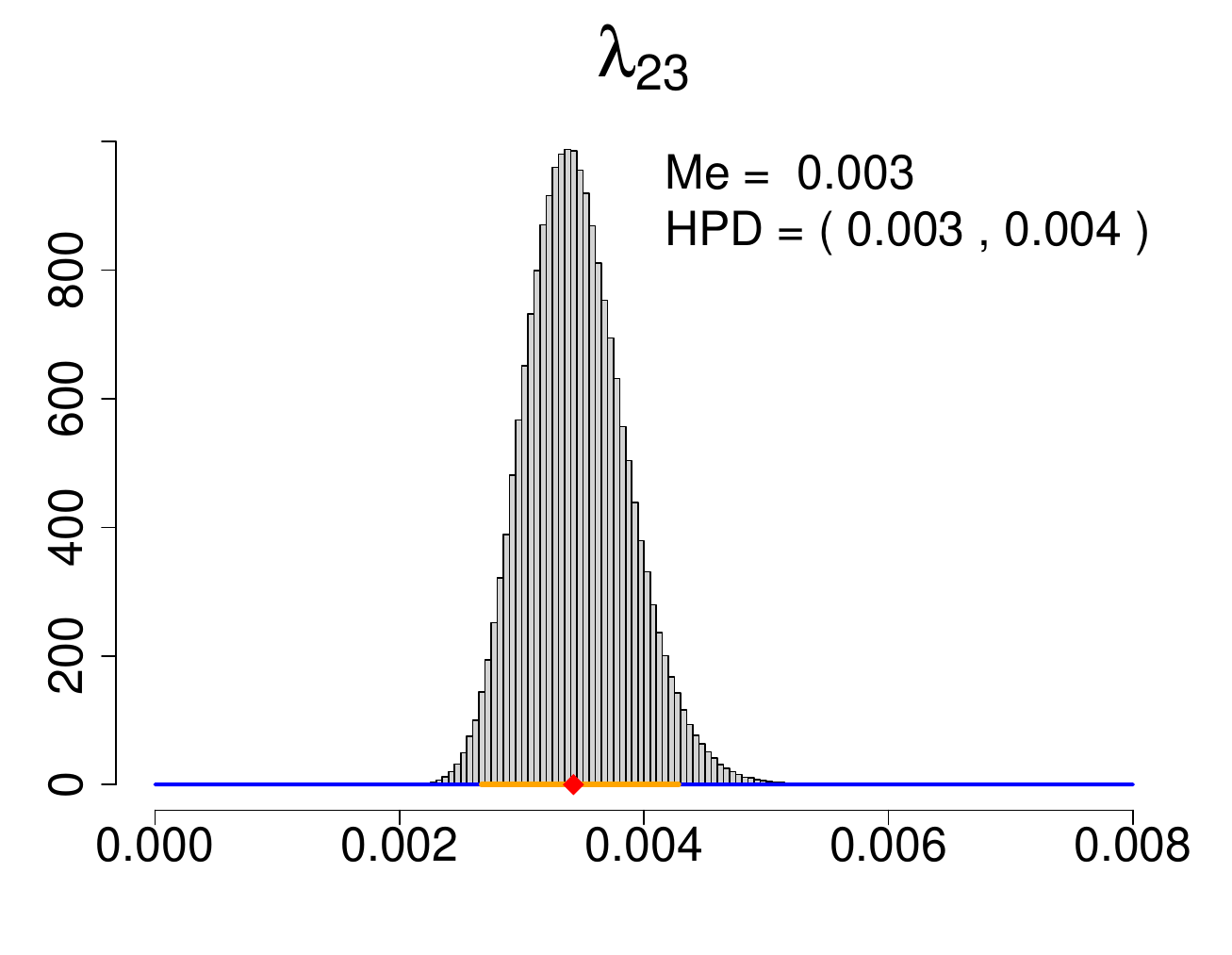}
		\caption{}
		\label{fig:real_l23}
	\end{subfigure}
	\caption{Posterior histograms and prior densities (lines) for the transition probabilities and $\lambda_{m,i}$ parameters in the empirical example, with the posterior medians (red markers) and 95\% HPD intervals (orange line).}
	\label{fig:lambdy_real}
\end{figure}

\begin{figure}[h]
	\centering
	\begin{subfigure}[b]{0.4\textwidth}
		\centering
		\includegraphics[width=\textwidth]{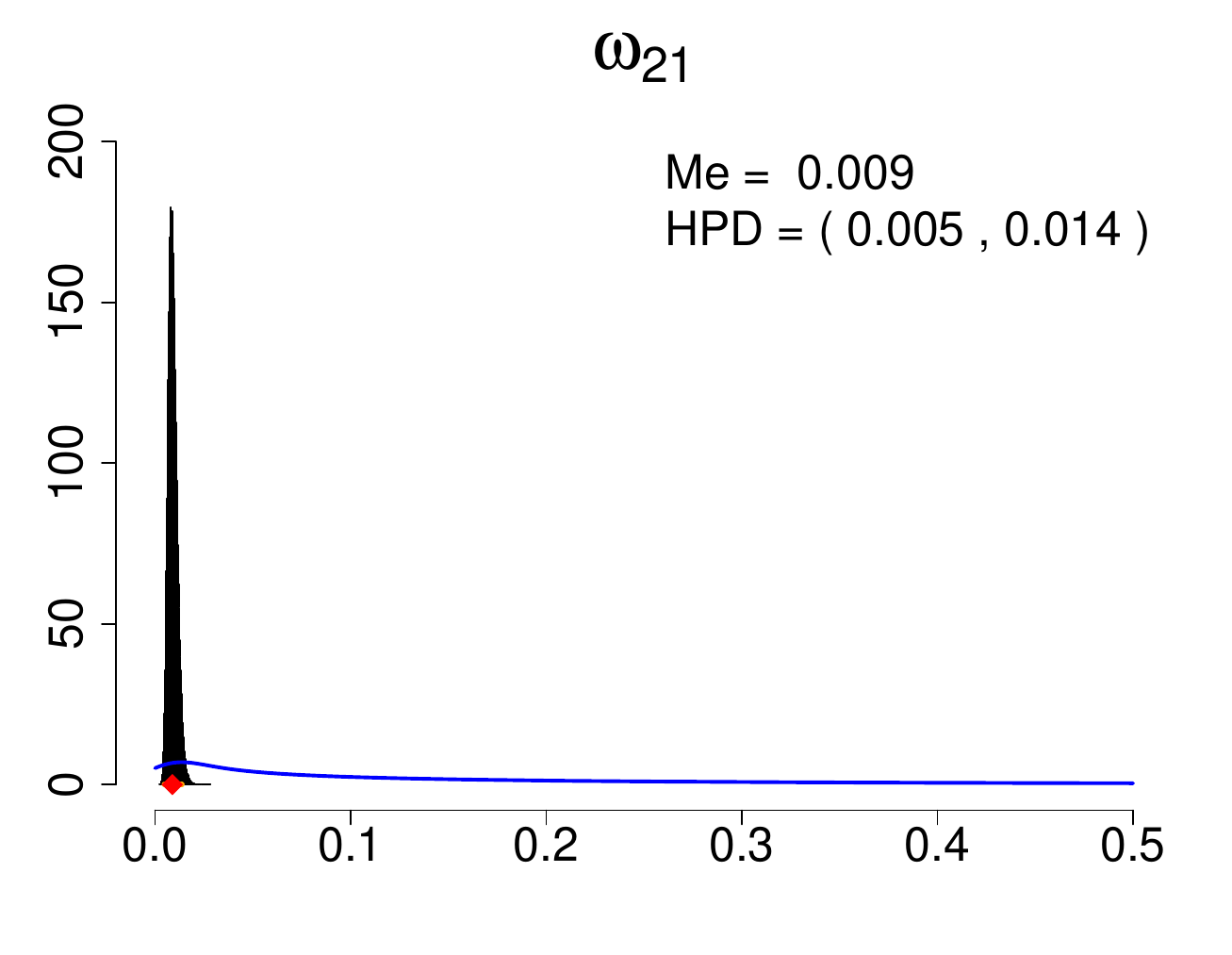}
		\caption{}
		\label{fig:real_o21}
	\end{subfigure}
	\hfill
	\begin{subfigure}[b]{0.4\textwidth}
		\centering
		\includegraphics[width=\textwidth]{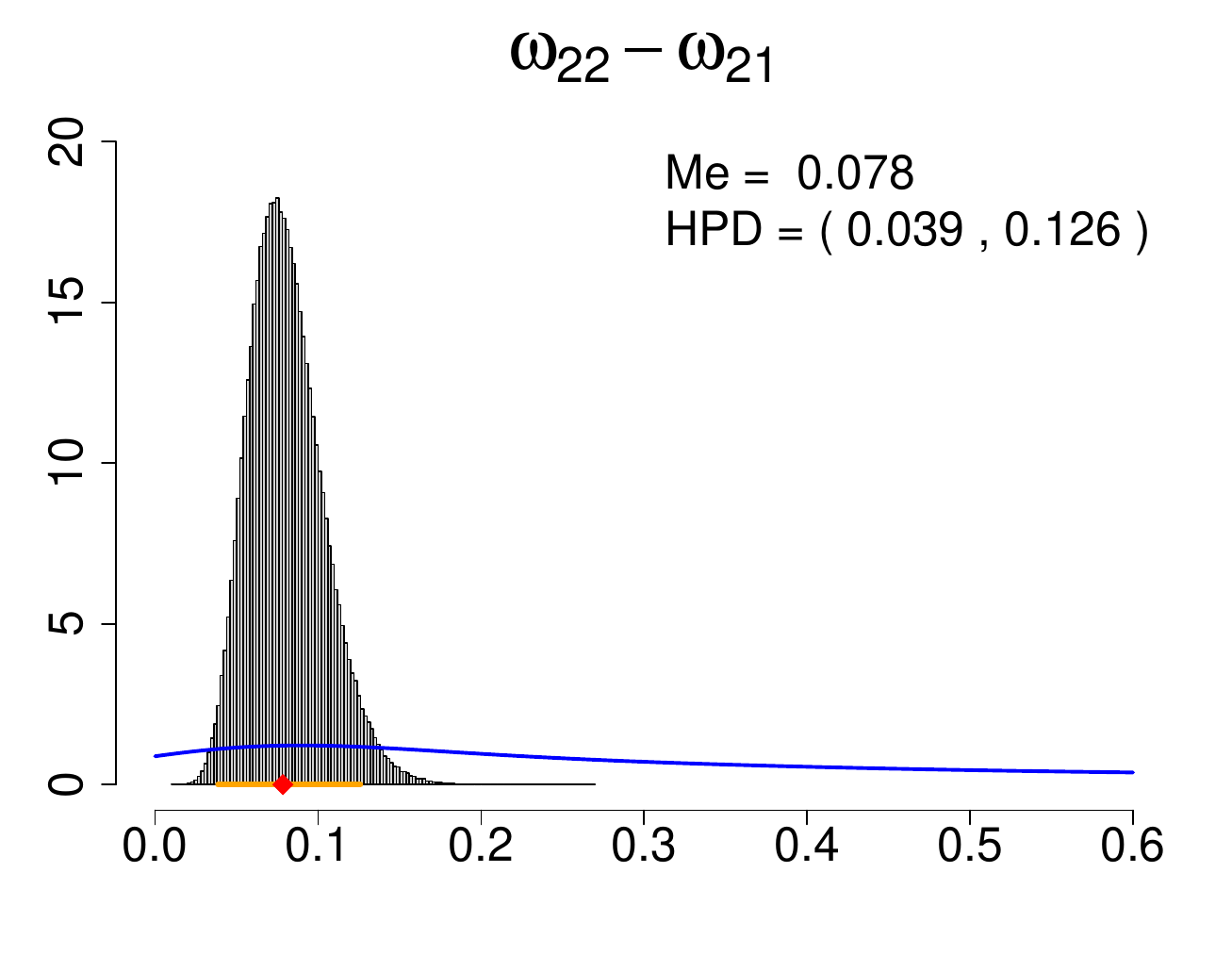}
		\caption{}
		\label{fig:real_c1}
	\end{subfigure}
	\vfill
	\begin{subfigure}[b]{0.4\textwidth}
		\centering
		\includegraphics[width=\textwidth]{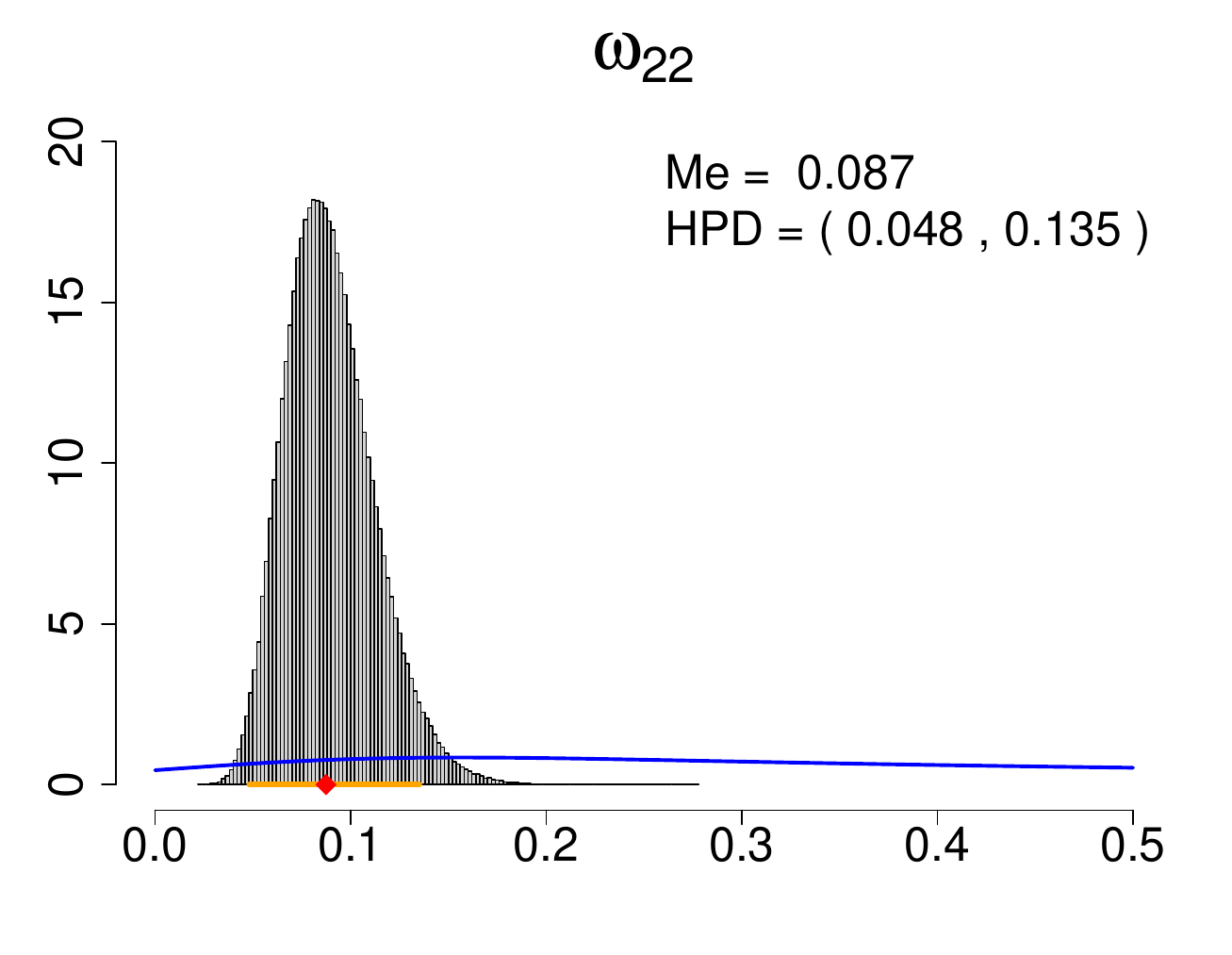}
		\caption{}
		\label{fig:real_o22}
	\end{subfigure}
	\hfill
	\begin{subfigure}[b]{0.4\textwidth}
		\centering
		\includegraphics[width=\textwidth]{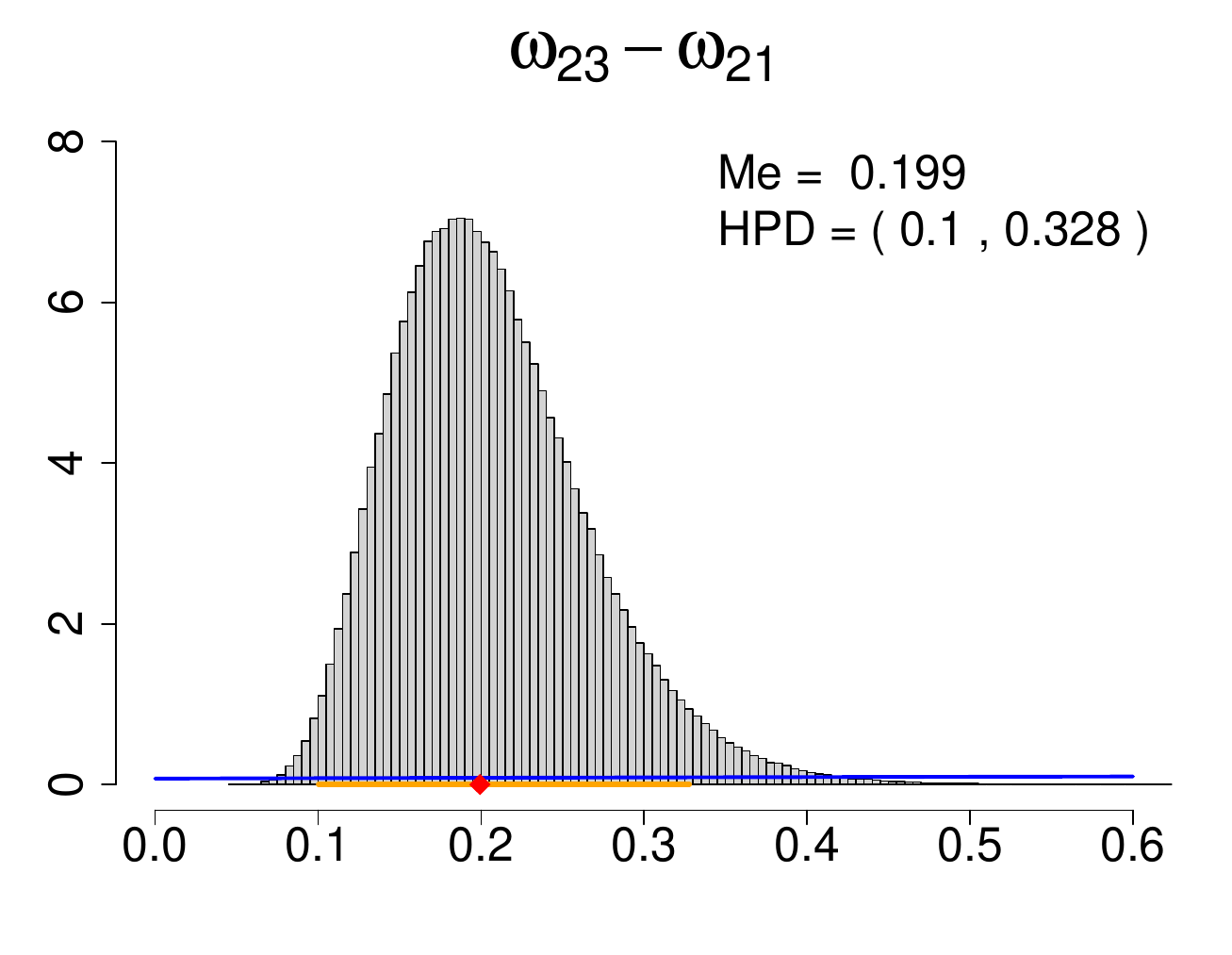}
		\caption{}
		\label{fig:real_c2}
	\end{subfigure}
	\vfill
	\begin{subfigure}[b]{0.4\textwidth}
		\centering
		\includegraphics[width=\textwidth]{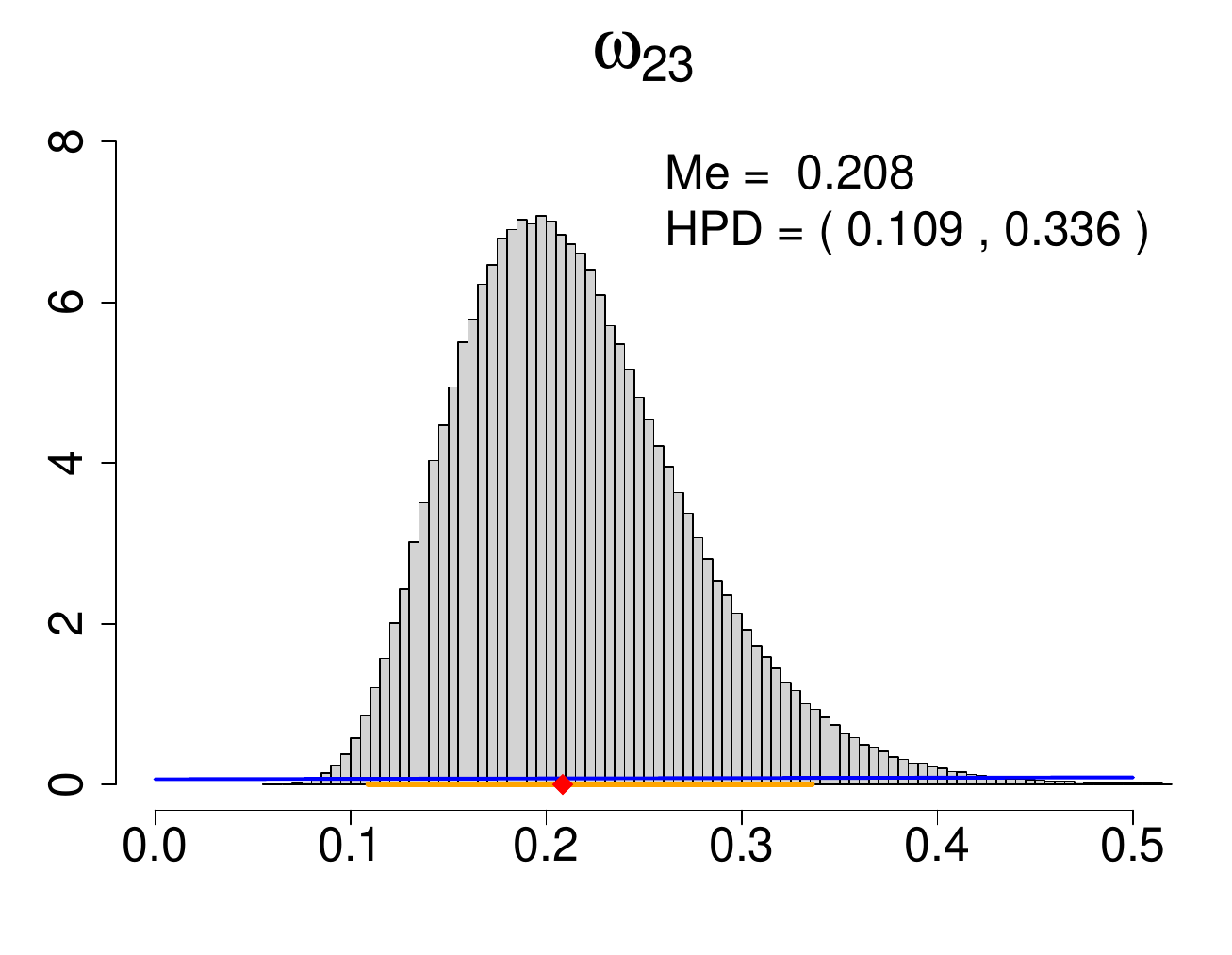}
		\caption{}
		\label{fig:real_o23}
	\end{subfigure}
	\hfill
	\begin{subfigure}[b]{0.4\textwidth}
		\centering
		\includegraphics[width=\textwidth]{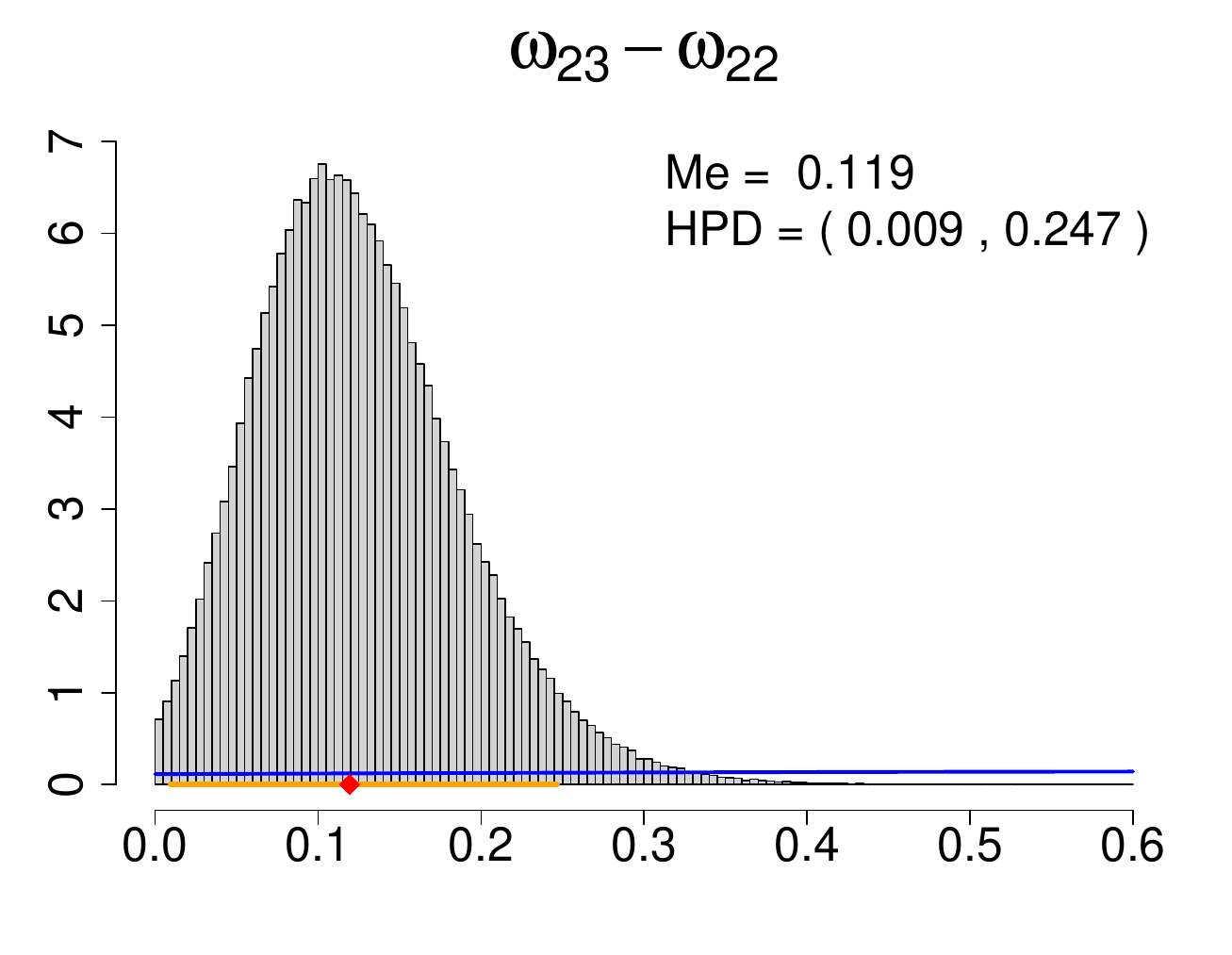}
		\caption{}
		\label{fig:real_c3}
	\end{subfigure}
	\caption{Posterior histograms and prior densities (lines) for the $\omega_{2,i}$ parameters and their contrasts in the empirical example, with the posterior medians (red markers) and 95\% HPD intervals (orange line).}
	\label{fig:omegi_real}
\end{figure}

\begin{figure}
	\begin{center}
		\begin{tabular}{ccc}
			{\small shock \#1 $\rightarrow$ $E$} &
			{\small shock \#2 $\rightarrow$ $E$} & 
			{\small shock \#3 $\rightarrow$ $E$} \\
			\includegraphics[width=0.33\textwidth]{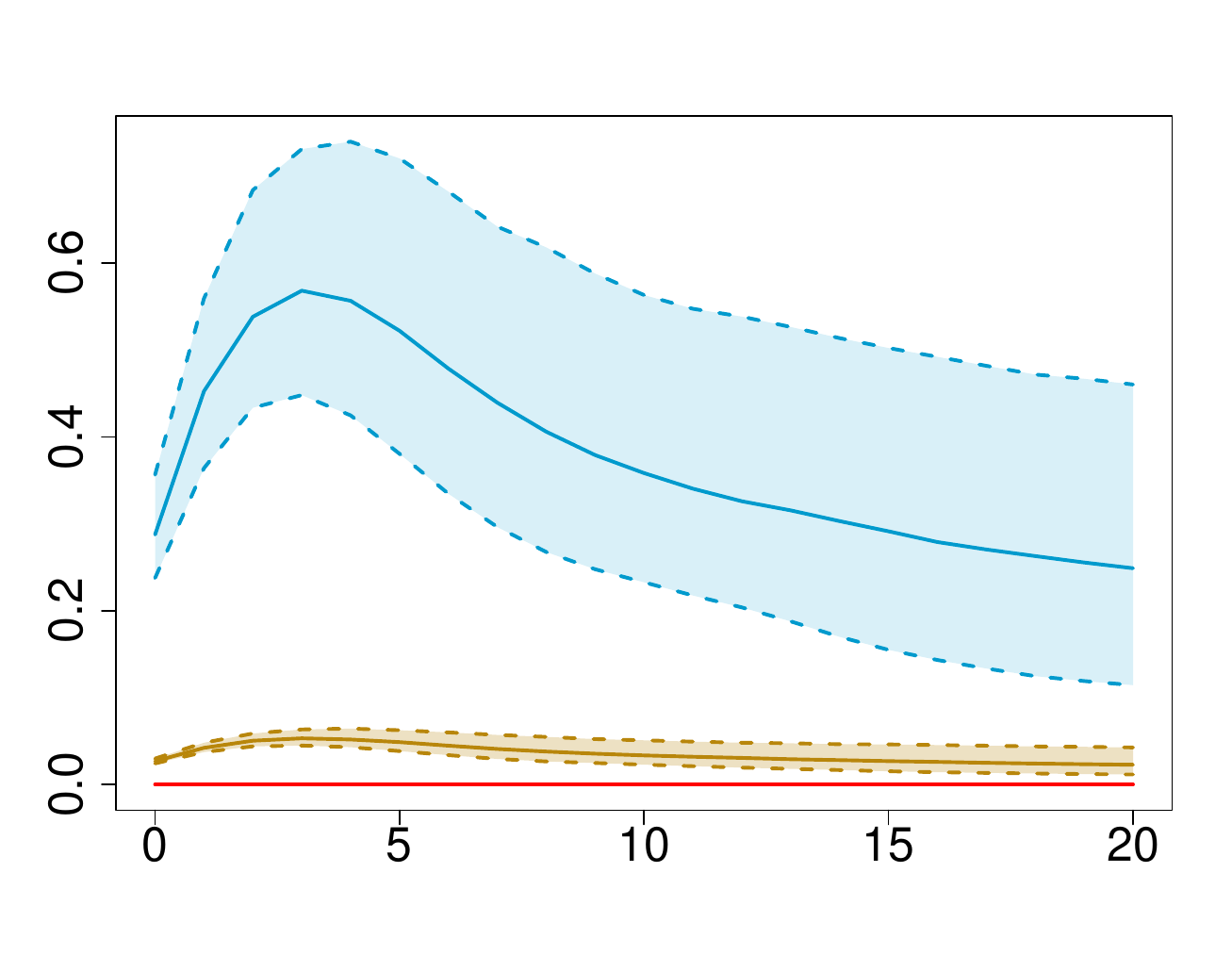} & \includegraphics[width=0.33\textwidth]{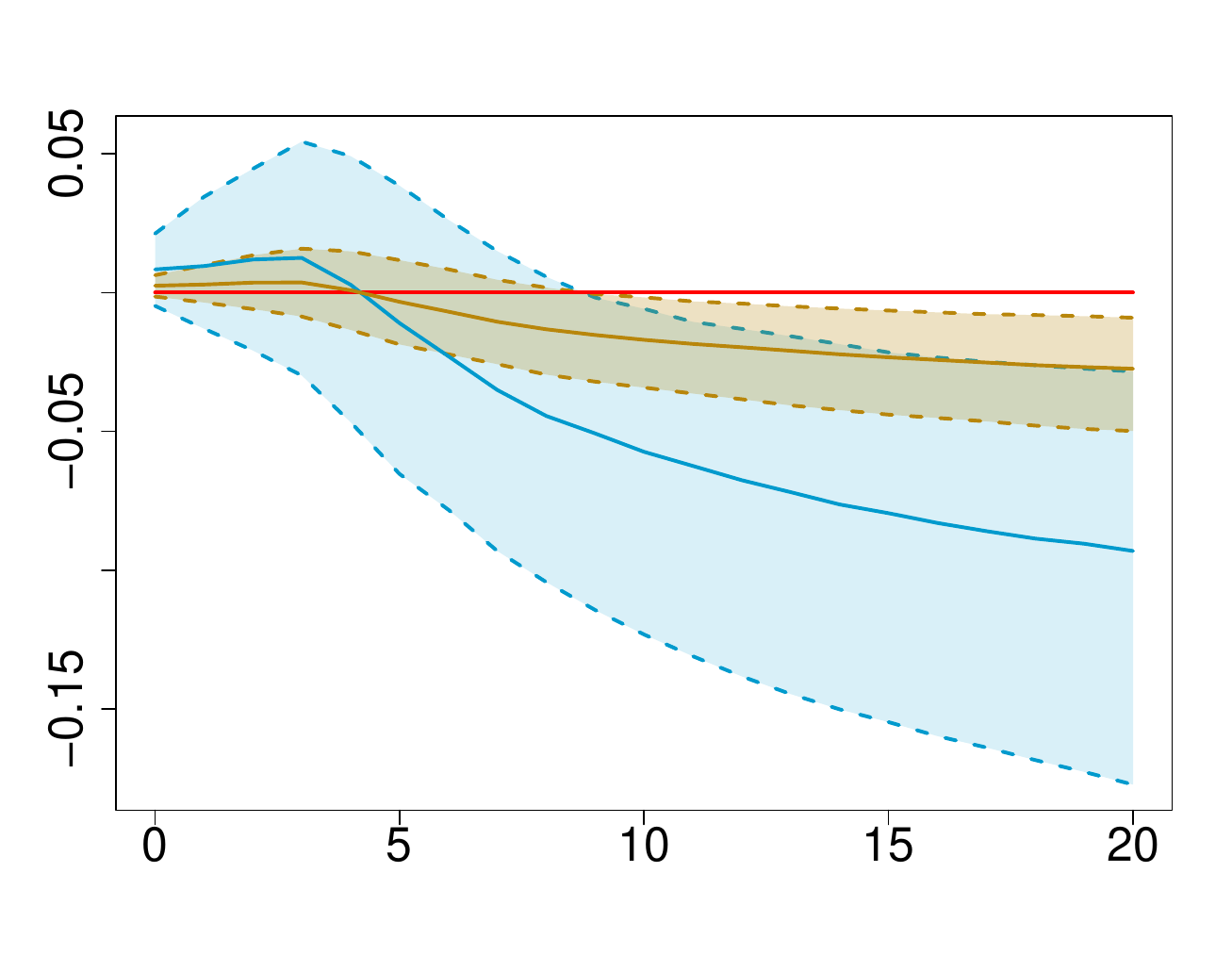} & \includegraphics[width=0.33\textwidth]{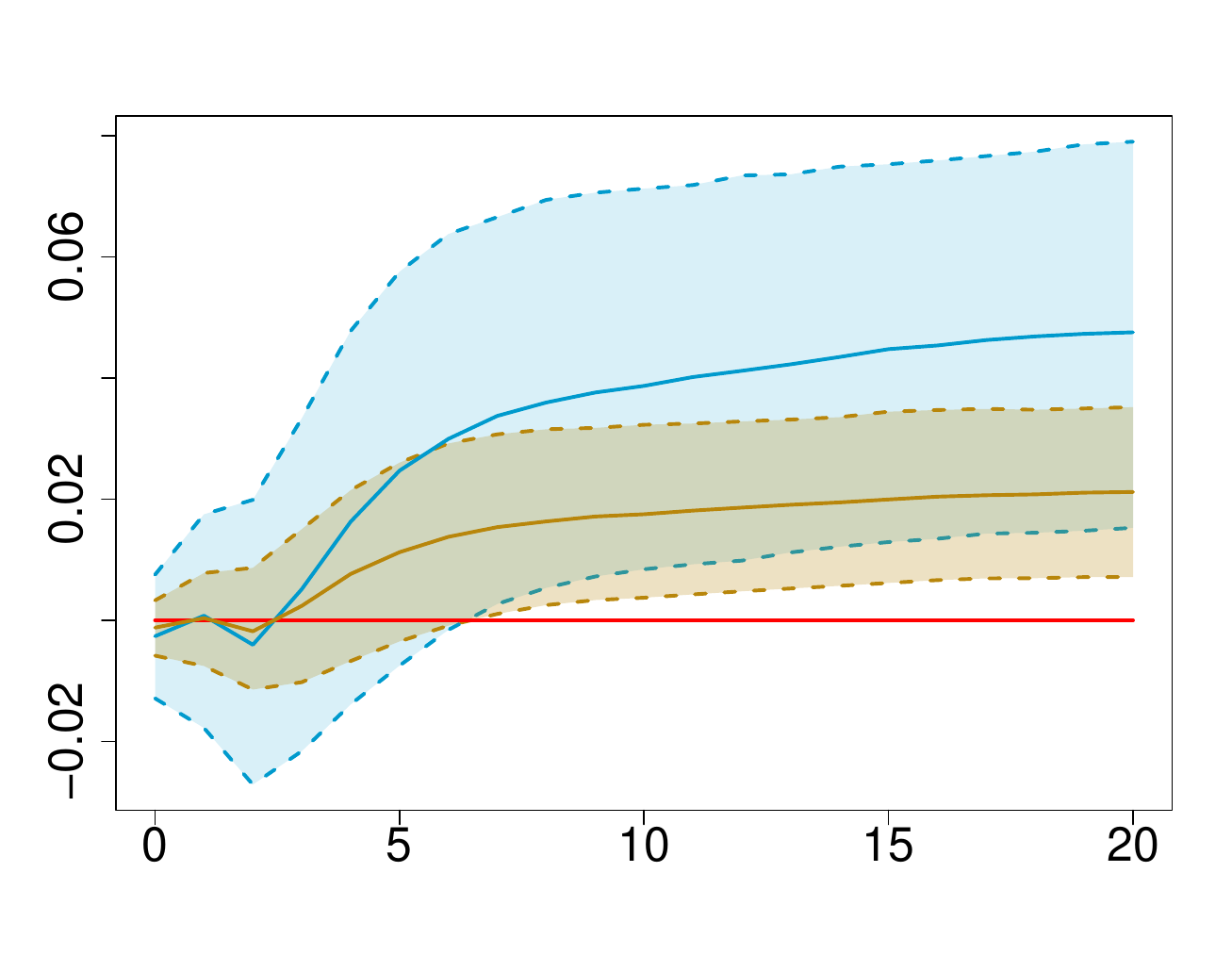}\\
			{\small shock \#1 $\rightarrow$ $r$} &
			{\small shock \#2 $\rightarrow$ $r$} & 
			{\small shock \#3 $\rightarrow$ $r$} \\
			\includegraphics[width=0.33\textwidth]{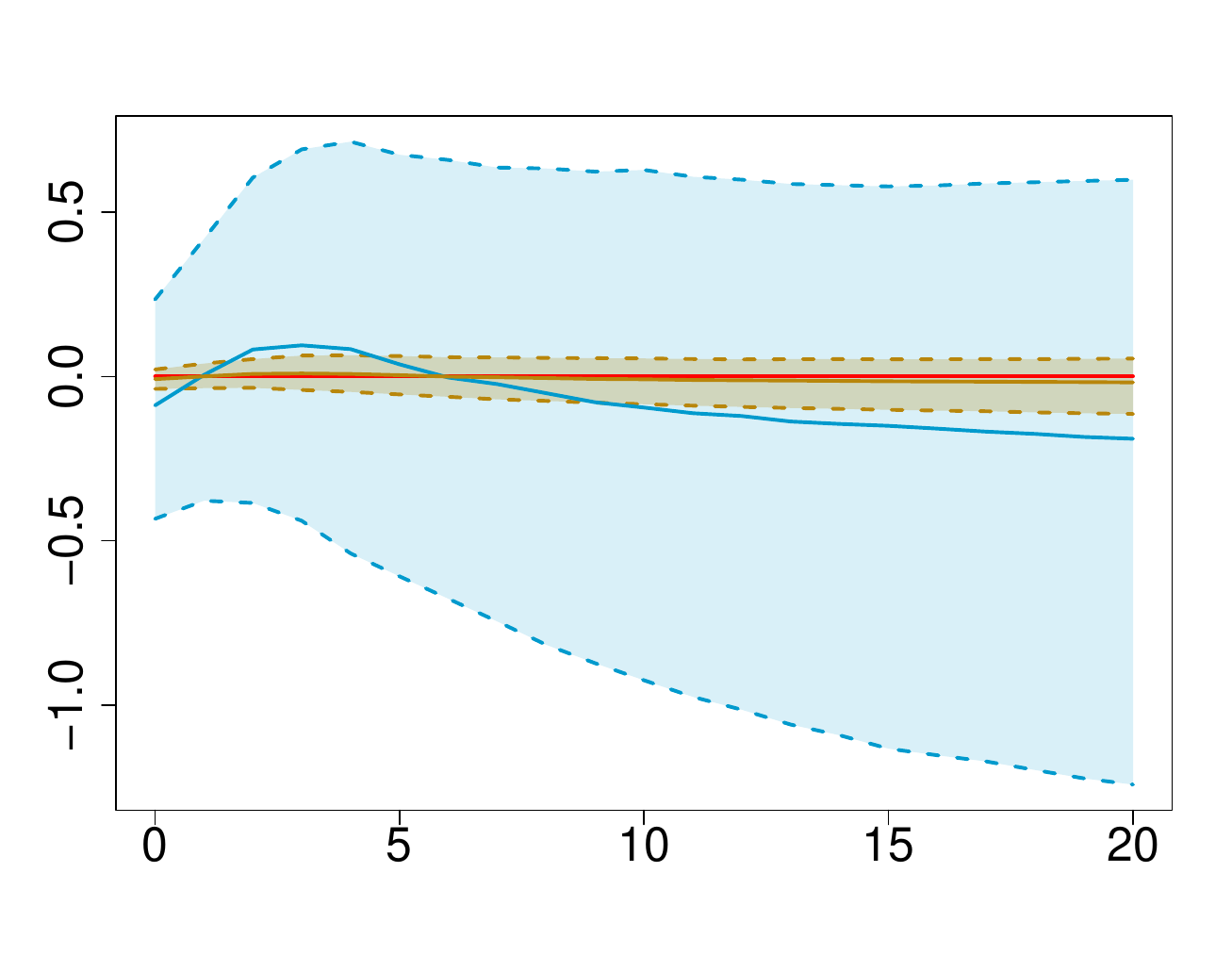} & \includegraphics[width=0.33\textwidth]{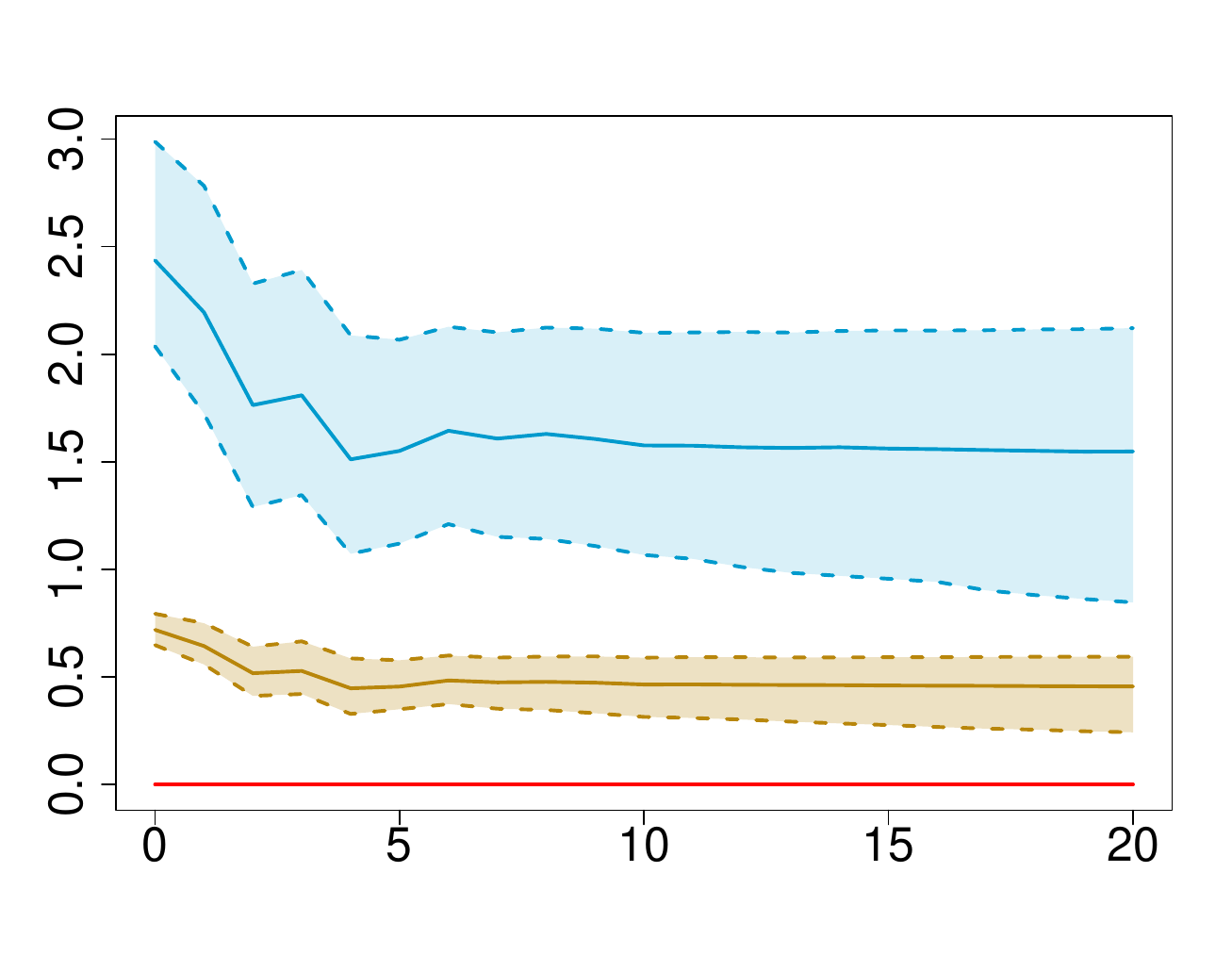} & \includegraphics[width=0.33\textwidth]{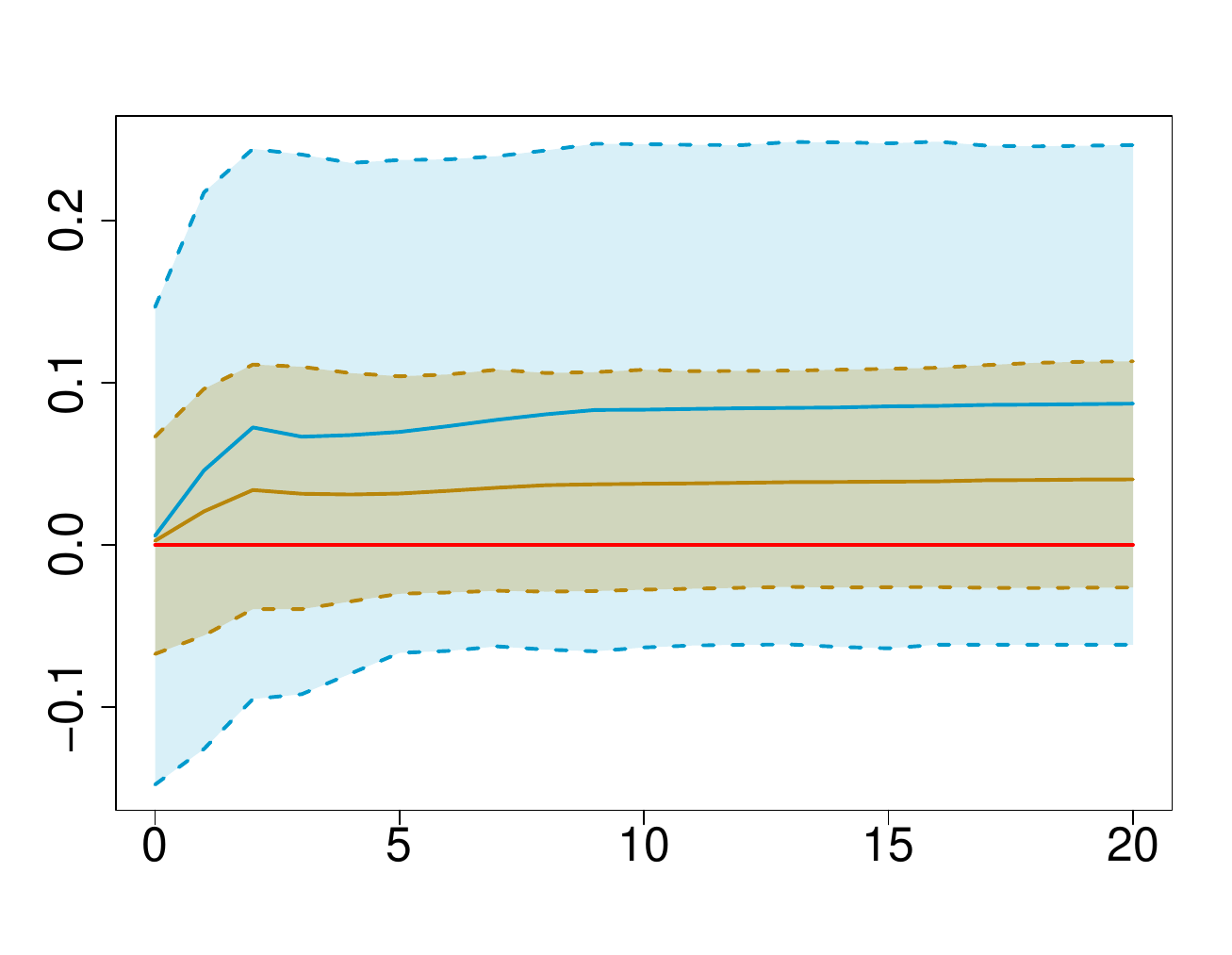}\\
			{\small shock \#1 $\rightarrow$ $s$} &
			{\small shock \#2 $\rightarrow$ $s$} & 
			{\small shock \#3 $\rightarrow$ $s$} \\
			\includegraphics[width=0.33\textwidth]{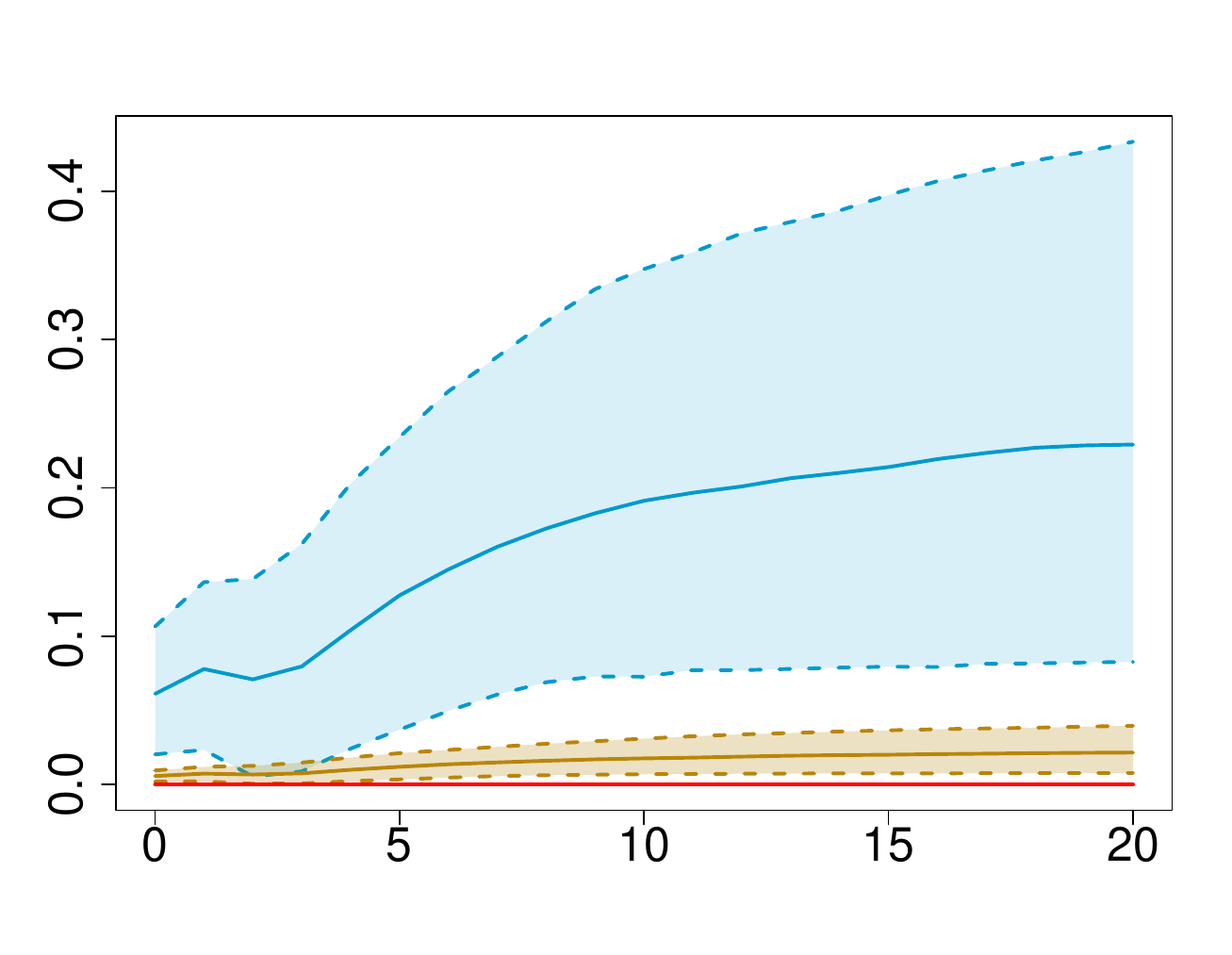} & \includegraphics[width=0.33\textwidth]{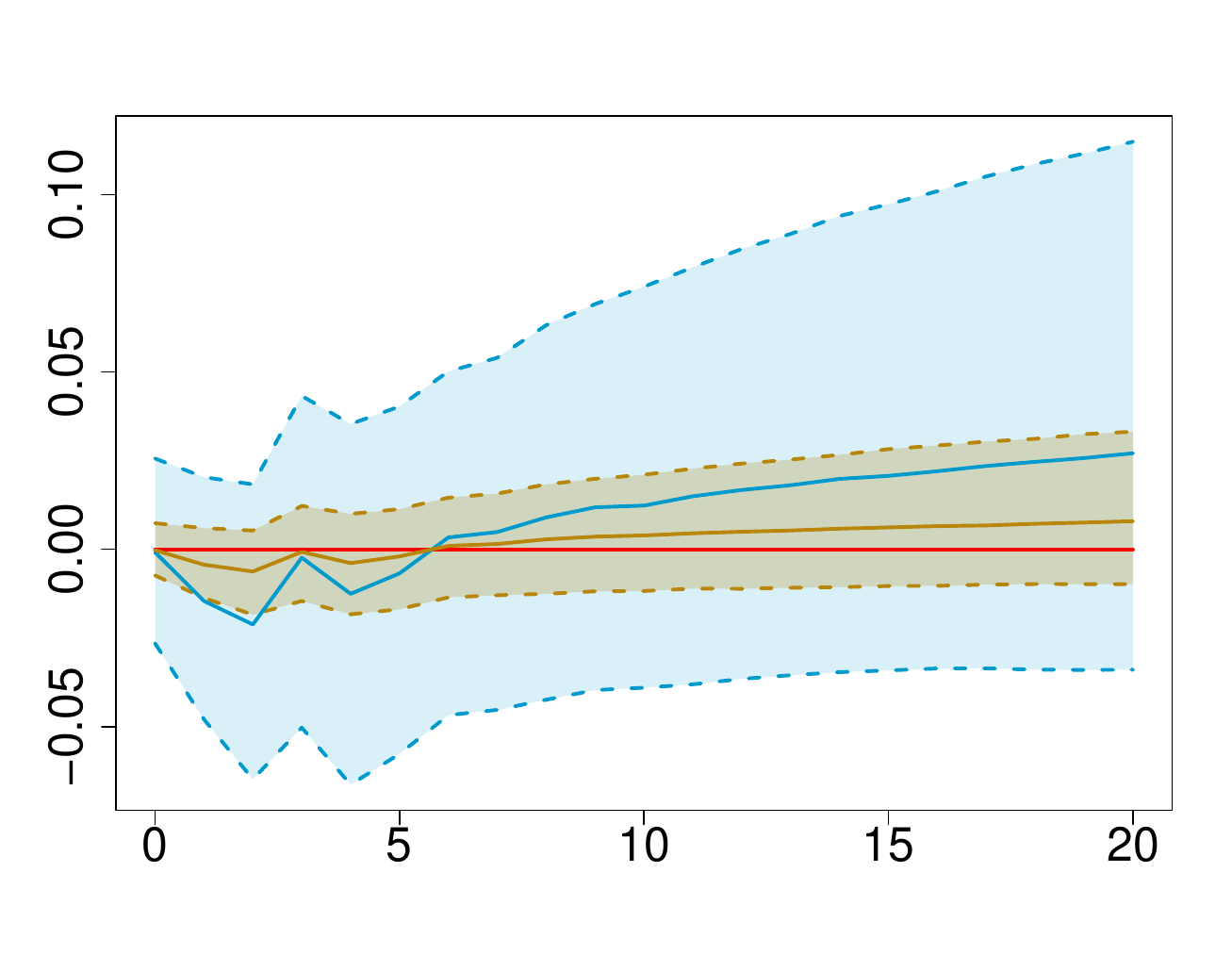} & \includegraphics[width=0.33\textwidth]{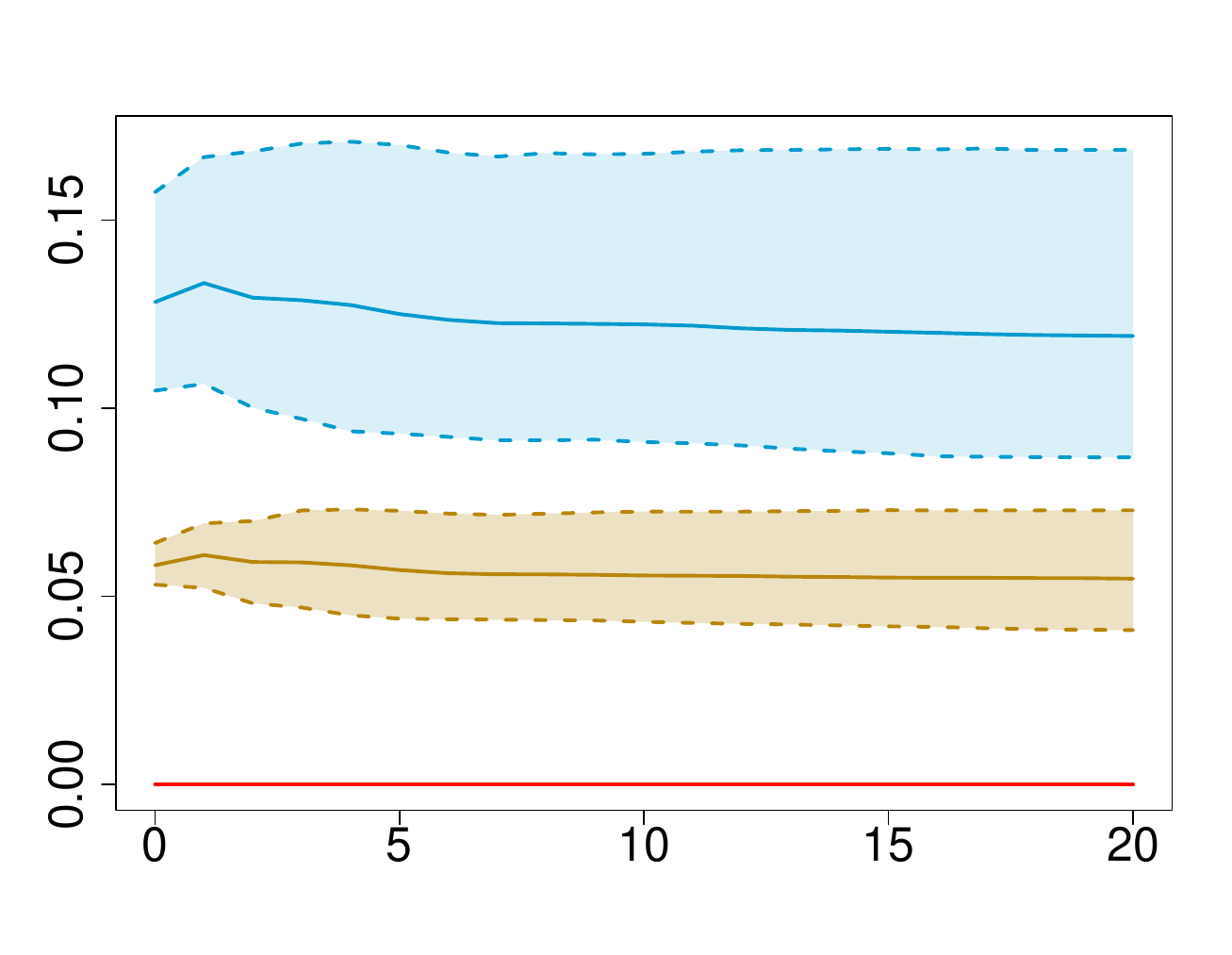}\\
		\end{tabular}    
	\end{center}
	\caption{Impulse response functions (IRFs) in state 1 (blue shading) and state 2 (golden shading). Solid lines indicates the posterior medians, whereas dashed lines represent the 16\% and 84\% posterior quantiles.}
	\label{fig:IRF}
\end{figure}

\begin{figure}
	\begin{center}
		\begin{tabular}{cc}
			\multicolumn{2}{c}{Forecast error of earnings}\\
			state 1 & state 2\\
			\includegraphics[width=0.48\textwidth]{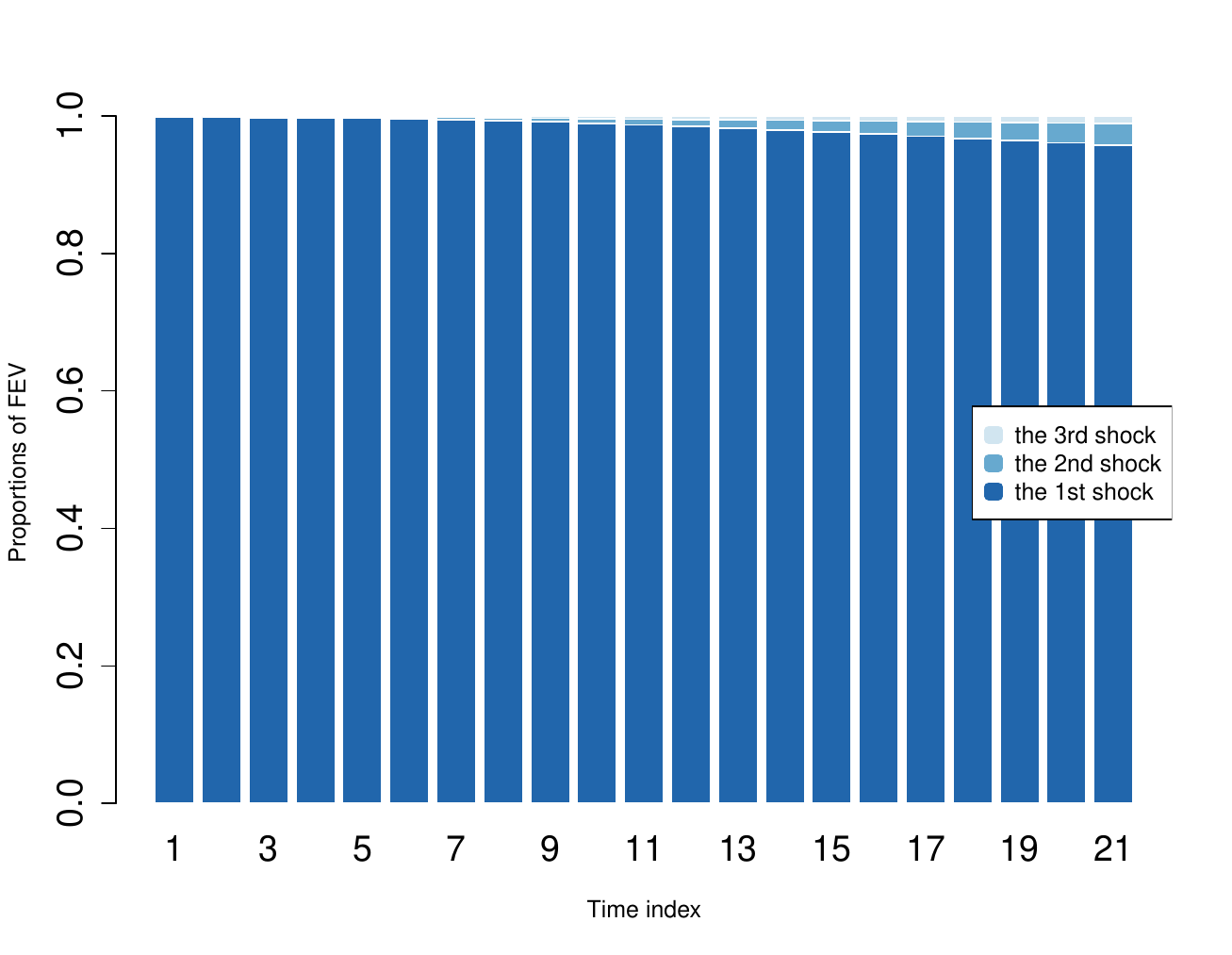} & \includegraphics[width=0.48\textwidth]{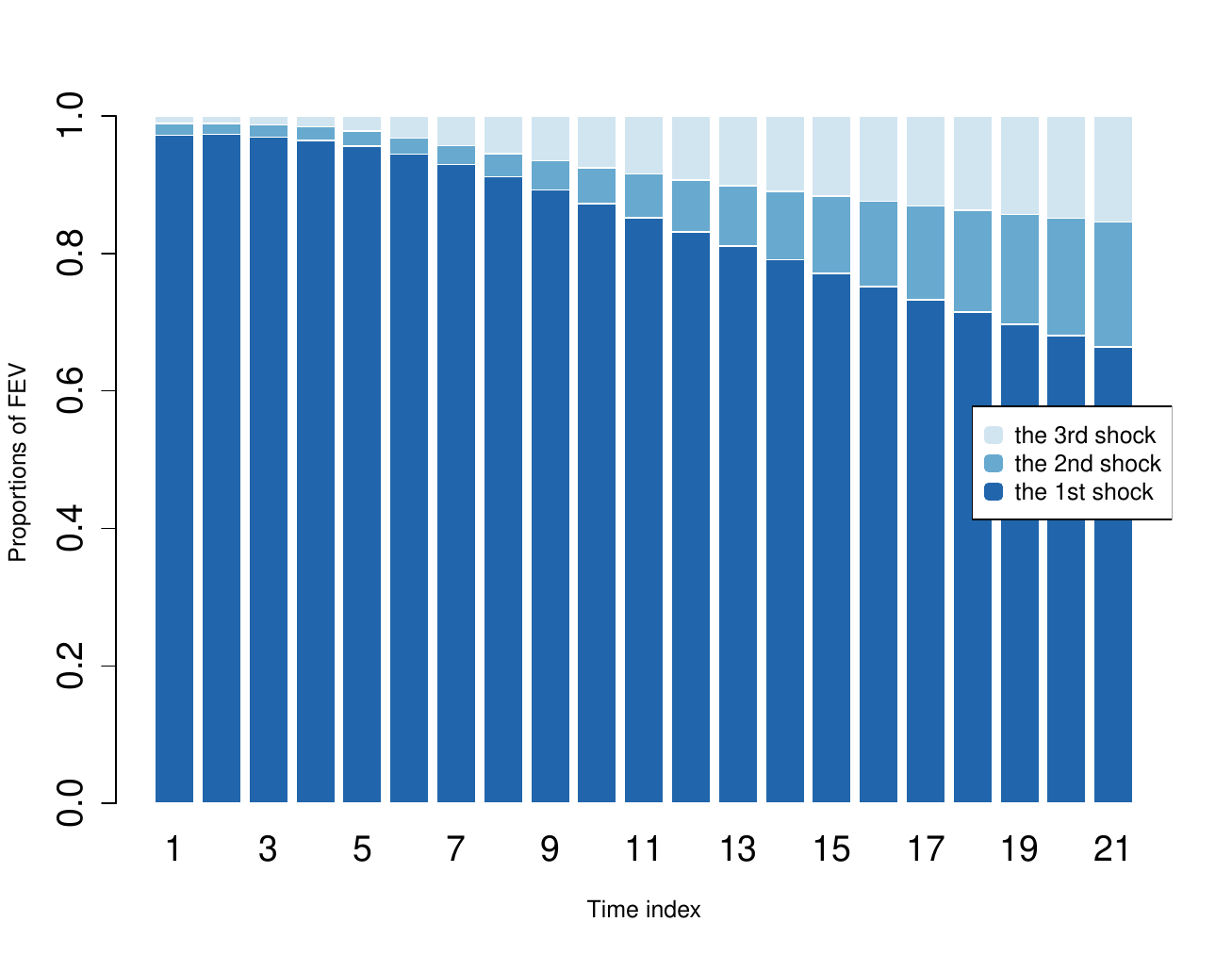}\\
			\multicolumn{2}{c}{Forecast error of interest rates}\\
			state 1 & state 2\\
			\includegraphics[width=0.48\textwidth]{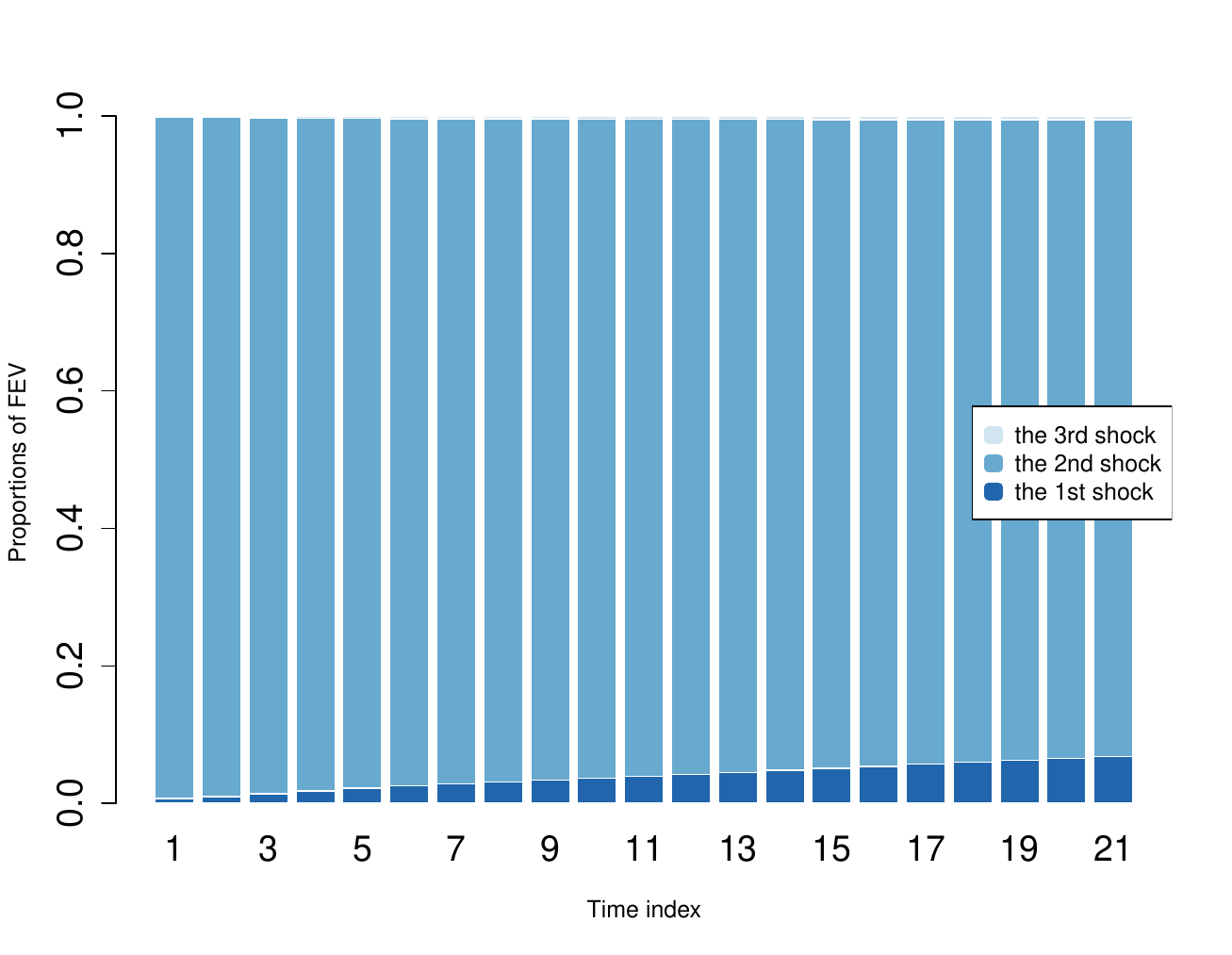} & \includegraphics[width=0.48\textwidth]{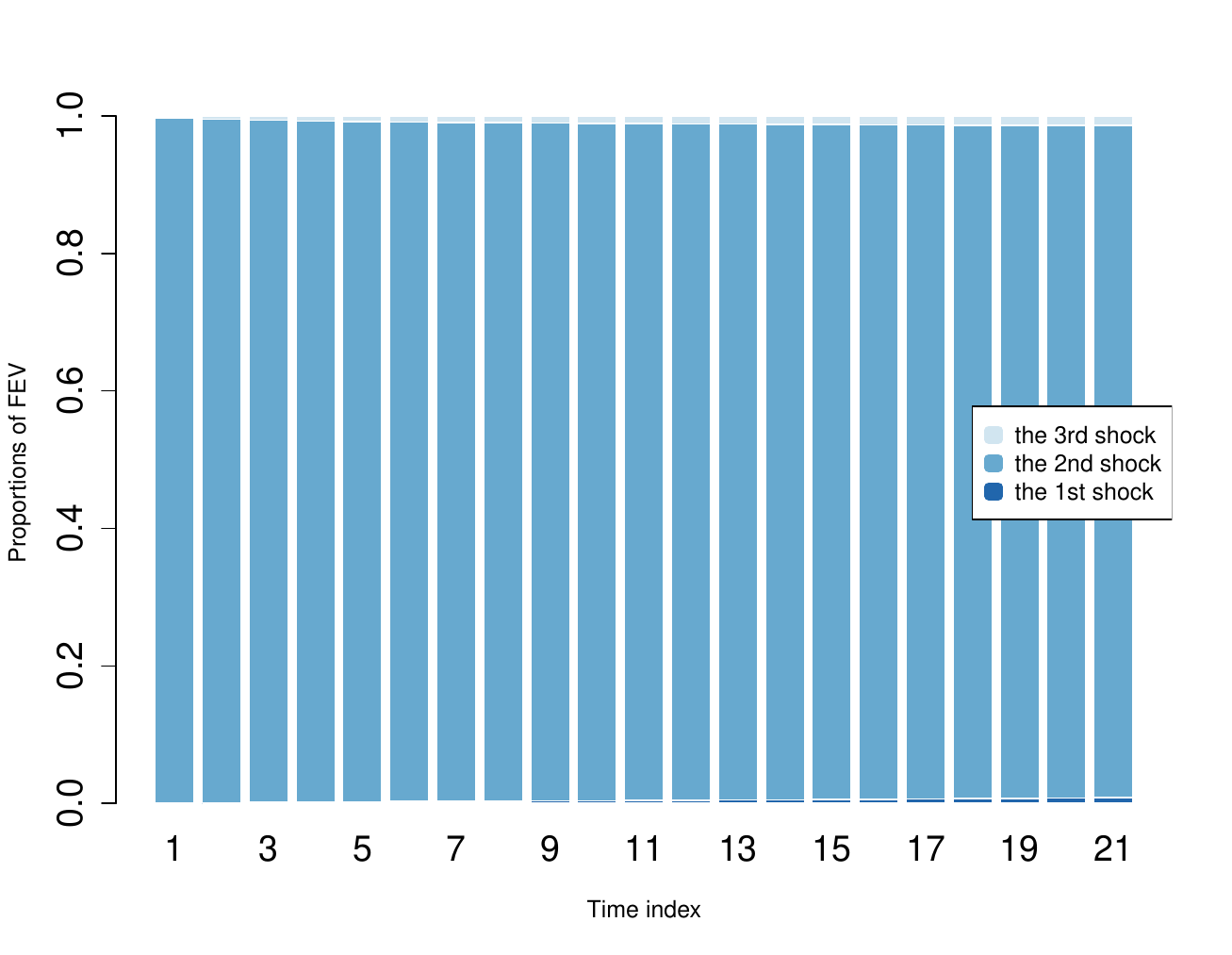}\\
			\multicolumn{2}{c}{Forecast error of stock prices}\\
			state 1 & state 2\\
			\includegraphics[width=0.48\textwidth]{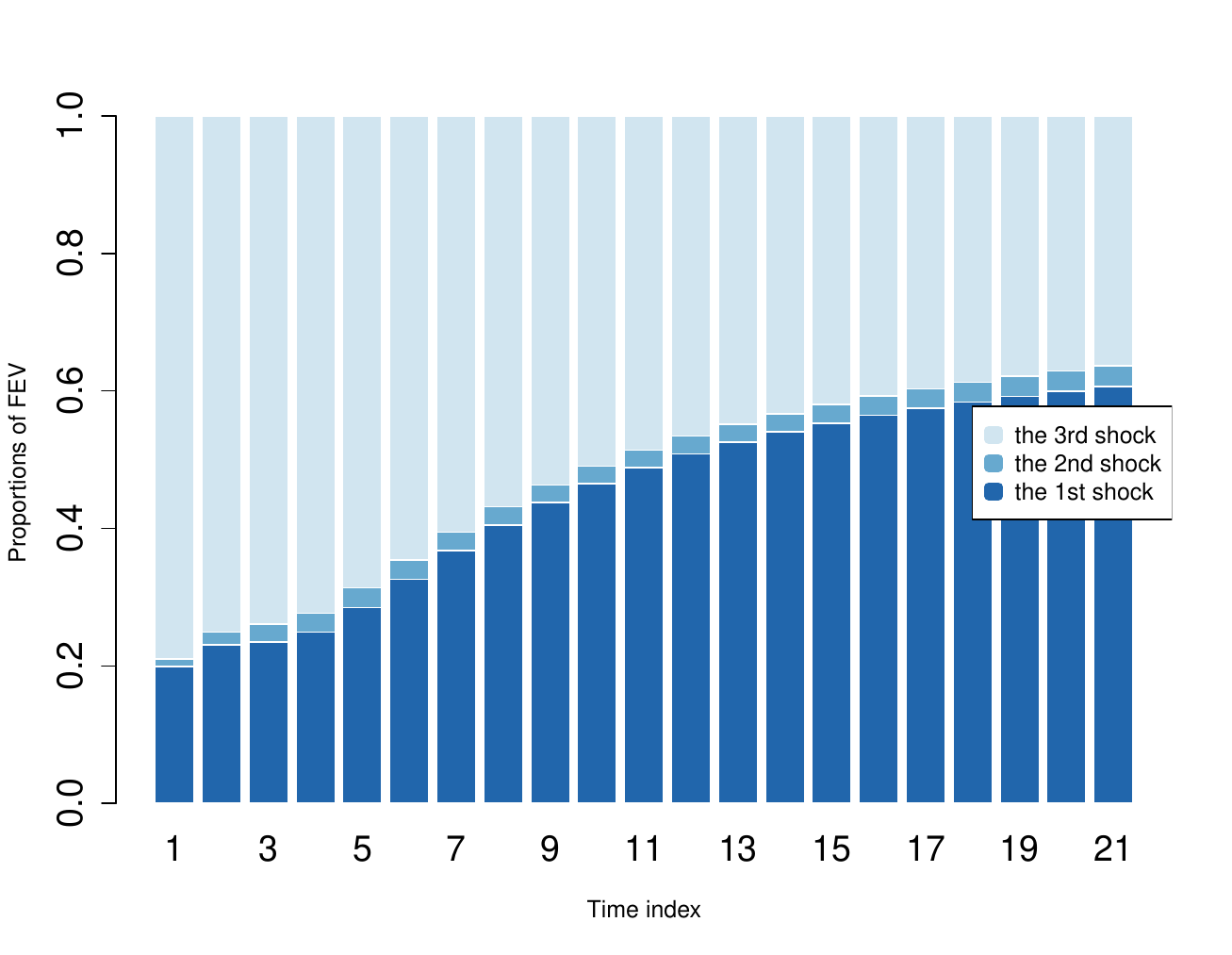} & \includegraphics[width=0.48\textwidth]{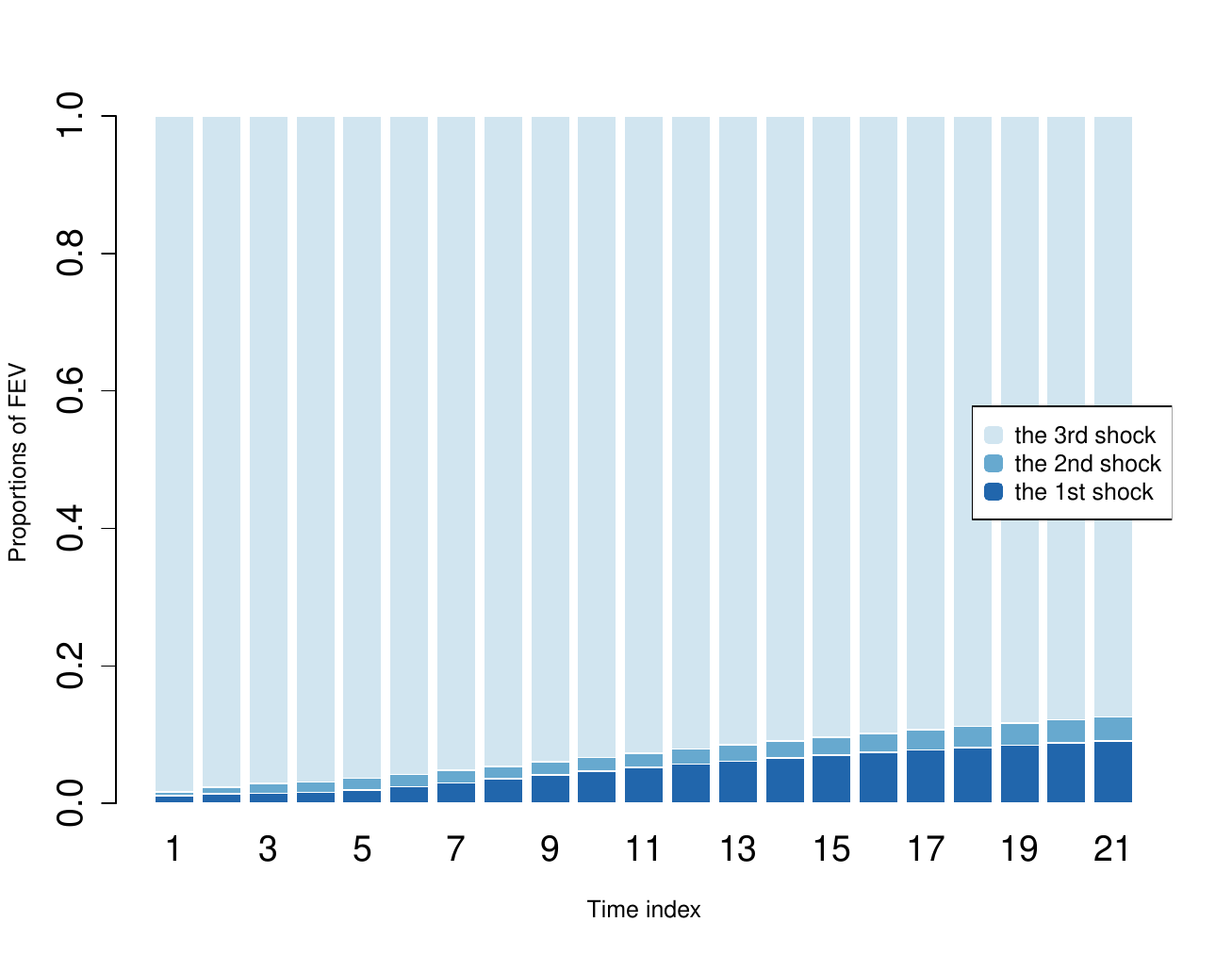}
		\end{tabular}    
	\end{center}
	\caption{Forecast error variance decomposition of the earnings/interest rates/stock prices system.}
	\label{fig:FEVD}
\end{figure}

\begin{figure}
	\centering
	\includegraphics[width=1\textwidth]{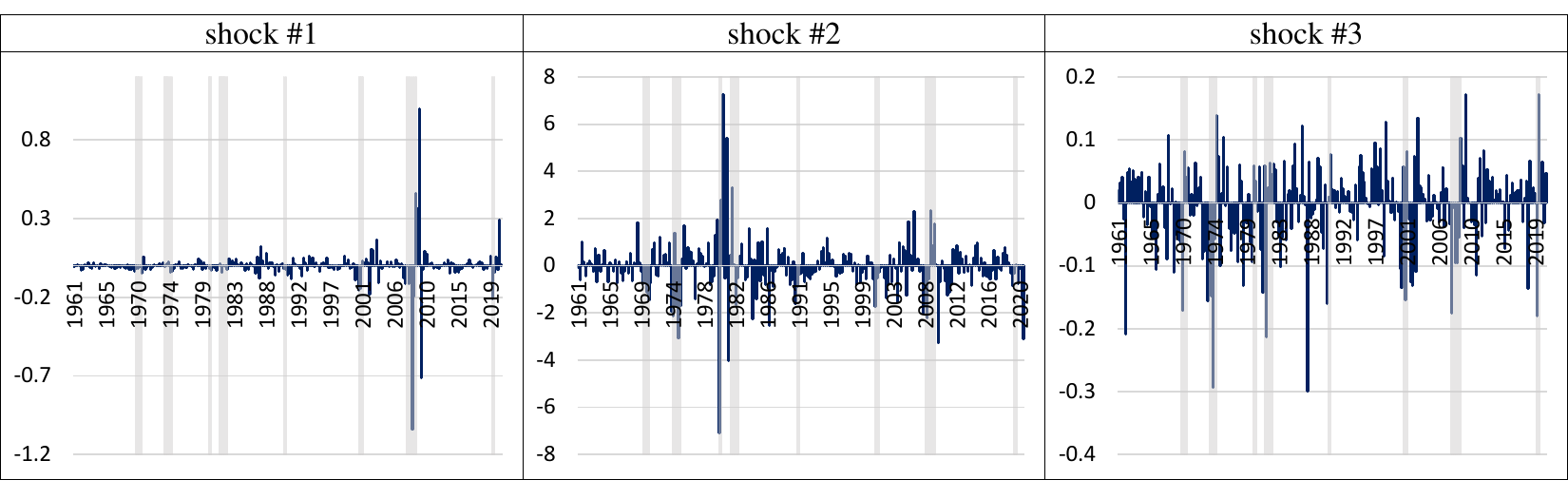}
	\caption{The point estimate of the shocks (blue bars) and NBER-dated recessions (grey shading).}
	\label{fig:shocks}
\end{figure}
	
\clearpage

\section{Supplementary material to the paper \textit{Identification of structural shocks in Bayesian VEC models with two-state Markov-switching heteroskedasticity}}

\subsection{Specification of the prior densities}
\label{Supp:sec:Prior}

The following list presents details about the prior densities of our choice:
\begin{itemize}
	\item a matrix normal distribution for the matrix of non-normalized adjustment coefficients: $\alpha_*\sim mN\left(0,I_r,\Omega_{a_*}\right)$,
	leading to $a_*=vec(\alpha_*')\sim N\left(0,\Omega_{a_*}\otimes I_r\right)$,
	\item a matrix normal distribution for the matrix of non-normalized cointegrating vectors: $\beta_*\sim mN\left(0,I_r,P\right)$, leading to $\beta\sim MACG\left(P\right)$ (a matrix angular central Gaussian distribution with an $\tilde{n}\times \tilde{n}$ matrix hyperparameter $P$; see \citet{chikuse1990matrix}, \citet{chikuse2012statistics}), and to $b_* = vec(\beta_*)\sim N\left(0,I_r\otimes P\right)$,
	\item a matrix normal distribution for the remaining  parameters of the conditional mean: $\Gamma\sim mN\left(\mu_{\Gamma},I_n,\Omega_{\Gamma}\right)$, which indicates that $\gamma=vec(\Gamma)\sim N\left(vec(\mu_{\Gamma}),I_n\otimes\Omega_{\Gamma}\right)$,
	\item a hierarchical normal distribution for $b$: $b|\nu_b\sim N^{d_b}\left(\mu_b,\nu_b\Omega_b\right)$ with $\nu_b\sim iG(n_{\nu_b},s_{\nu_b})$,\footnote{By $iG(a,b)$ we denote an inverse gamma distribution with mean $\frac{b}{a-1}$ (under $a>1$), and variance $\frac{b^2}{(a-1)^2(a-2)}$ (under $a>2$).}
	\item hierarchical inverse gamma distributions for the variances of structural shocks: $\lambda_{m,i}|s^{\lambda}_{m,i}~\sim iG\left(n^{\lambda}_{m,i},s^{\lambda}_{m,i}\right)$ with $s^{\lambda}_{m,i}\sim G\left(n_{m,i},s_{m,i}\right)$, $i=1,2,...,n$, and $m = 1,2$,\footnote{By $G(a,b)$ we denote a gamma distribution with mean $ab$ and variance $ab^2$.}
	\item beta distributions for the transition probabilities: $p_{mm}\sim Beta\left(c_{m}, d_{m}\right)$, $m=1,2$.
\end{itemize}

\subsection{MCMC Sampler}
\label{Supp:sec:MCMCsampler}

To obtain a pseudo-random sample from the joint posterior distribution of the SVEC-MSH model developed in our paper, we design a sampler that combines two standard MCMC techniques: the Gibbs sampler and Metropolis-Hastings (MH) algorithm, both relying on the full conditional posterior distributions. As presented below, only the posterior conditional of $b$ is some unknown type of distribution, thereby necessitating an MH step within the Gibbs routine.

The algorithm proceeds as follows. After setting the initial values: $b^{(0)}$, $\lambda_{m,i}^{(0)}$ ($i=1,2,\dots,n; m=1,2$), $\Gamma^{(0)}$, $\alpha_{*}^{(0)}$, $\beta_{*}^{(0)}$, and $S_{0:T}^{(0)}=(S_0^{(0)}, S_1^{(0)}, \ldots, S_T^{(0)})$, the following steps are reiterated for $s=1, 2, ..., M+N$, where $M$ and $N$ stand for the \textit{burn-in} and posterior MCMC sample sizes, respectively:
\begin{enumerate}
	\item Draw $p_{mm}^{(s)}$ ($m=1, 2$) from $p_{mm}|\cdot,y\sim Beta(c_{m}+N_{m,m}(S_{0:T}), d_{m}+N_{m,3-m}(S_{0:T}))$, where $N_{m,l}(S_{0:T})$ denotes the number of one-step transitions from state $m\in\{1,2\}$ to state $l\in\{1,2\}$ in the sequence $S_{0:T}=(S_0, S_1,...,S_T)$.
	\item Draw $S_{0:T}^{(s)}$ using the \textit{Forward-Filtering-Backward-Sampling} (FFBS) algorithm designed by \citet{carter1996markov} and \citet{chib1996calculating}.
	\item Draw $s^{\lambda\ (s)}_{m,i}$ $(i=1,2,\dots,n;\,m=1,2)$ from $s^{\lambda}_{m,i}|\cdot,y\sim G(n_{m,i}+n_{m,i}^{\lambda},\frac{s_{m,i}\lambda_{m,i}}{s_{m,i}+\lambda_{m,i}})$.
	\item Draw $\lambda_{m,i}^{(s)}$ $(i=1,2,\dots,n;\,m=1,2)$ from $\lambda_{m,i}|\cdot,y\sim iG(n_{m,i}^{\lambda}+\frac{T_m}{2},s_{m,i}^{\lambda}+\frac{1}{2}d^{\lambda}_{m,ii})$, where $d^{\lambda}_{m,ii}$ is the $i^{th}$ diagonal element of the matrix $D_m^{\lambda} = B^{-1}E^{(m)'}E^{(m)}(B')^{-1}$. If a relevant restriction for the identification of states (e.g., $(\lambda_{1,l}^{(s)}>\lambda_{2,l}^{(s)})$ for a given $l\in\{1, 2, ..., n\}$) is not met, then permute the state labels of all regime-linked quantities drawn so far (see \citet{fruhwirth2001permutation} for details on the so-called permutation sampler).
	\item Calculate $\omega_{2,i}^{(s)} = \lambda_{2,i}^{(s)}/\lambda_{1,i}^{(s)}$, for $i=1,2,\dots,n$. If a relevant order restriction (e.g., $\omega_{2,1}^{(s)}\le\omega_{2,2}^{(s)}\le\dots\le\omega_{2,n}^{(s)}$) is not satisfied, then reject the draw and return to Step 4.\\
	Notice that it is not possible to employ a permutation sampler to ensure the restriction, despite apparent appeals of the approach. The reason for this is that permuting the relative variance changes, $\omega_{2,i}^{(s)}$ ($i=1,2,\dots,n$), may break the identification of the states ensured by the previous step.            
	\item Draw $\nu_b^{(s)}$ from $\nu_b|\cdot,y\sim iG\left(n_{\nu_b}+\frac{d_b}{2},s_{\nu_b}+\frac{1}{2}(b-\mu_b)'\Omega_b^{-1}(b-\mu_b)\right)$.
	\item Draw $b^{(s)}$ from $b|\cdot,y$, with the kernel of the density assuming the form:
	
	$\begin{aligned}
		& p(b|\cdot,y)\propto |B|^{-T}\times\\
		&\times \exp\left\{-\frac{1}{2}\left[(b-\mu_b)'\left(\nu_b\Omega_b\right)^{-1}(b-\mu_b)+\sum_{m=1}^M tr\left((B\Lambda^{(m)}B')^{-1}E^{(m)'}E^{(m)}\right)\right]\right\}.\end{aligned}$
	
	To that end, a Random-Walk Metropolis-Hastings algorithm can be employed (we use a normal proposal distribution with an online recalibration of the covariance matrix after each a given, arbitrarily preset number of draws within the \textit{burn-in} phase).
	\item Draw $\gamma^{(s)}=vec(\Gamma^{(s)})$ from $\gamma|\cdot,y\sim N(\overline{\mu}_{\gamma},\overline{\Omega}_{\gamma})$, where\\ $\overline{\Omega}_{\gamma}=\left[\left(I_n\otimes\frac{1}{\nu_{\Gamma}}\Omega_{\Gamma}^{-1}\right)+\tilde{z}_2'\tilde{\Sigma}^{-1}\tilde{z}_2\right]^{-1}$, and\\ $\overline{\mu}_{\gamma}=\overline{\Omega}_{\gamma}\left[\left(I_n\otimes\frac{1}{\nu_{\Gamma}}\Omega_{\Gamma}^{-1}\right)vec(\mu_{\gamma})+\tilde{z}_2'\tilde{\Sigma}^{-1}\left(\tilde{z}_0-\tilde{z}_1vec(\beta_*\alpha_*')\right)\right]$; reshape the draw into $\Gamma^{(s)}$.
	\item Draw $a_{*}^{(s)}=vec(\alpha_{*}^{(s)})$ from $a_*|\cdot,y\sim N\left(\overline{\mu}_{a_*},\overline{\Omega}_{a_*}\right)$, where\\ $\overline{\Omega}_{a_*} = \left[\left(\Omega_{a_*}^{-1}\otimes I_r\right)+(I_n\otimes\beta_*')\tilde{z}_1'\tilde{\Sigma}^{-1}\tilde{z}_1(I_n\otimes \beta_*)\right]^{-1}$, and\\
	$\overline{\mu}_{a_*} = \overline{\Omega}_{a_*}(I_n\otimes \beta_*')\tilde{z}_1'\tilde{\Sigma}^{-1}(\tilde{z}_0-\tilde{z}_2\gamma)$;
	reshape the draw into $\alpha_{*}^{(s)}$.
	\item Draw $b_{*}^{(s)}=vec(\beta_{*}^{(s)})$ from $b_*|\cdot,y\sim N\left(\overline{\mu}_{b_*},\overline{\Omega}_{b_*}\right)$, where\\ $\overline{\Omega}_{b_*} = \left[\left(I_r\otimes P^{-1}\right)+(\alpha_*'\otimes I_{\tilde{n}})\tilde{z}_1'\tilde{\Sigma}^{-1}\tilde{z}_1(\alpha_*\otimes I_{\tilde{n}})\right]^{-1}$, and\\
	$\overline{\mu}_{b_*} = \overline{\Omega}_{b_*}(\alpha_*'\otimes I_{\tilde{n}})\tilde{z}_1'\tilde{\Sigma}^{-1}(\tilde{z}_0-\tilde{z}_2\gamma)$; reshape the draw into $\beta_{*}^{(s)}$.
	\item Check the non-explosiveness condition: if satisfied, then keep the draws and proceed with the algorithm. Otherwise, rerun Steps 8-10 (note that rerunning the earlier stages is not required).
	\item Optionally, if one intends to obtain the posterior estimates also of the cointegration relations and adjustment coefficients, calculate $\beta^{(s)}=\beta_{*}^{(s)}(\beta_{*}^{(s)'}\beta_{*}^{(s)})^{-\frac{1}{2}}$ and $\alpha^{(s)}=\alpha_{*}^{(s)}(\beta_{*}^{(s)'}\beta_{*}^{(s)})^{\frac{1}{2}}$. The transformation ensures that the columns of $\beta^{(s)}$ are orthonormal and the cointegration space spanned by them is uniquely identified. However, for a complete identification of the cointegrating vectors and adjustment coefficients, further transformations of matrix $\beta^{(s)}$ are required to minimise its distance from the point estimate of $\beta$; see \citet{assmann2016bayesian} and \citet{wroblewska2023bayesian} for more details in the context of, correspondingly, Bayesian factor models and Bayesian VEC models.
\end{enumerate}

\subsection{Simulation study}
\label{Supp:sec:Simulation_study}

Here we examine the methodology developed in the paper through two illustrative studies based on data sets simulated from known data generating processes (DGP) based on a two-state VEC-MSH formulated for a two-dimensional VAR(2) (thus, $p=2$). The key difference between the two cases consists in the size of the contrast between the relative (to state 1) structural variances, i.e. $\omega_{2,2}-\omega_{2,1}$, where 
$\omega_{2,2}=\lambda_{2,2}/\lambda_{1,2}$ and
$\omega_{2,1}=\lambda_{2,1}/\lambda_{1,1}$. Our basic conjecture is that assuming only 'small' true values of the contrasts would result in a relatively high posterior uncertainty thereof, thus hindering a precise identification of structural shocks. Conversely, under fairly 'large' values of the contrasts, the posterior uncertainty should reduce, enabling one to distill the shocks in a more clear-cut way. Therefore, in what follows we refer to those two variants of simulated data as the 'small contrast' (SC) case and the 'large contrast' (LC) case.

Both DGPs under study share a group of parameters for which the same values are set. In particular, the non-normalized adjustment coefficients and cointegrating vector are: $\alpha_*'=(\begin{array}{cc} -0.1 & 0.3 \end{array})$ and $\beta_*'=(\begin{array}{cc} 1 & -1 \end{array})$. The constant term and parameters by the first lagged differences in the VEC equation are given as, correspondingly, $\Phi'=(\begin{array}{cc} 0.1 & 0.2 \end{array})$ and $\Gamma_1=\left(\begin{array}{cc}
	0.24 & -0.08  \\
	0.1 & -0.31 
\end{array} \right)$. In both cases, we specify $p_{11}=p_{22}=0.97$ so as to mimic regime persistence fairly typical to macroeconomic time series. The true values of all the remaining, key parameters are provided in Table~\ref{tab:true_parameters}, where the originally set values of: the $B$ matrix, the vectors of state-dependent variances, $\lambda_m=(\lambda_{m,1}\,\, \lambda_{m,2})$, $m=1,\, 2$, and the resulting relative volatility changes, $\omega_{2,i}$, $i=1,\, 2$, are superscripted with '(1)', referring to the first, original solution, while parameters corresponding to the other solution are indicated with '(2)'. Since both solutions are observationally equivalent, they yield the same state-specific conditional covariance matrices in the reduced form, i.e. $\Sigma_1$ and $\Sigma_2$ (hence, with no additional superscripts), equal 
$\Sigma_1=\left(\begin{array}{cc}
	1.028 & 0.36  \\
	0.36 & 0.95 
\end{array} \right)$ and
$\Sigma_2=\left(\begin{array}{cc}
	0.204 & 0.08  \\
	0.08 & 0.15 
\end{array} \right)$
in the SC case, and
$\Sigma_1=\left(\begin{array}{cc}
	2.56 & 0.95  \\
	0.95 & 2.125 
\end{array} \right)$ and
$\Sigma_2=\left(\begin{array}{cc}
	2.0096 & 0.952  \\
	0.952 & 0.74 
\end{array} \right)$
in the LC scenario. Notice that the SC and LC values of the contrast ($-0.057$ and $-0.64$, respectively) differ by over an order of magnitude, which is expected to yield qualitatively different posterior results for the identification of matrices $B^{(1)}$ and $B^{(2)}$ (shared by the SC and LC cases). Moreover, the two solutions are uniquely identified only through the sign of the contrast between $\omega_{2,2}$ and $\omega_{2,1}$, since $\omega_{2,2}^{(1)}-\omega_{2,1}^{(1)}=-(\omega_{2,2}^{(2)}-\omega_{2,1}^{(2)})$.

In both cases under study, we simulated time series of $T=200$ observations, preceded by 100 data points, discarded to limit the influence of initial conditions. Incidentally, we also performed a similar analysis as the one presented below but increasing the sample size to $T=400$, which naturally yielded more compelling results, as one might have expected. We do not present them in the paper (although they are available upon request), not only for the sake of brevity, but mainly to retain our focus on the case based on a more realistic sample size.

As regards the prior distribution, the following values of hyperparameters are set (both here as well as in the empirical analysis presented in Section 4 in the main text):

\begin{itemize}
	\item $b|\nu_b \sim N^{d_b}\left(0,\nu_b I_{d_b}\right)$ with $\nu_b\sim iG(3,2)$, 
	\item $\lambda_{m,i}|s^{\lambda}_{m,i}\sim iG\left(1,s^{\lambda}_{m,i}\right)$ with $s^{\lambda}_{m,i}\sim G\left(1,1\right)$, $i=1,2,...,n$, $m=1,2$, 
	\item $\alpha_*\sim mN\left(0,I_r,0.1I_n\right)$, 
	\item $\beta_*\sim mN\left(0,I_r,\frac{1}{\tilde{n}}I_{\tilde{n}}\right)$, which leads to a uniform distribution for the cointegration space,
	\item $\Gamma\sim mN\left(0,I_n,\Omega_{\Gamma}\right)$, with $\Omega_{\Gamma} = \frac{1}{2}\ diag\left(I_n,\frac{1}{4}I_n,\dots,\frac{1}{(p-1)^2}I_n\right)$, 
	\item $p_{mm}\sim Beta(1,1) \equiv U[0,1]$ (for $m=1,2$), with $U[0,1]$ denoting a uniform distribution over the interval $[0,1]$. 
\end{itemize}

All results presented below are based on $500\,000$ MCMC draws, preceded by as many \textit{burn-in} iterations. To ensure the identification of the regimes, a restriction is imposed of the form: $\lambda_{11}>\lambda_{21}$, i.e. the variance of the first structural shock in the first state is higher than the one of the same shock in the second state: $Var(\varepsilon_{t1}|S_t=1, \theta)>Var(\varepsilon_{t1}|S_t=2, \theta)$.

\begin{figure}[h]
	\centering
	\begin{tabular}{|c|c|}
		\hline
		With the uniqueness restriction  &  Without the uniqueness restriction\\
		\hline\hline
		\includegraphics[width=0.4\textwidth]{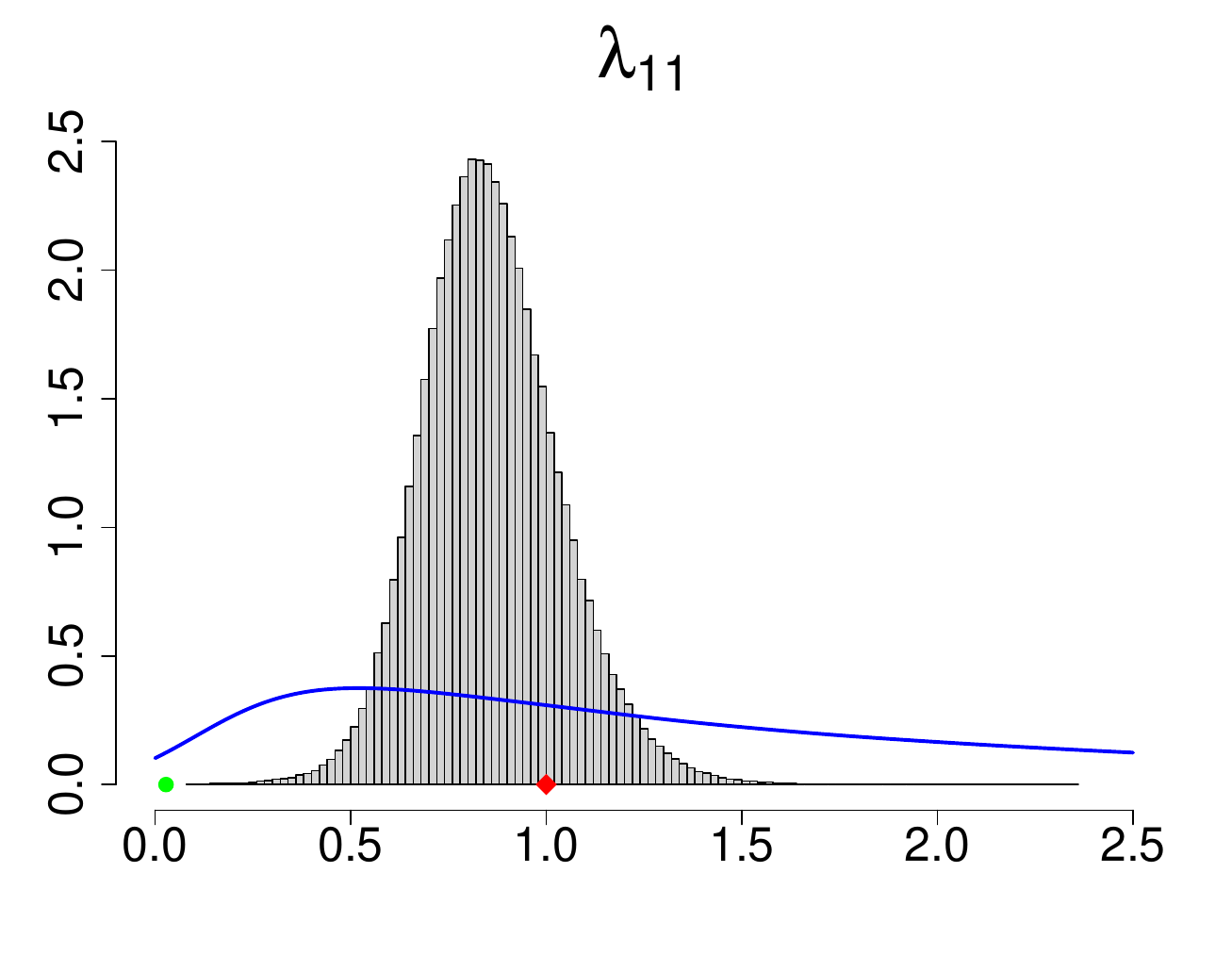} & \includegraphics[width=0.4\textwidth]{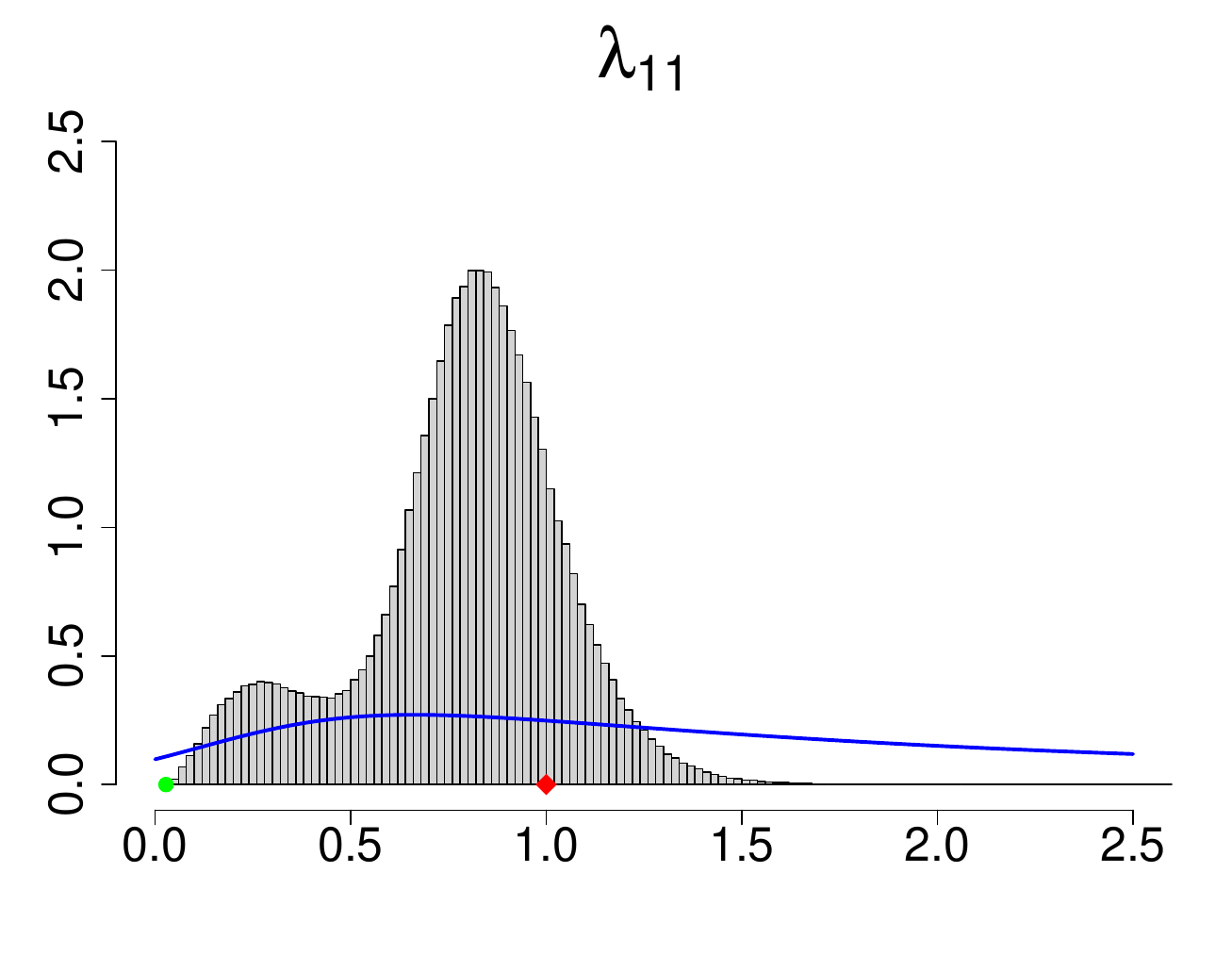}\\
		\hline
		\includegraphics[width=0.4\textwidth]{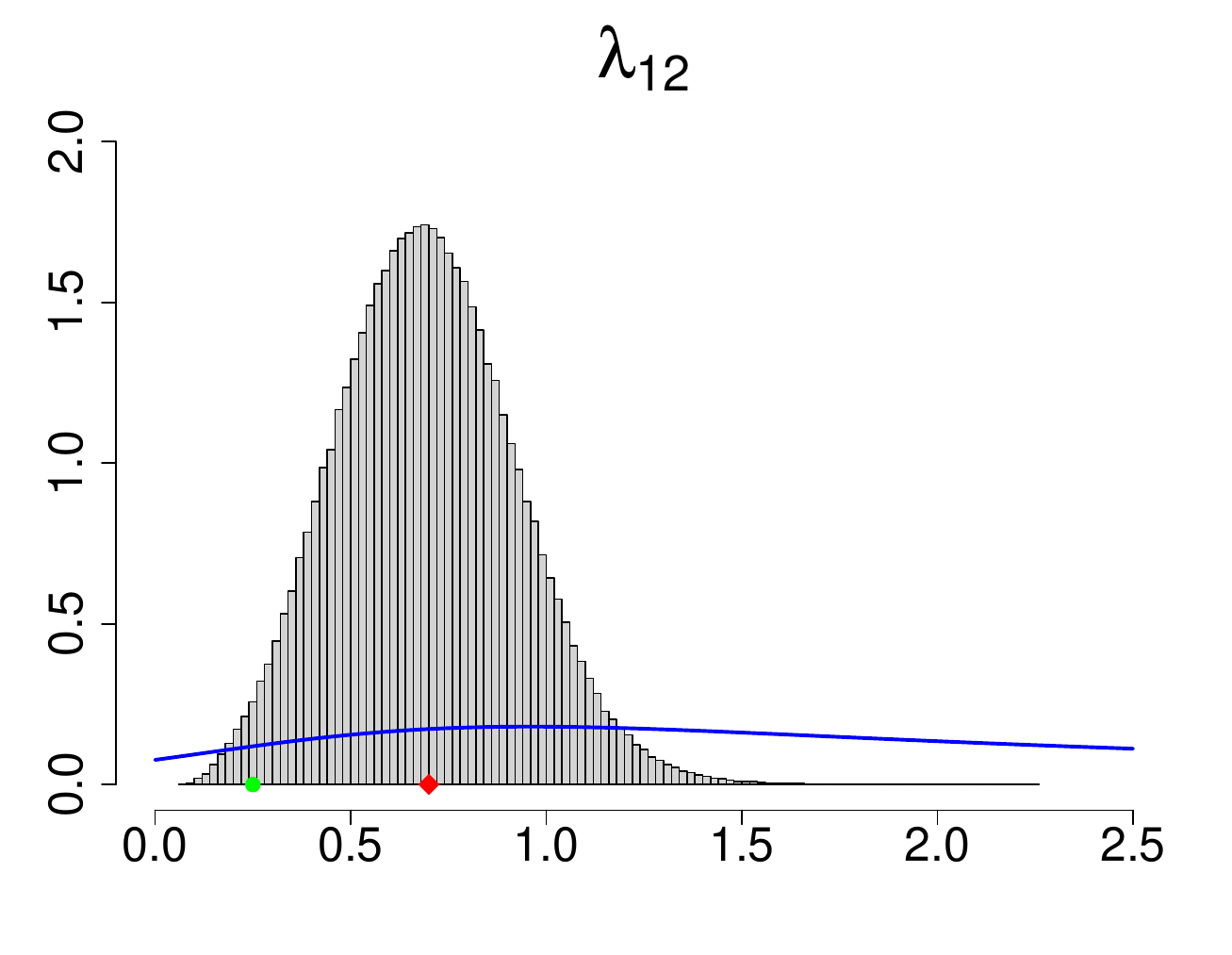} & \includegraphics[width=0.4\textwidth]{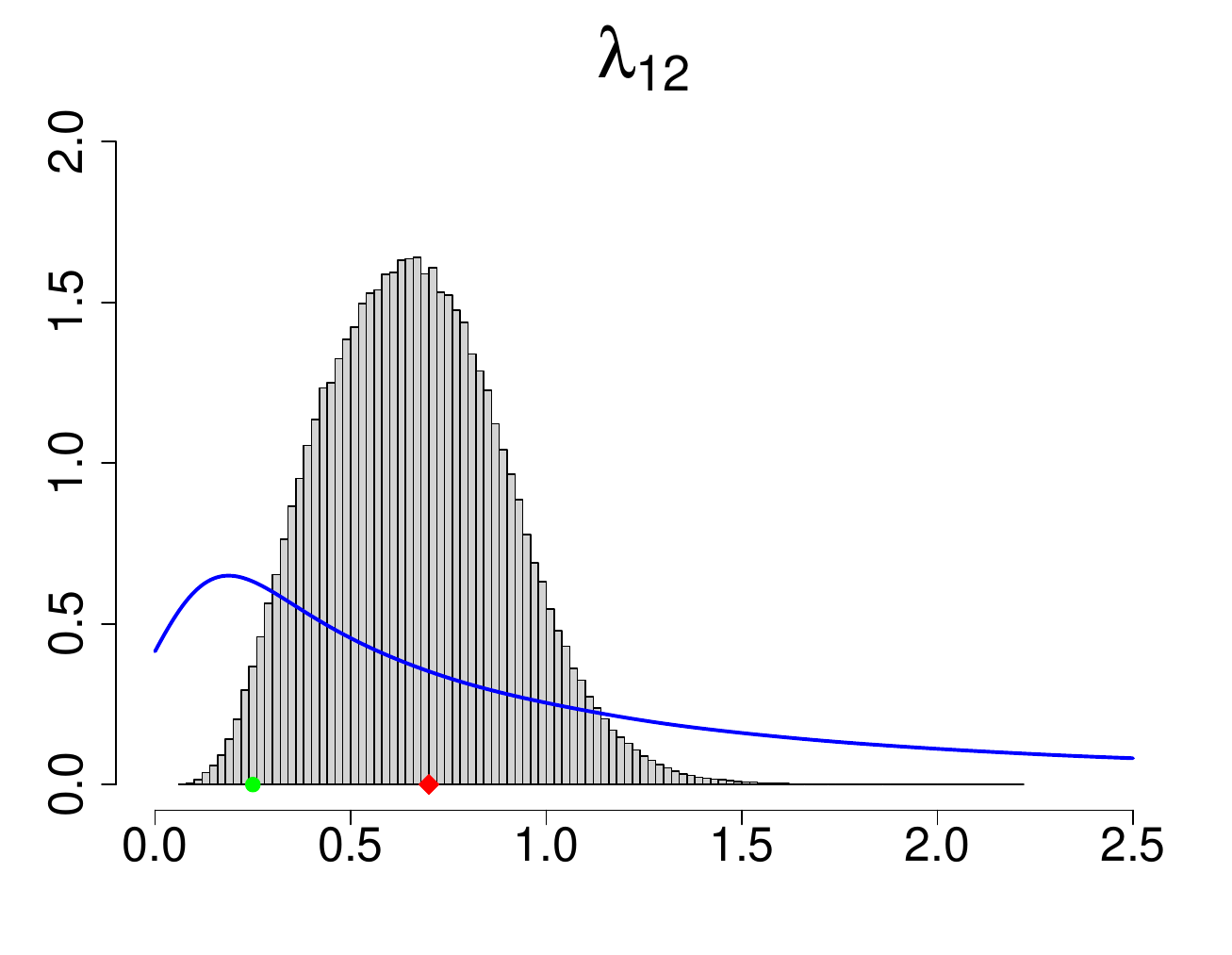}\\
		\hline
		\includegraphics[width=0.4\textwidth]{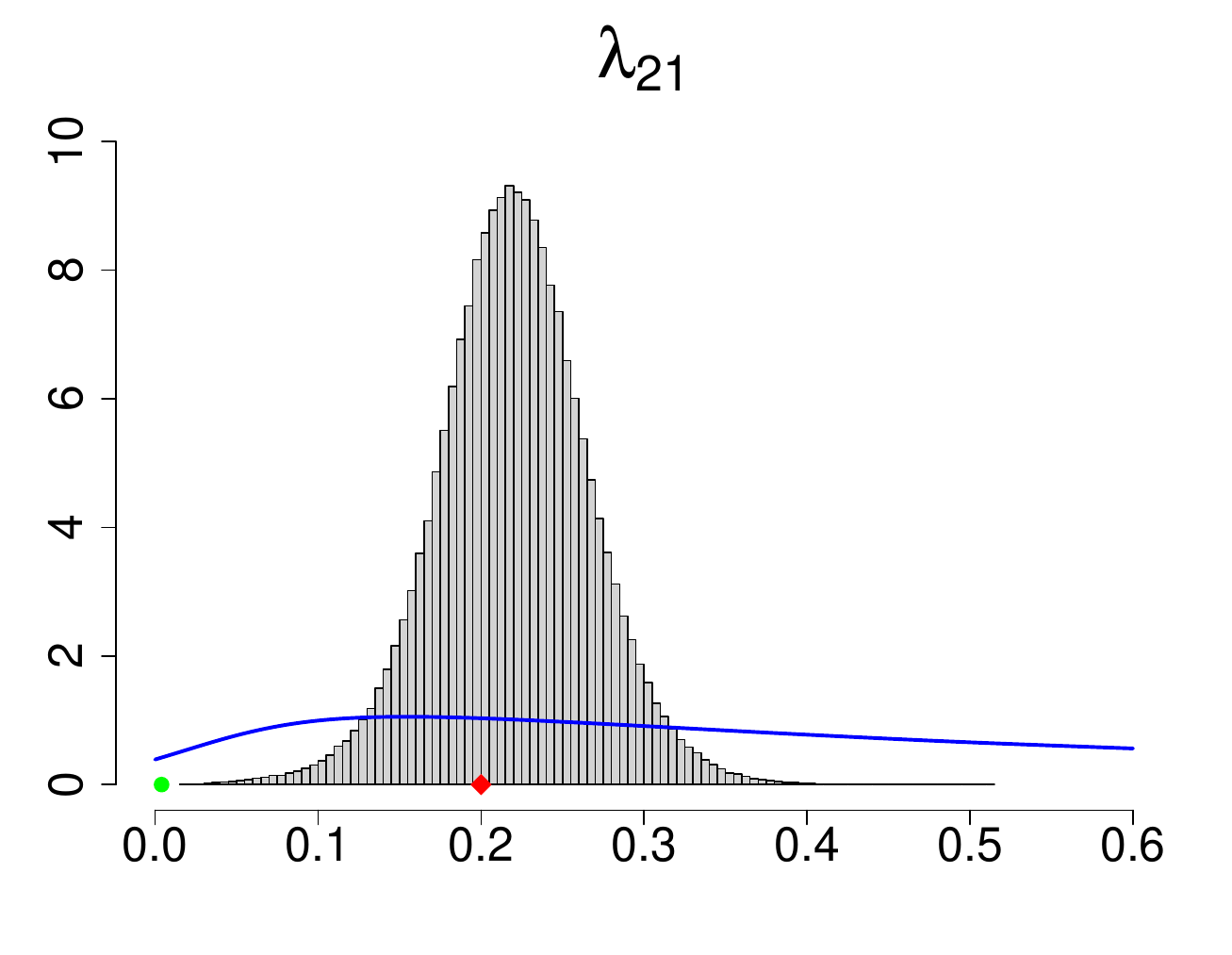} & \includegraphics[width=0.4\textwidth]{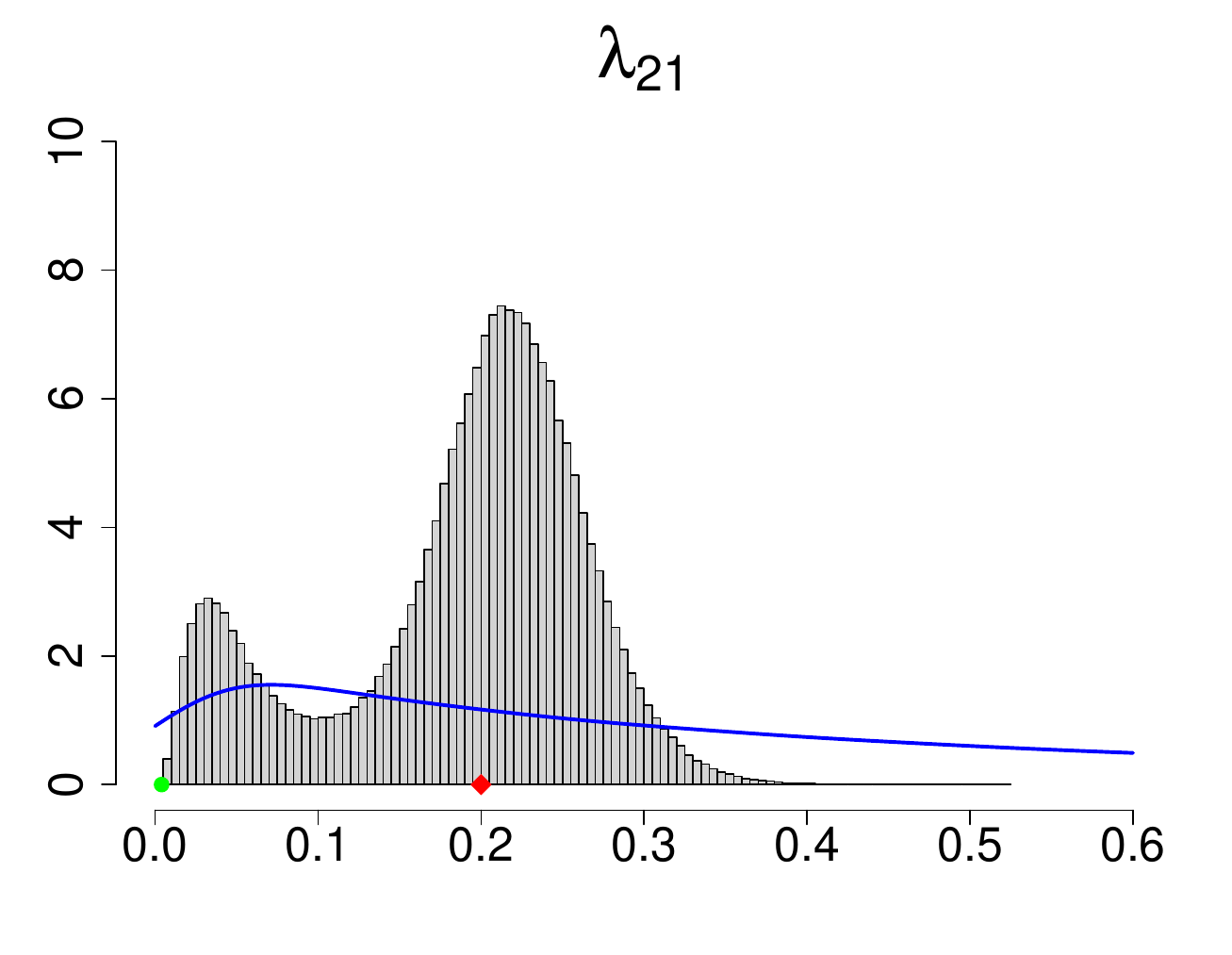}\\
		\hline
		\includegraphics[width=0.4\textwidth]{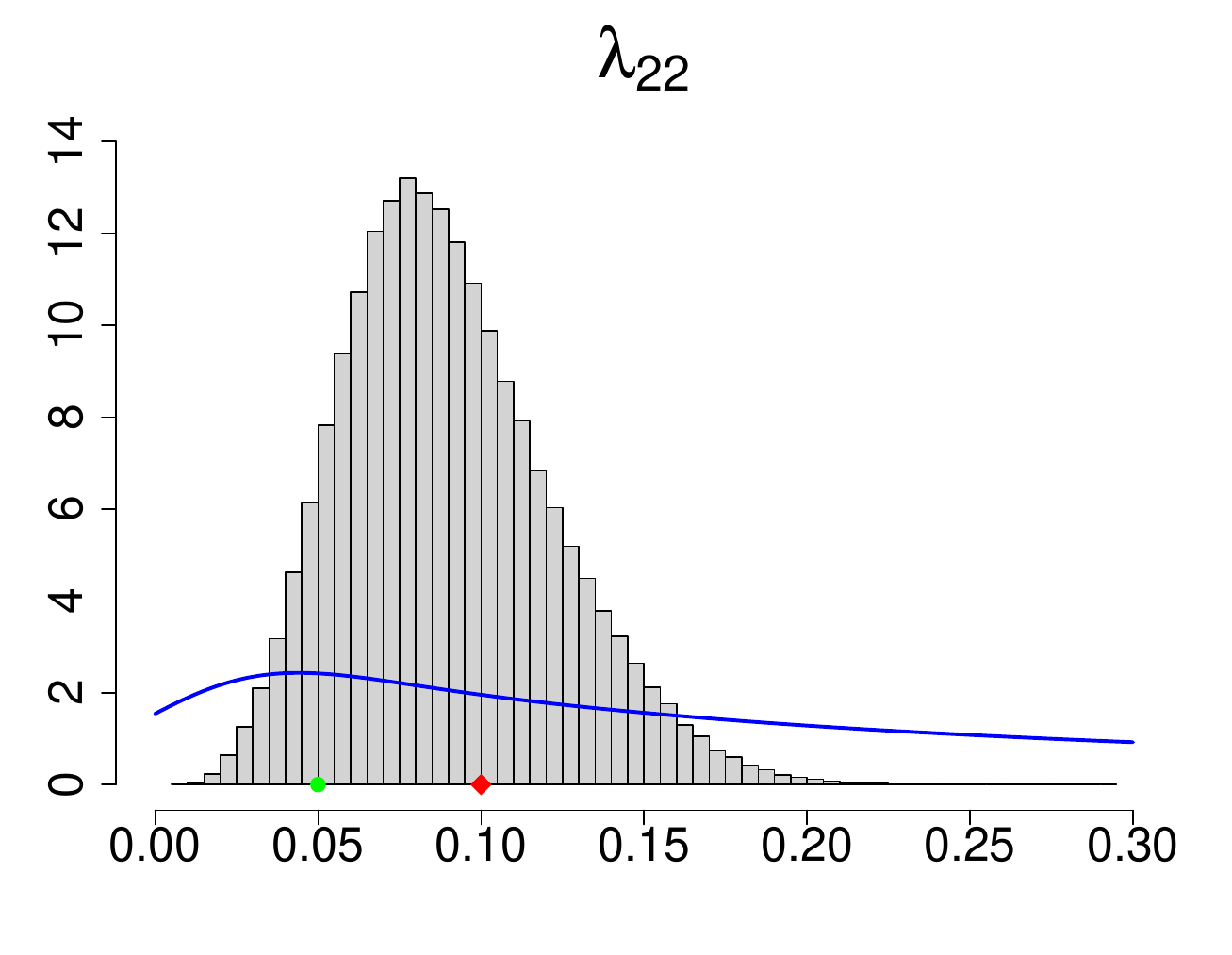} & \includegraphics[width=0.4\textwidth]{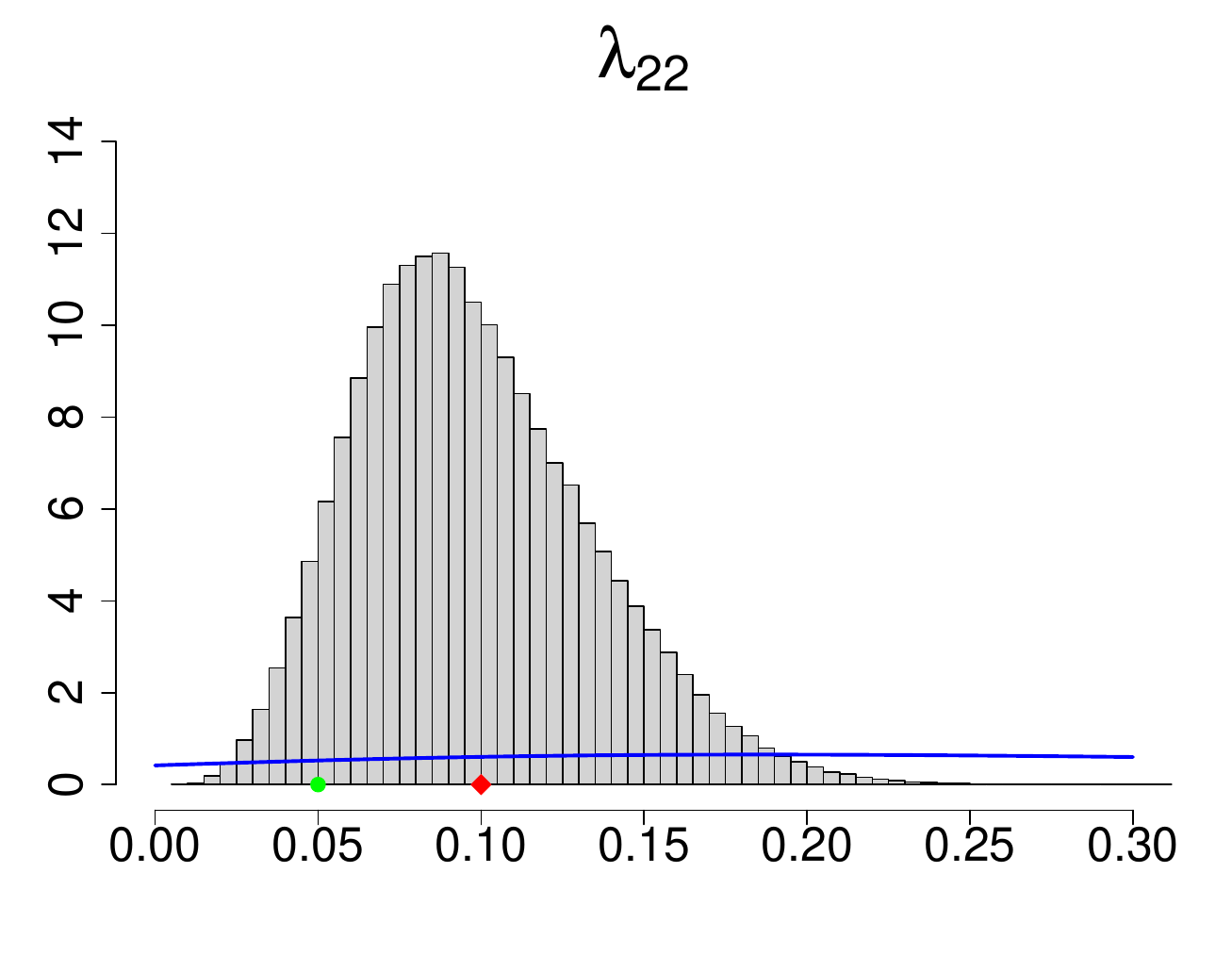}\\
		\hline
	\end{tabular}
	\caption{Posterior histograms and prior densities (lines) for the $\lambda_{m,i}$ parameters in the SC simulation study, with red and green markers indicating their original (underlying the DGP) and second-solution values, respectively.}
	\label{fig:SC_lambdas}
\end{figure}

\begin{figure}[h]
	\centering
	\begin{tabular}{|c|c|}
		\hline
		With the uniqueness restriction  &  Without the uniqueness restriction\\
		\hline\hline
		\includegraphics[width=0.4\textwidth]{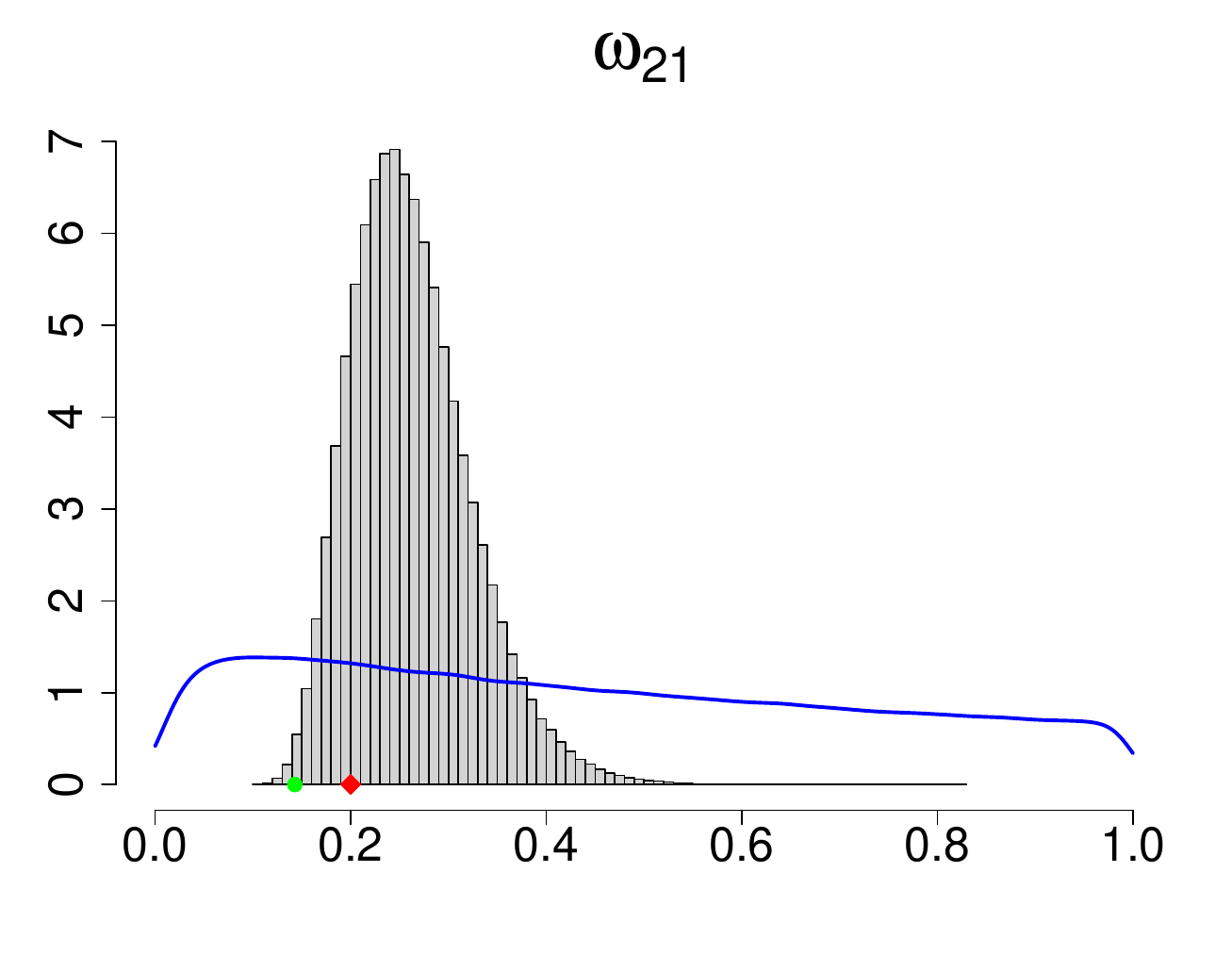} & \includegraphics[width=0.4\textwidth]{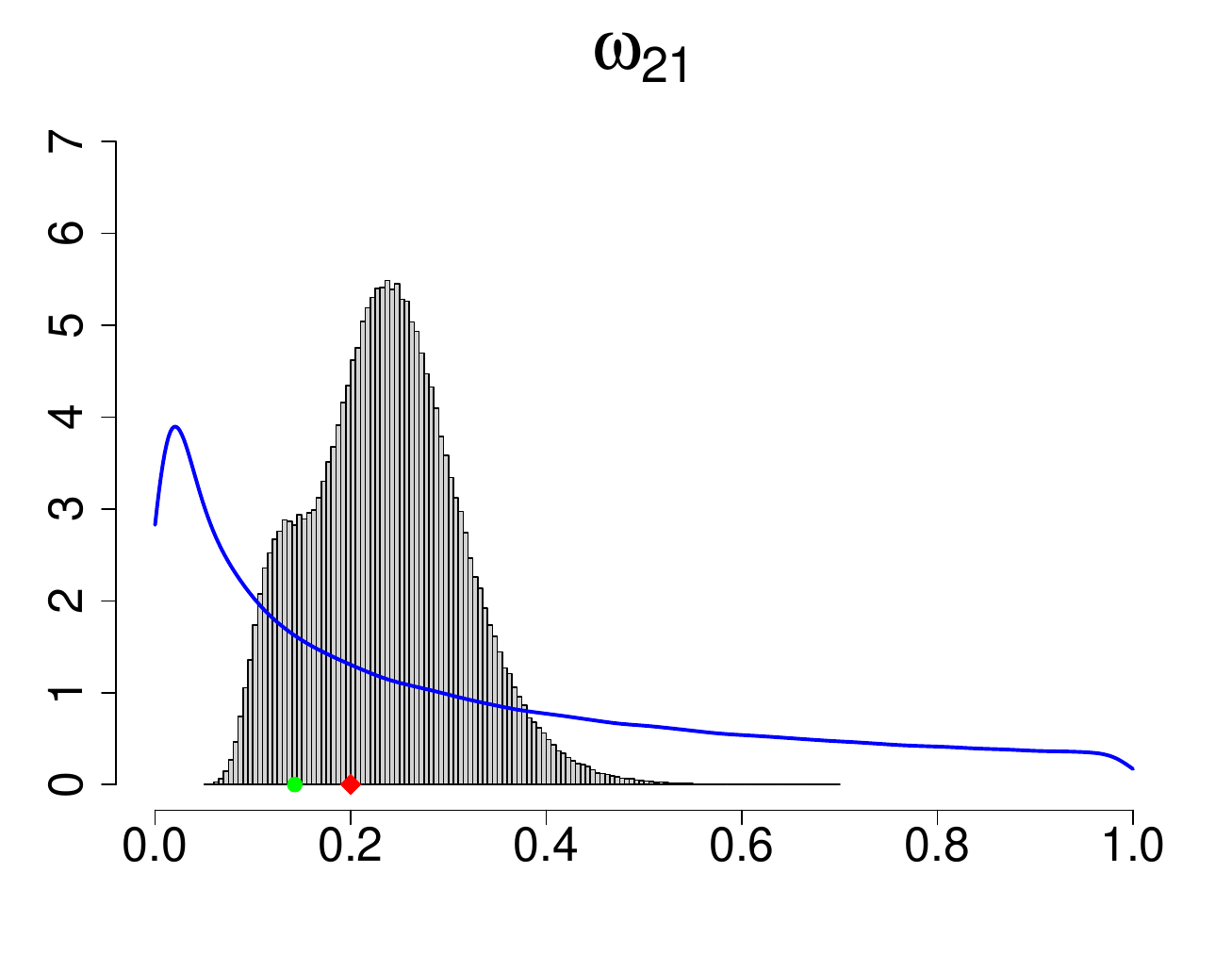}\\
		\hline
		\includegraphics[width=0.4\textwidth]{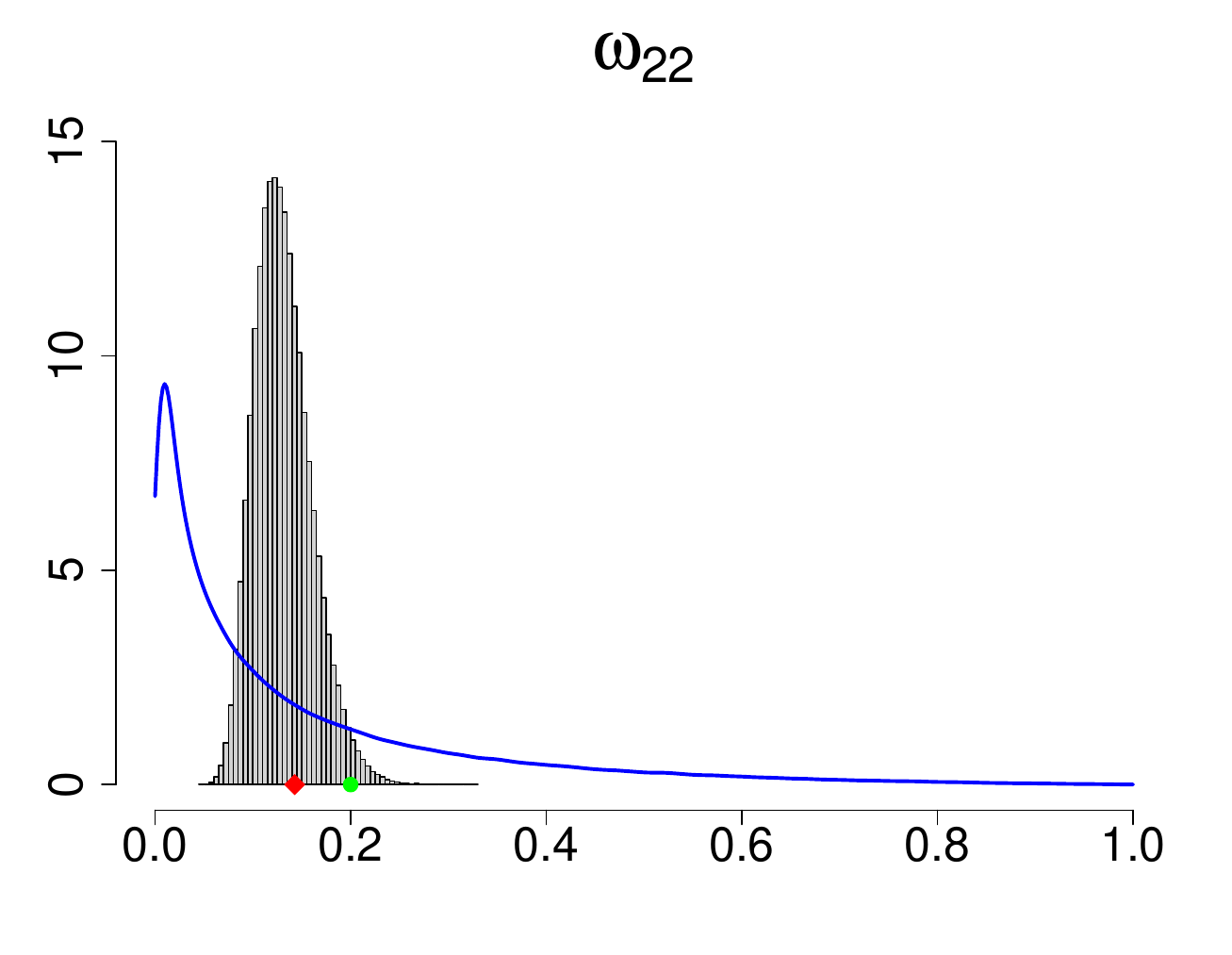} & \includegraphics[width=0.4\textwidth]{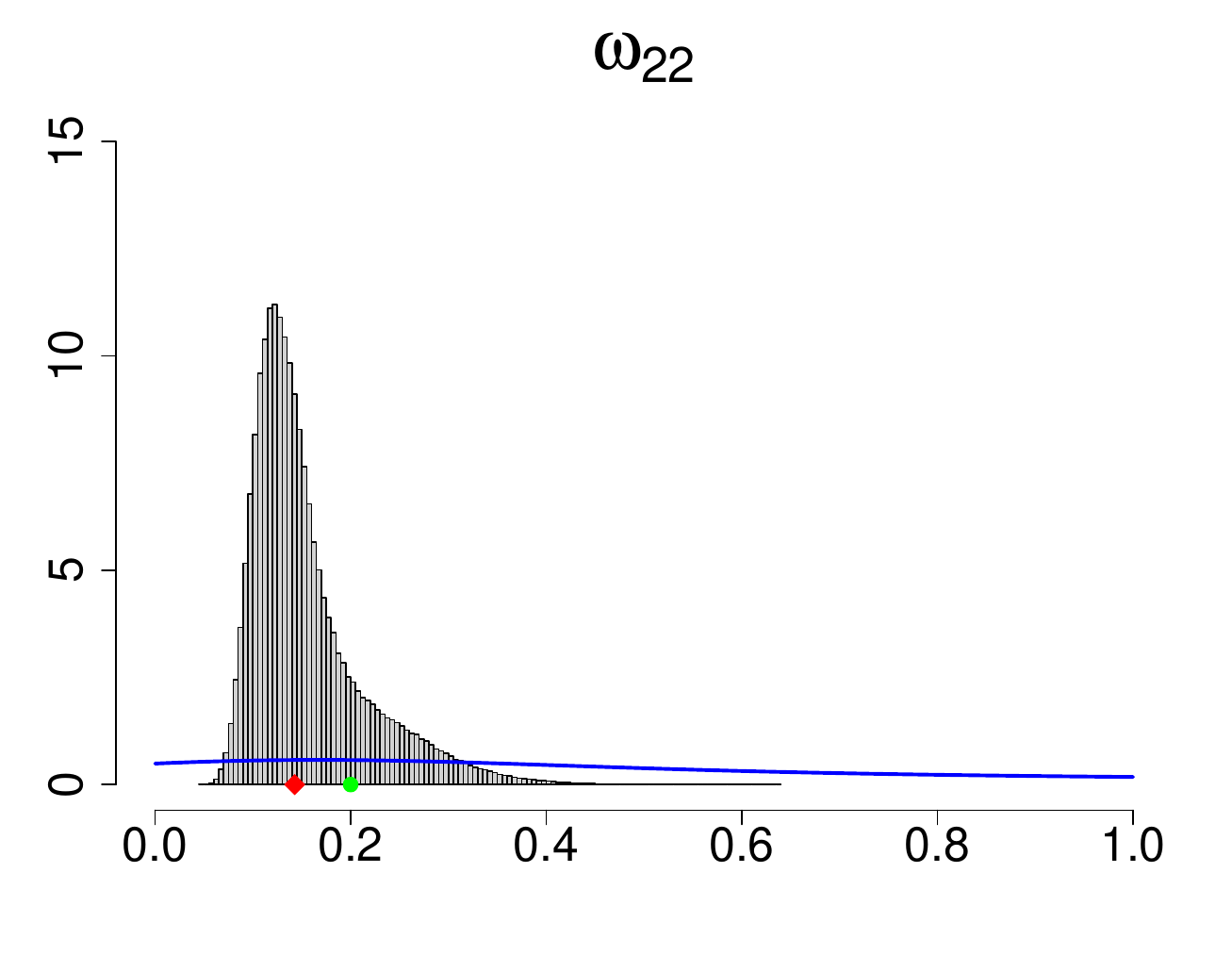}\\
		\hline
		\includegraphics[width=0.4\textwidth]{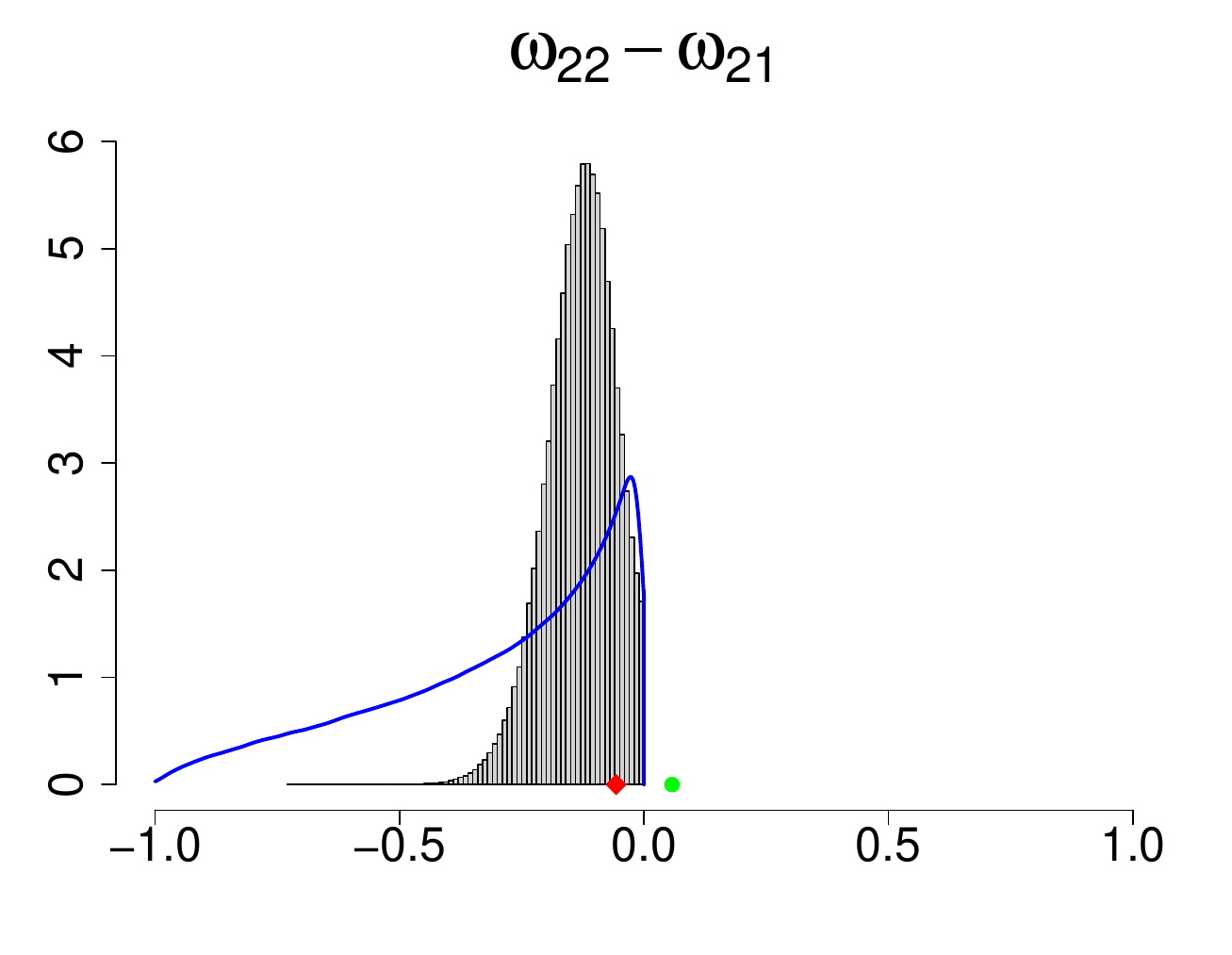} & \includegraphics[width=0.4\textwidth]{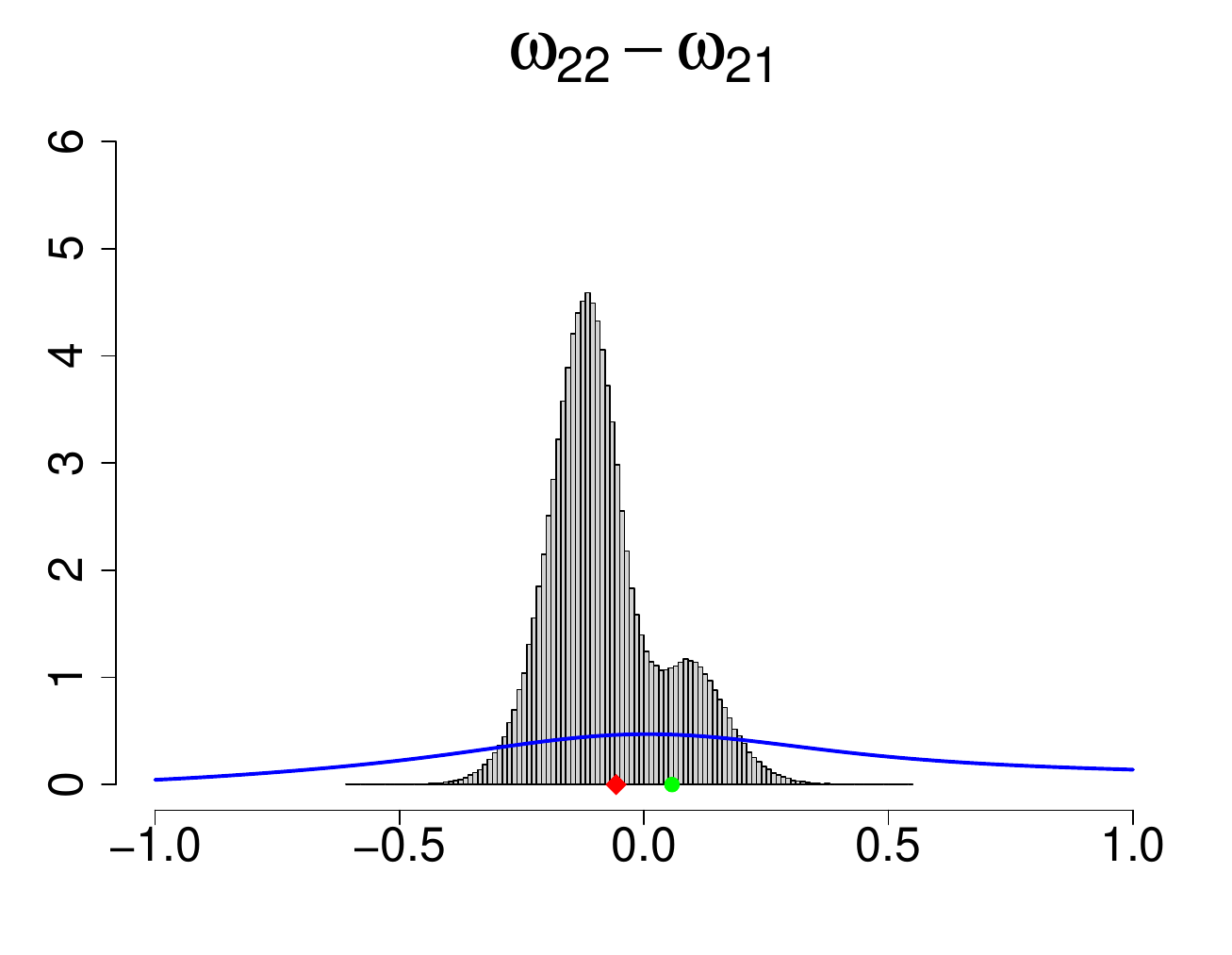}\\
		\hline
	\end{tabular}
	\caption{Posterior histograms and prior densities (lines) for the $\omega_{2,i}$ parameters and their contrast in the SC simulation study, with red and green markers indicating their original (underlying the DGP) and second-solution values, respectively.}
	\label{fig:SC_omegas}
\end{figure}

\begin{figure}[h]
	\centering
	\begin{tabular}{|c|c|}
		\hline
		With the uniqueness restriction  &  Without the uniqueness restriction\\
		\hline\hline
		\includegraphics[width=0.4\textwidth]{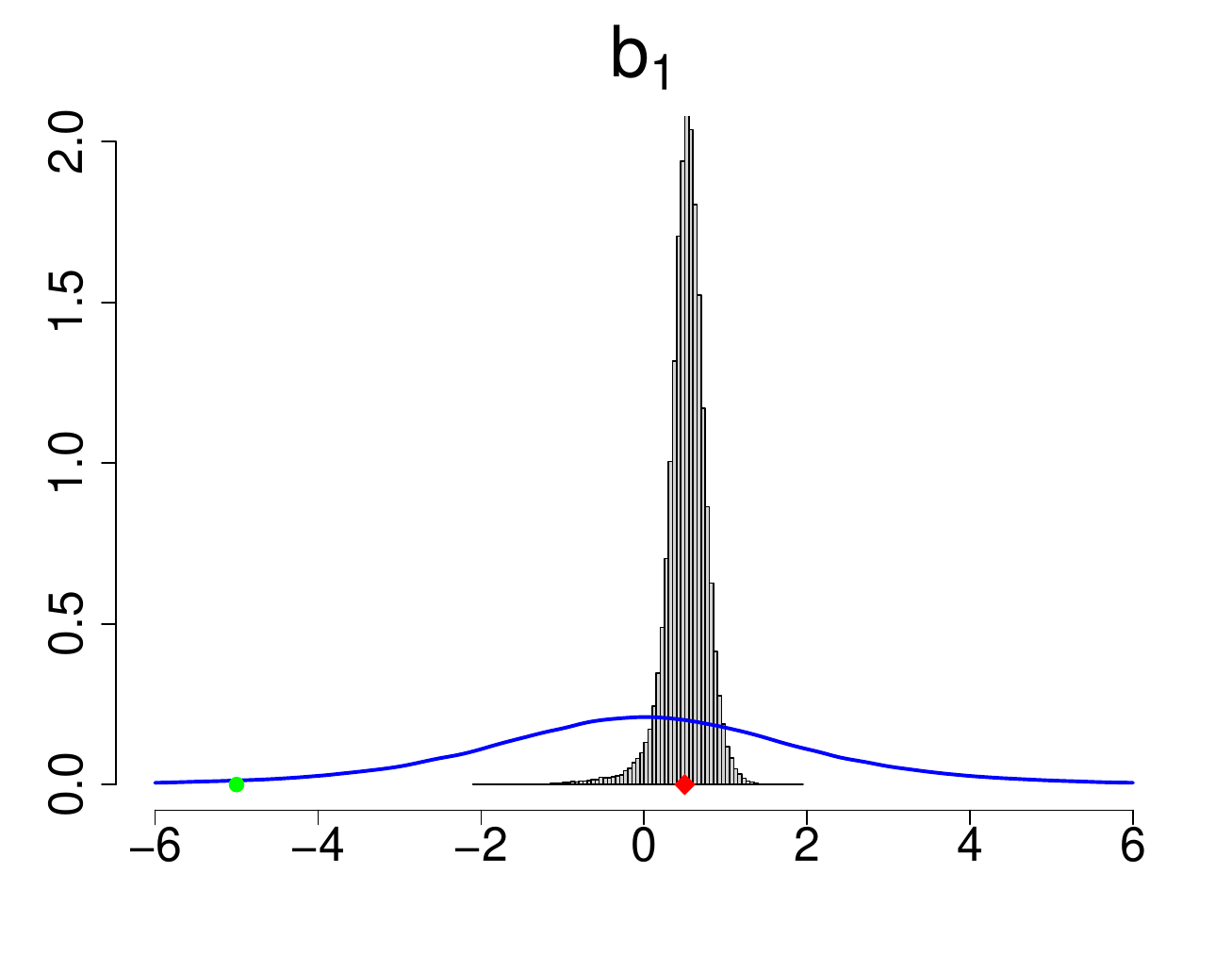} & \includegraphics[width=0.4\textwidth]{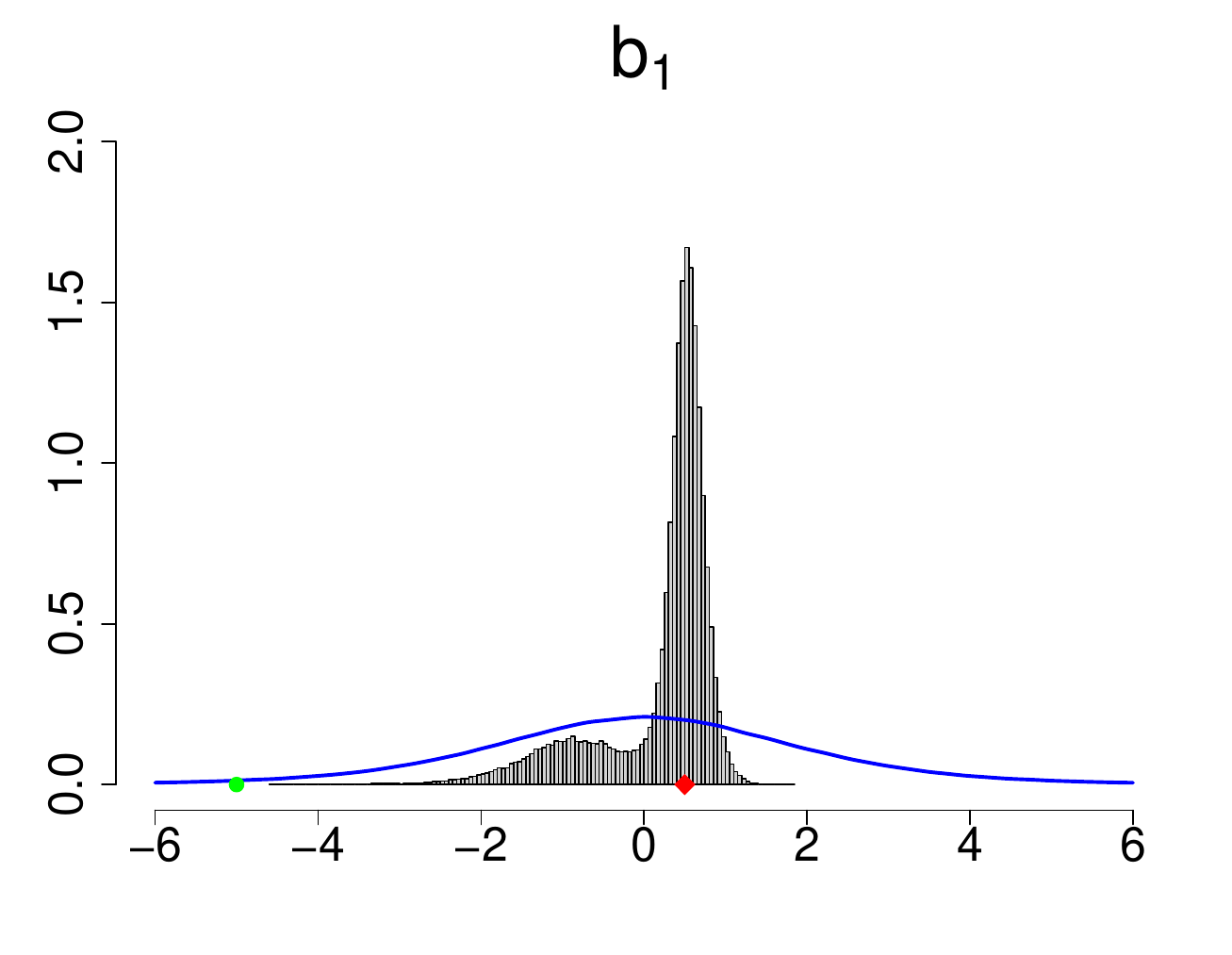}\\
		\hline
		\includegraphics[width=0.4\textwidth]{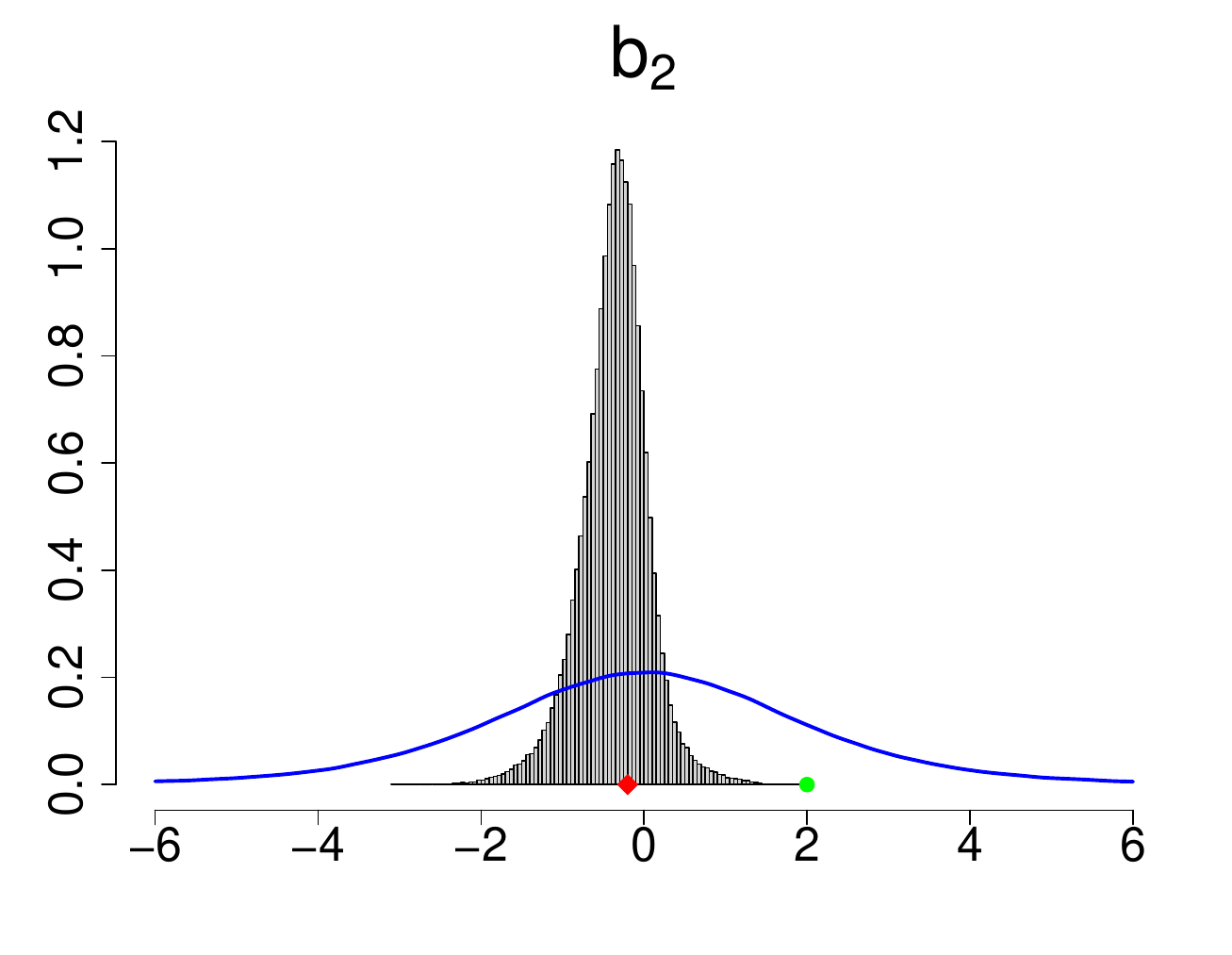} & \includegraphics[width=0.4\textwidth]{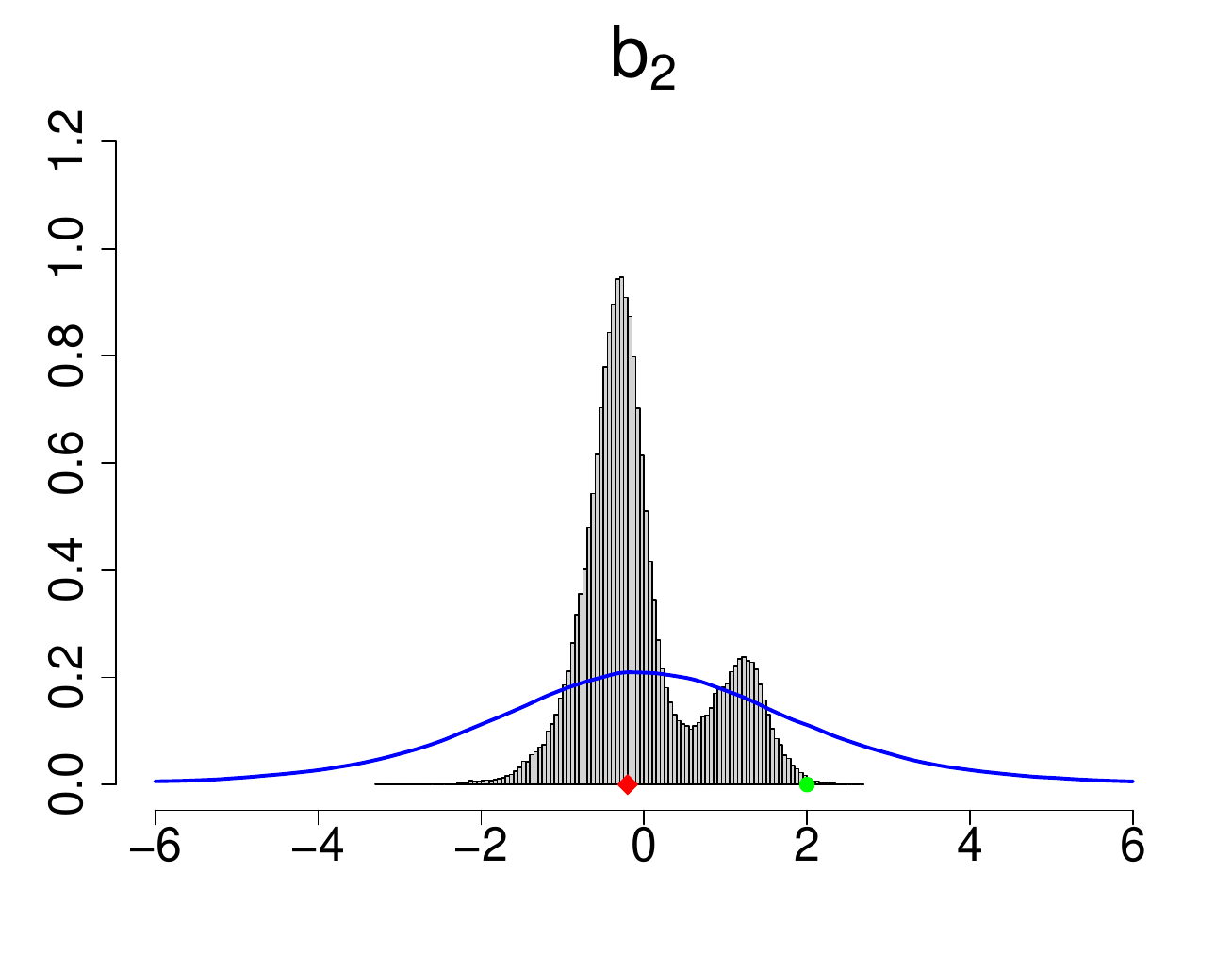}\\
		\hline
	\end{tabular}
	\caption{Posterior histograms and prior densities (lines) for the elements of the $B$ matrix in the SC simulation study, with red and green markers indicating their original (underlying the DGP) and second-solution values, respectively.}
	\label{fig:SC_bs}
\end{figure}

In Figures~\ref{fig:SC_lambdas}-\ref{fig:SC_bs}, pertaining to the SC case, we present the posterior histograms for the parameters key to identification, along with their prior densities, with red and green markers indicating their original (underlying the DGP) and second-solution values, respectively. In all those figures, the results are split into two columns, depending on whether or not the uniqueness restriction has been imposed: $\omega_{2,1}>\omega_{2,2}$. 

As seen in the right-hand-side columns of Figures~\ref{fig:SC_lambdas}-\ref{fig:SC_bs}, failing to impose the uniqueness restriction leads to most of the marginal posterior distributions displaying a second mode, conceivably stemming from the existence of an alternative solution. Note, however, that the other solution for $\lambda_{11}$, $\lambda_{21}$, and $b_2$ is only hardly covered by the second modal region, while the one for $b_1$ is already subdued a priori. The posteriors for the other parameters are also affected by the alternative solution, which manifests in either a more or less pronounced hump ($\lambda_{12}$, $\omega_{2,1}$ and $\omega_{2,2}$) or a slightly more 'squat' histogram ($\lambda_{22}$). However, introducing the uniqueness restriction into the model removes all the anomalies and regularises all the posteriors.

\begin{table}[htbp]
	\centering
	\begin{tabular}{|c||c|c|}
		\hline
		\multicolumn{3}{|c|}{The first (original) solution} \\
		\hline\hline
		Parameter & Small contrast & Large contrast\\
		\hline
		&&\\
		$B^{(1)}$ & $\left(\begin{array}{cc}1 & -0.2\\0.5 & 1\end{array} \right)$ & $\left(\begin{array}{cc}1 & -0.2\\0.5 & 1\end{array} \right)$\\
		&&\\
		$\lambda_1^{(1)}$&$\left(\begin{array}{cc}1 & 0.7\end{array} \right)$&$\left(\begin{array}{cc}2.5 & 1.5\end{array} \right)$\\
		&&\\
		$\lambda_2^{(1)}$&$\left(\begin{array}{cc}0.2 & 0.1\end{array} \right)$&$\left(\begin{array}{cc}2 & 0.24\end{array} \right)$\\
		&&\\
		$\omega_2^{(1)'}$&$\left(\begin{array}{cc}\frac{1}{5} & \frac{1}{7}\end{array} \right)$&$\left(\begin{array}{cc}0.8 & 0.16\end{array} \right)$\\
		&&\\
		$\omega_{2,2}^{(1)}-\omega_{2,1}^{(1)}$&$-\frac{2}{35}\approx -0.057$&$-0.64$\\
		&&\\
		\hline\hline
		\multicolumn{3}{|c|}{The second solution}\\
		\hline\hline
		Parameter & Small contrast & Large contrast\\
		\hline
		&&\\
		$B^{(2)}$ & $\left(\begin{array}{cc}1 & 2\\-5 & 1\end{array} \right)$ & $\left(\begin{array}{cc}1 & 2\\-5 & 1\end{array} \right)$\\
		&&\\
		$\lambda_1^{(2)}$&$\left(\begin{array}{cc}0.028 & 0.25\end{array} \right)$&$\left(\begin{array}{cc}0.06 & 0.625\end{array} \right)$\\
		&&\\
		$\lambda_2^{(2)}$&$\left(\begin{array}{cc}0.004 & 0.05\end{array} \right)$&$\left(\begin{array}{cc}0.0096 & 0.5\end{array} \right)$\\
		&&\\
		$\omega_2^{(2)'}$&$\left(\begin{array}{cc}\frac{1}{7} & \frac{1}{5}\end{array} \right)$&$\left(\begin{array}{cc}0.16 & 0.8\end{array} \right)$\\
		&&\\
		$\omega_{2,2}^{(2)}-\omega_{2,1}^{(2)}$&$\frac{2}{35}\approx 0.057$&$0.64$\\
		&&\\
		\hline\hline
	\end{tabular}
	\caption{The true values of parameters in the simulated-data examples}
	\label{tab:true_parameters}
\end{table}

\subsection{Empirical illustration: Additional results}

\subsubsection{The number of cointegration relations}

For choosing the number of cointegration relationships, we resort to the Savage-Dickey density ratio (\textit{SDDR}; see, e.g., \citet{verdinelli1995computing}, \citet{mulder2022generalization}), which is the Bayes factor of a constrained model ($M_c$) and the unconstrained model ($M_u$), calculated as a ratio of the posterior and prior densities at the restriction:
\begin{equation}
	B_{cu} = \frac{p(y|M_c)}{p(y|M_u)} = \frac{p(\alpha = 0, \Gamma = 0|y, M_u)}{p(\alpha = 0, \Gamma = 0|M_u)}
\end{equation}
For the unconstrained model, we actually consider four different specifications featuring various rank of the cointegration matrix ($r\in \{0, 1, 2, 3\}$), under all of which imposing $\alpha = 0, \Gamma = 0$ yields a very same multivariate random walk with a Markov-switching covariance matrix. This constrained model serves here as the base structure, to which all of the unconstrained models ($r\in \{0, 1, 2, 3\}$) are compared. Therefore, it is more conducive here to analyze the reciprocals of $B_{cu}$: $B_{uc}=1/B_{cu}$, the values of which are presented in Table~\ref{tab:SDDR} (under equal prior probabilities of all the unconstrained models, $Pr(M_u(r))=\frac{1}{4}$ for $r\in\{0,1,2,3\}$).
\begin{table}[h!]
	\centering
	\begin{tabular}{|c||c|c|c|c|}
		\hline
		$r$ & 0 & 1 & 2 & 3 \\
		\hline\hline
		$log_{10}(B_{uc})$ &  22.638 & 29.265 & 28.882 & 27.347 \\
		\hline
	\end{tabular}
	\caption{Decimal logs of the inverse Savage-Dickey ratios for comparison of the unconstrained models with a given cointegration rank $r\in\{0,1,2,3\}$, against the base model (random walk with a Markov-switching covariance matrix).}
	\label{tab:SDDR}
\end{table}

The table indicates a huge superiority of the unconstrained models, with the specification featuring only one cointegration relation outperforming the others (although $r=2$ is a very close second), and therefore being selected here for further analysis.

\subsubsection{Impulse responses}
\begin{figure}
	\begin{center}
		\begin{tabular}{|>{\centering\arraybackslash}m{.33\linewidth} |>{\centering\arraybackslash}m{.33\linewidth} |>{\centering\arraybackslash}m{.33\linewidth}|}
			\hline
			\large{1} &
			\includegraphics[width=0.33\textwidth]{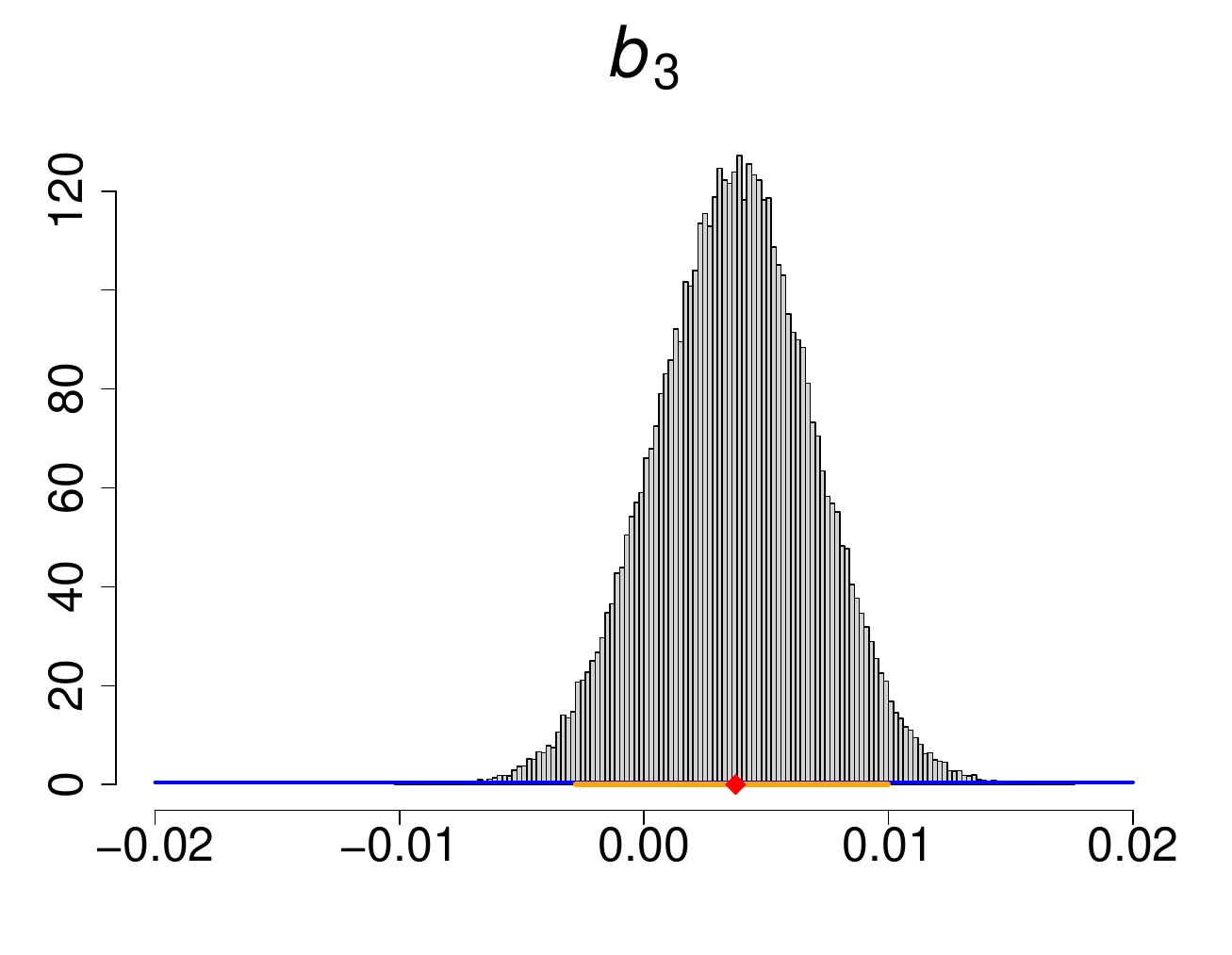} \tiny{$Me = 0.004,\ HPD = (-0.003, 0.010)$}&
			\includegraphics[width=0.33\textwidth]{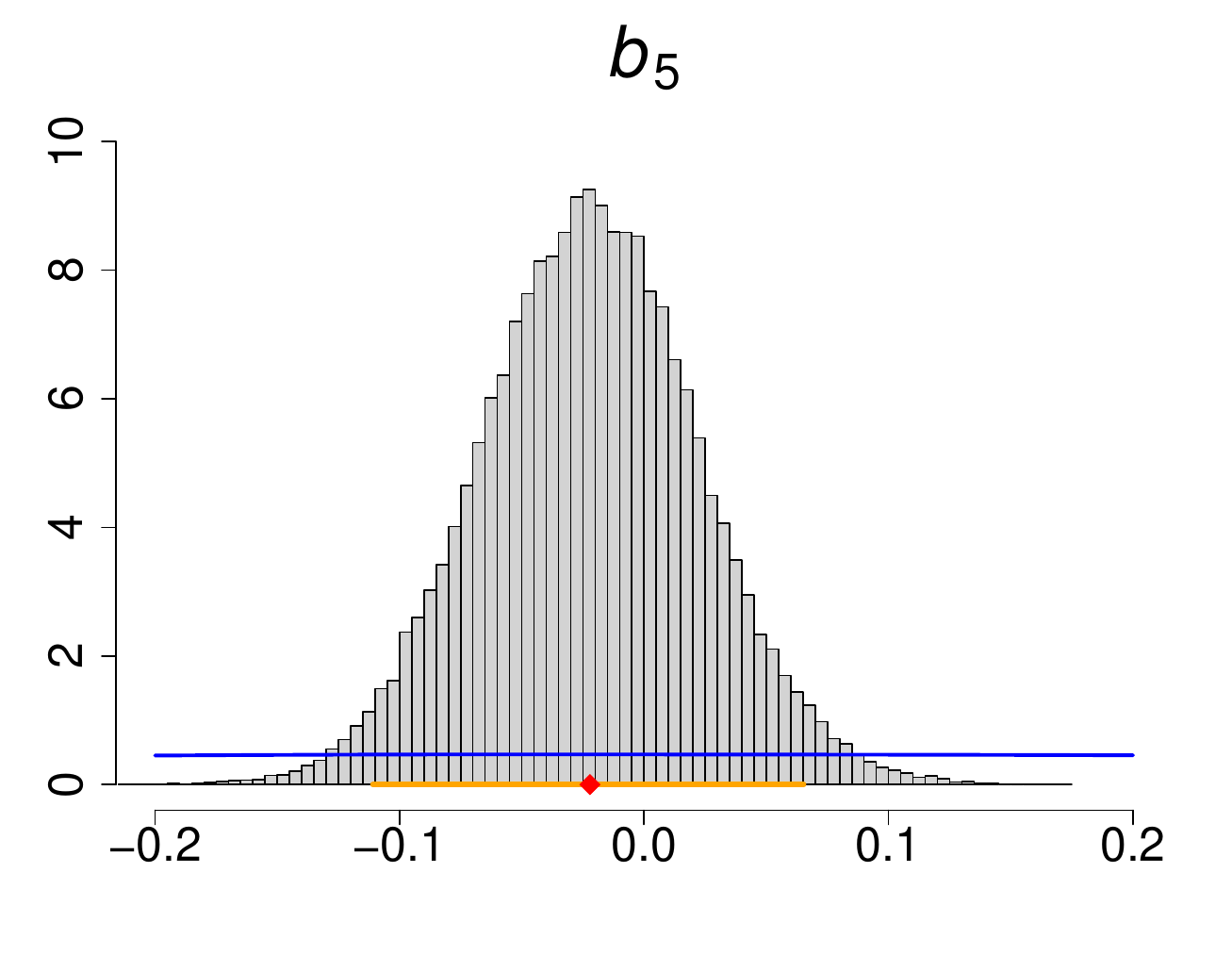} \tiny{$Me = -0.022,\ HPD = (-0.111, 0.065)$}\\
			\hline
			\includegraphics[width=0.33\textwidth]{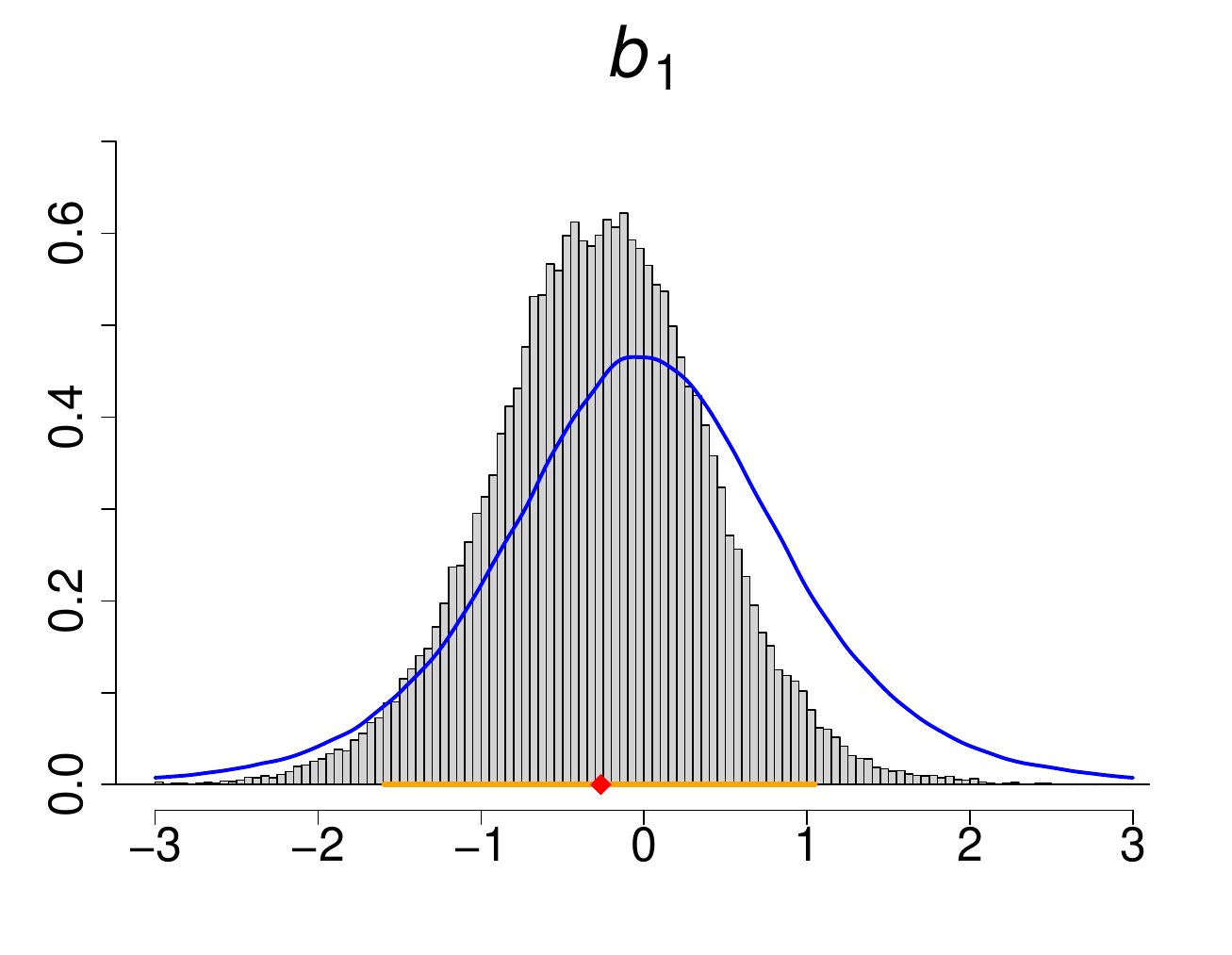} \tiny{$Me = -0.265,\ HPD = (-1.595, 1.048)$} & \large{1} &
			\includegraphics[width=0.33\textwidth]{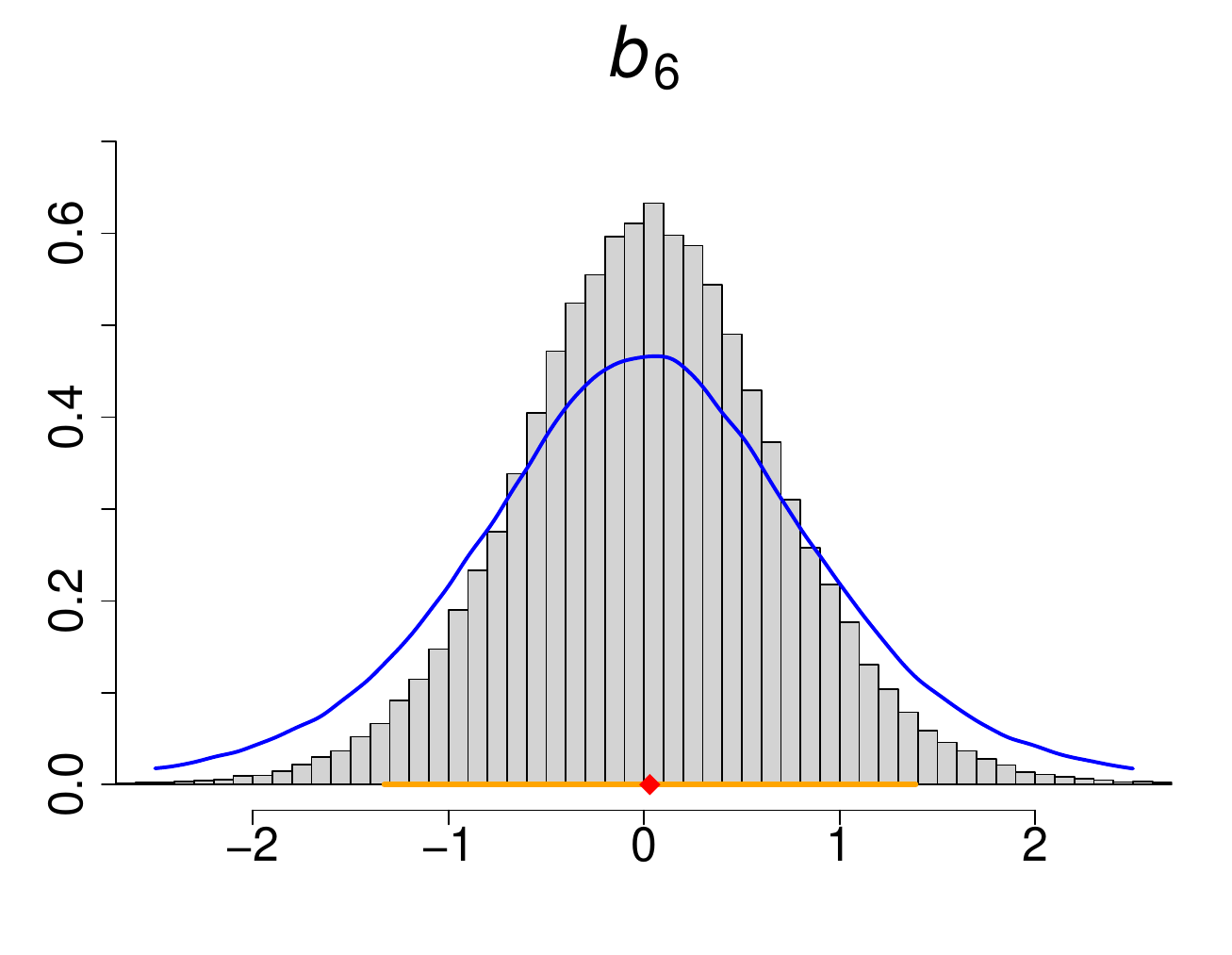} \tiny{$Me = 0.029,\ HPD = (-1.328, 1.389)$}\\
			\hline
			\includegraphics[width=0.33\textwidth]{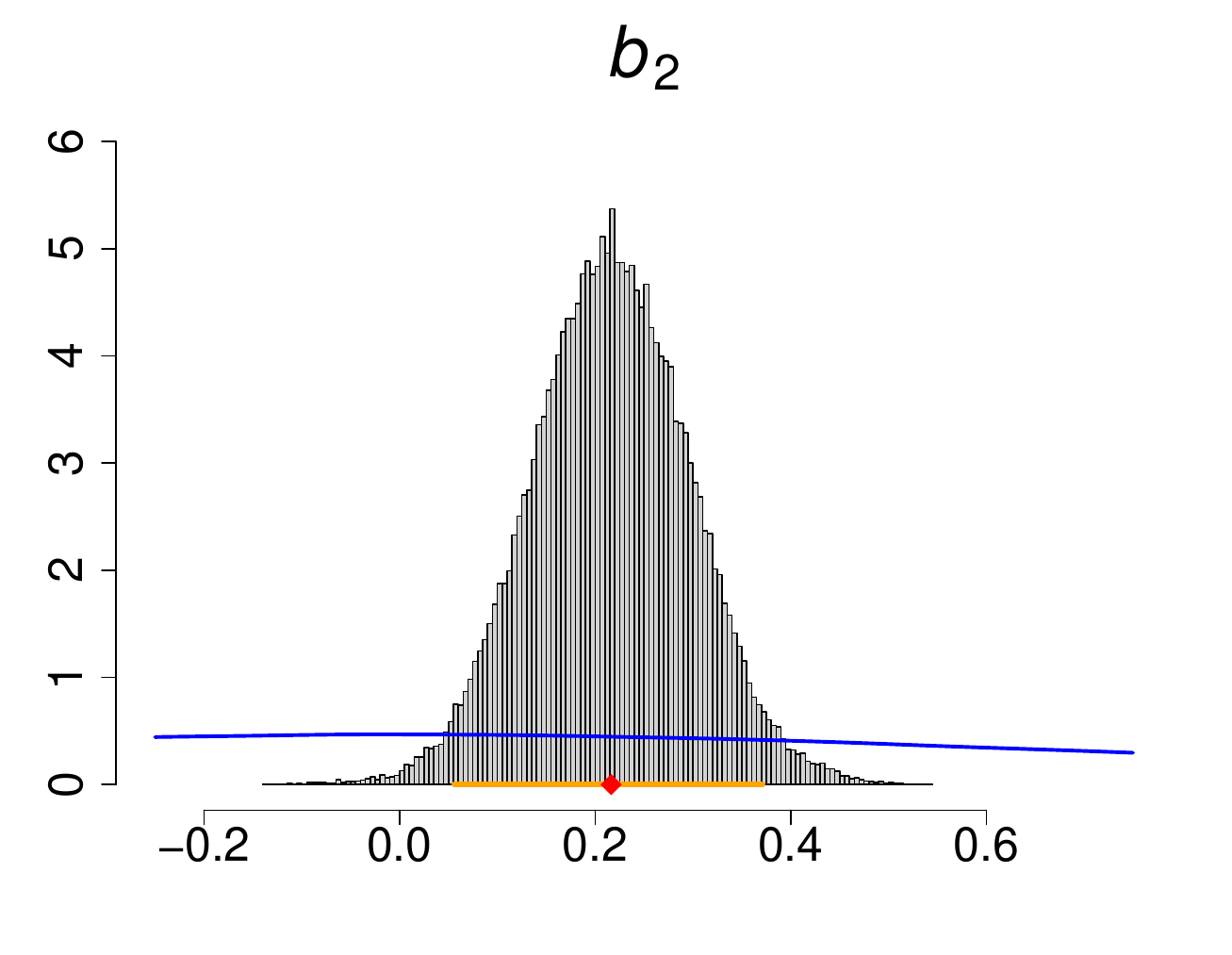} \tiny{$Me = 0.216,\ HPD = (0.056, 0.371)$} & \includegraphics[width=0.33\textwidth]{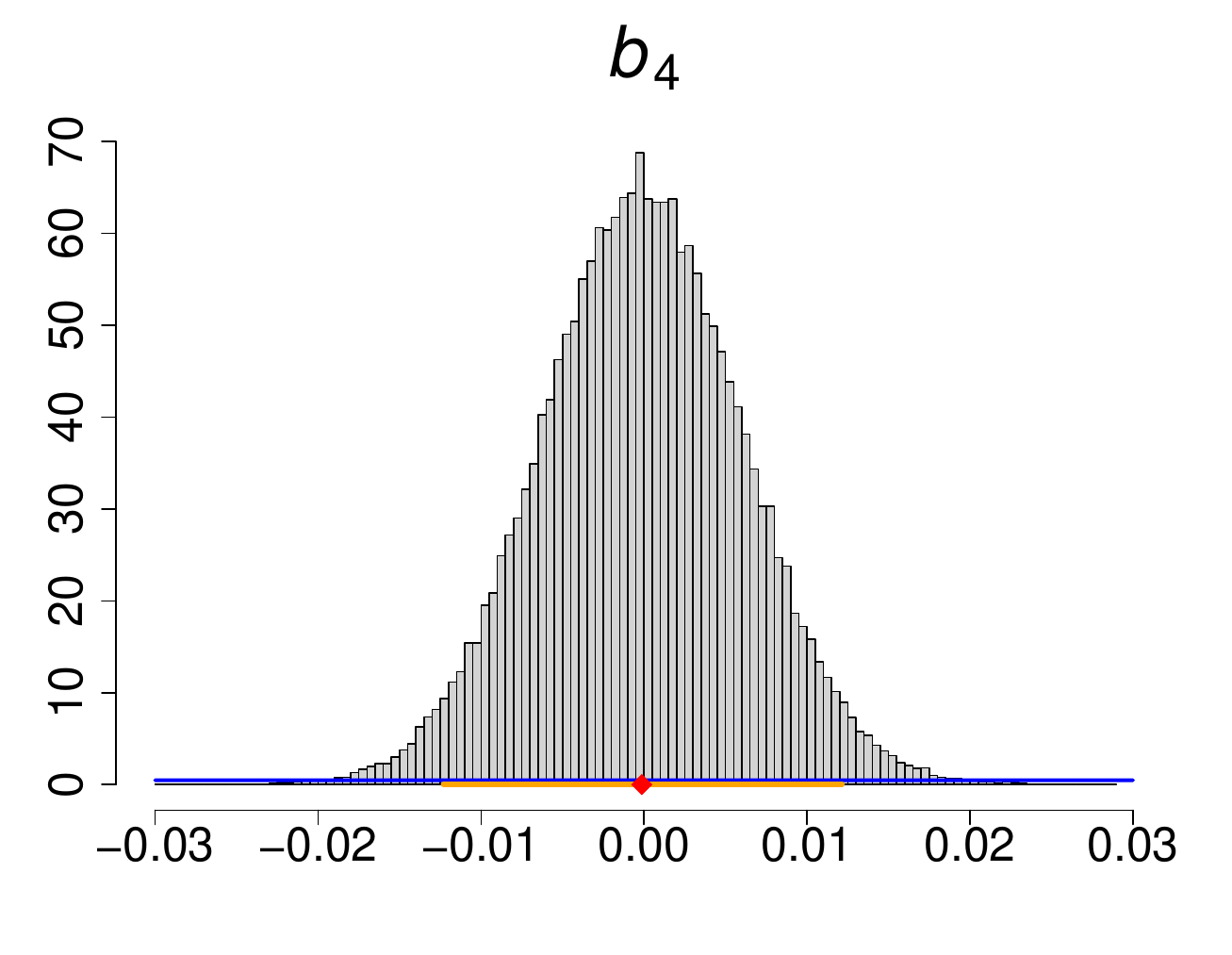} \tiny{$Me = 0.000,\ HPD = (-0.012, 0.012)$} & \large{1}\\
			\hline
		\end{tabular}    
	\end{center}
	\caption{Posterior histograms and prior densities (lines) for the elements of the $B$ matrix in the empirical example, with the posterior medians (red markers) and 95\% HPD intervals (orange line).}
	\label{fig:B_real}
\end{figure}

\begin{figure}
	\begin{center}
		\begin{tabular}{|>{\centering\arraybackslash}m{.33\linewidth} |>{\centering\arraybackslash}m{.34\linewidth} |>{\centering\arraybackslash}m{.33\linewidth}|}
			\hline
			\includegraphics[width=0.33\textwidth]{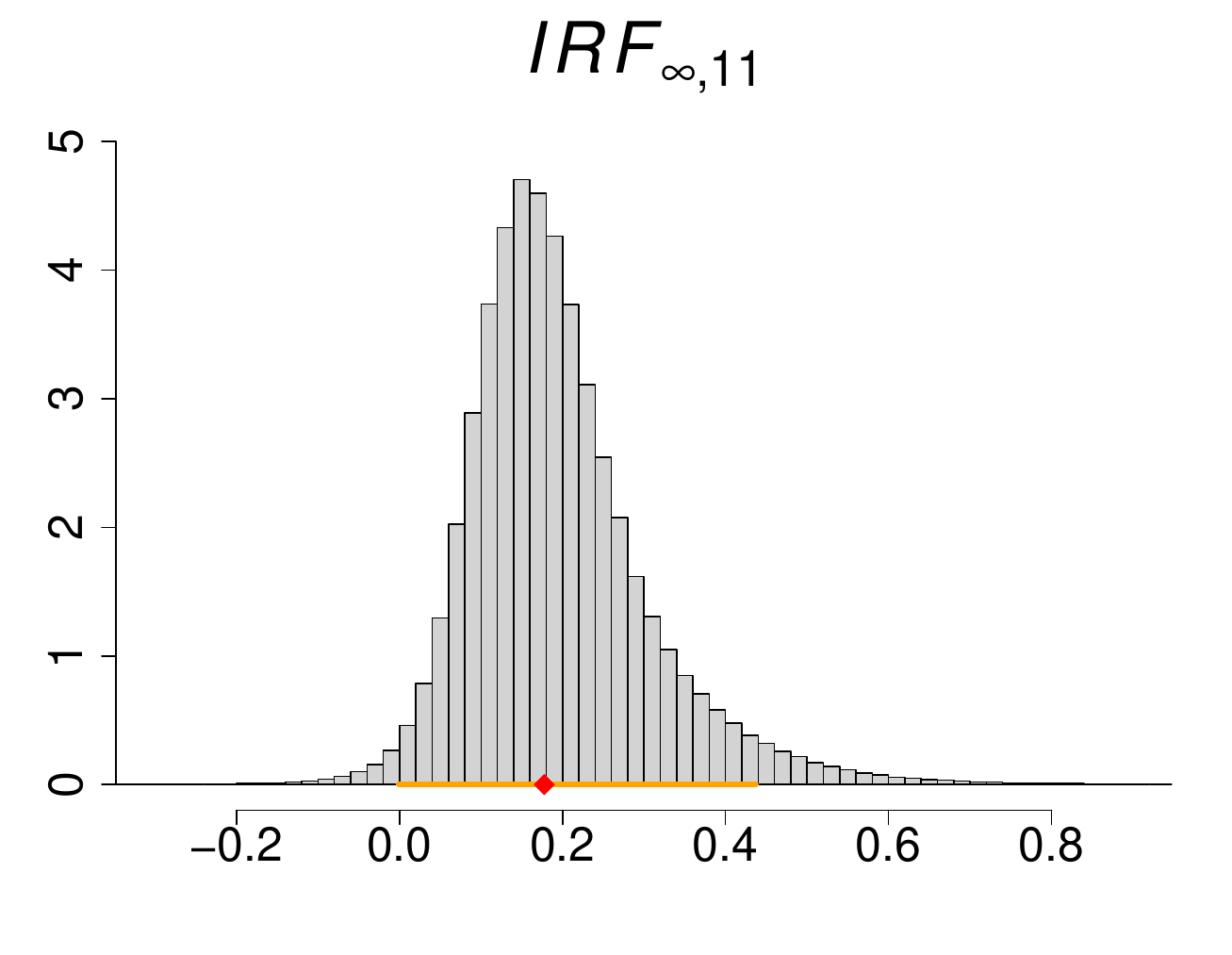} \tiny{$Me = 0.177,\ HPD = (-0.001, 0.437)$} &
			\includegraphics[width=0.33\textwidth]{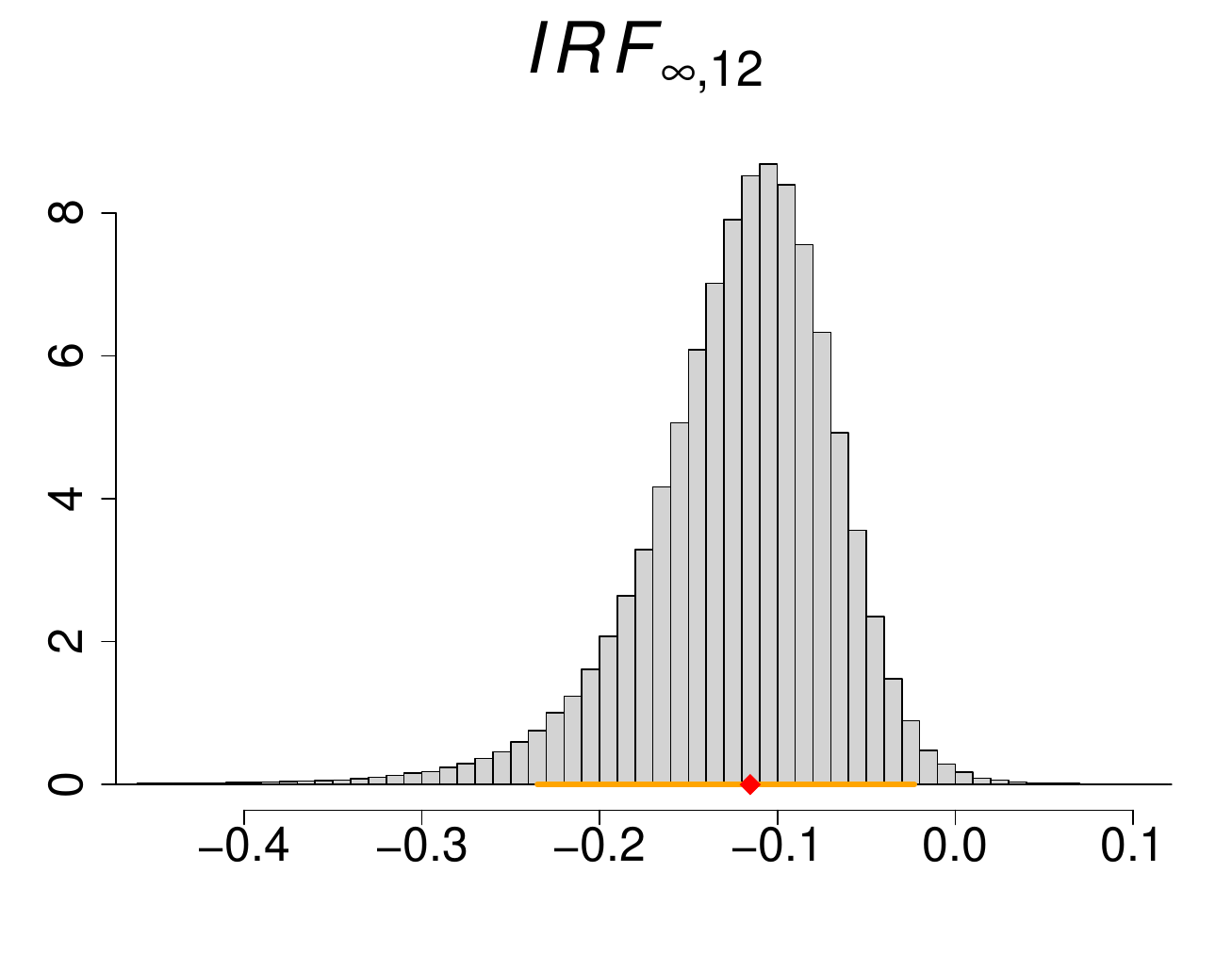} \tiny{$Me = -0.115,\ HPD = (-0.235, -0.023)$}&
			\includegraphics[width=0.33\textwidth]{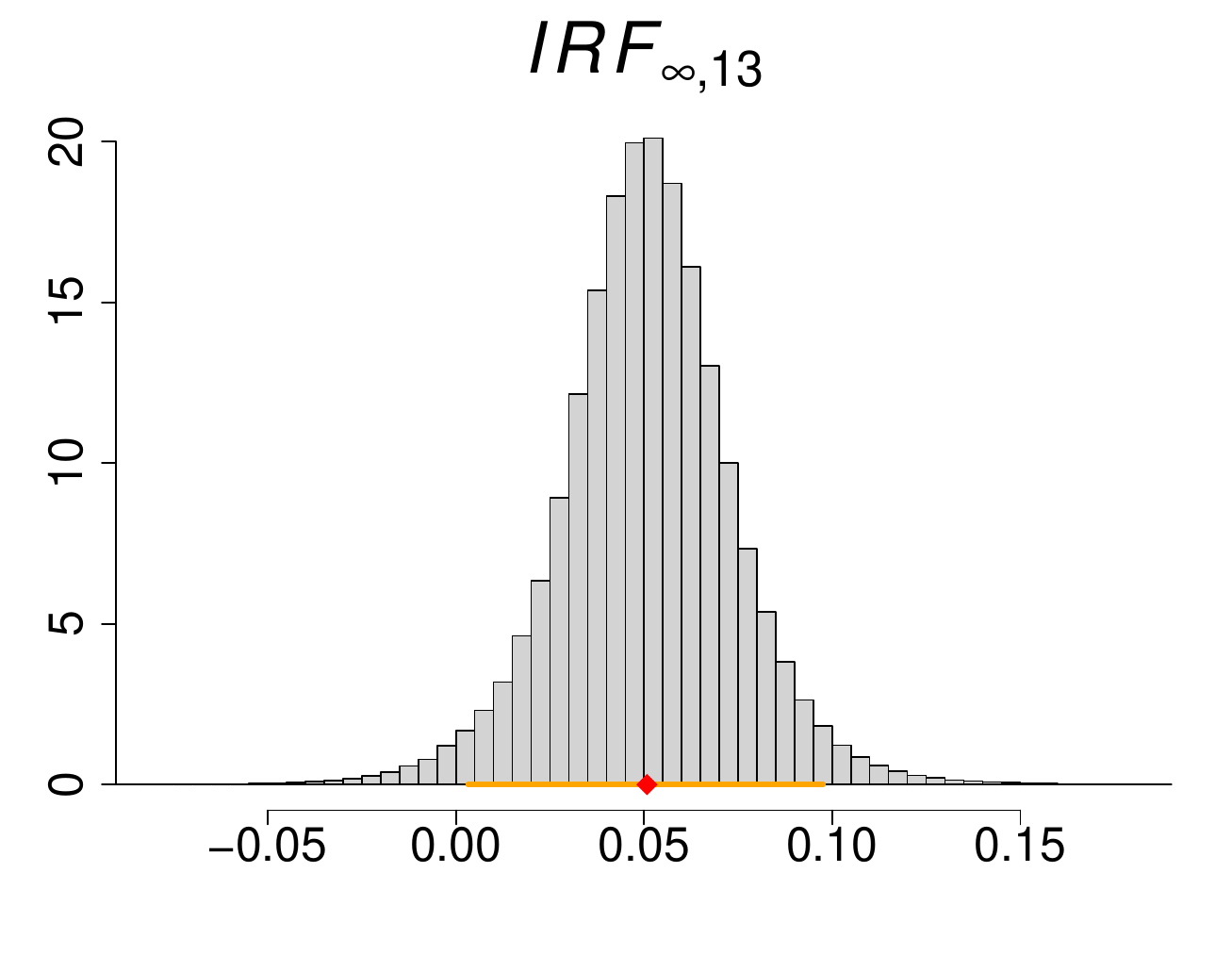} \tiny{$Me = 0.051,\ HPD = (0.003, 0.098)$}\\
			\hline
			\includegraphics[width=0.33\textwidth]{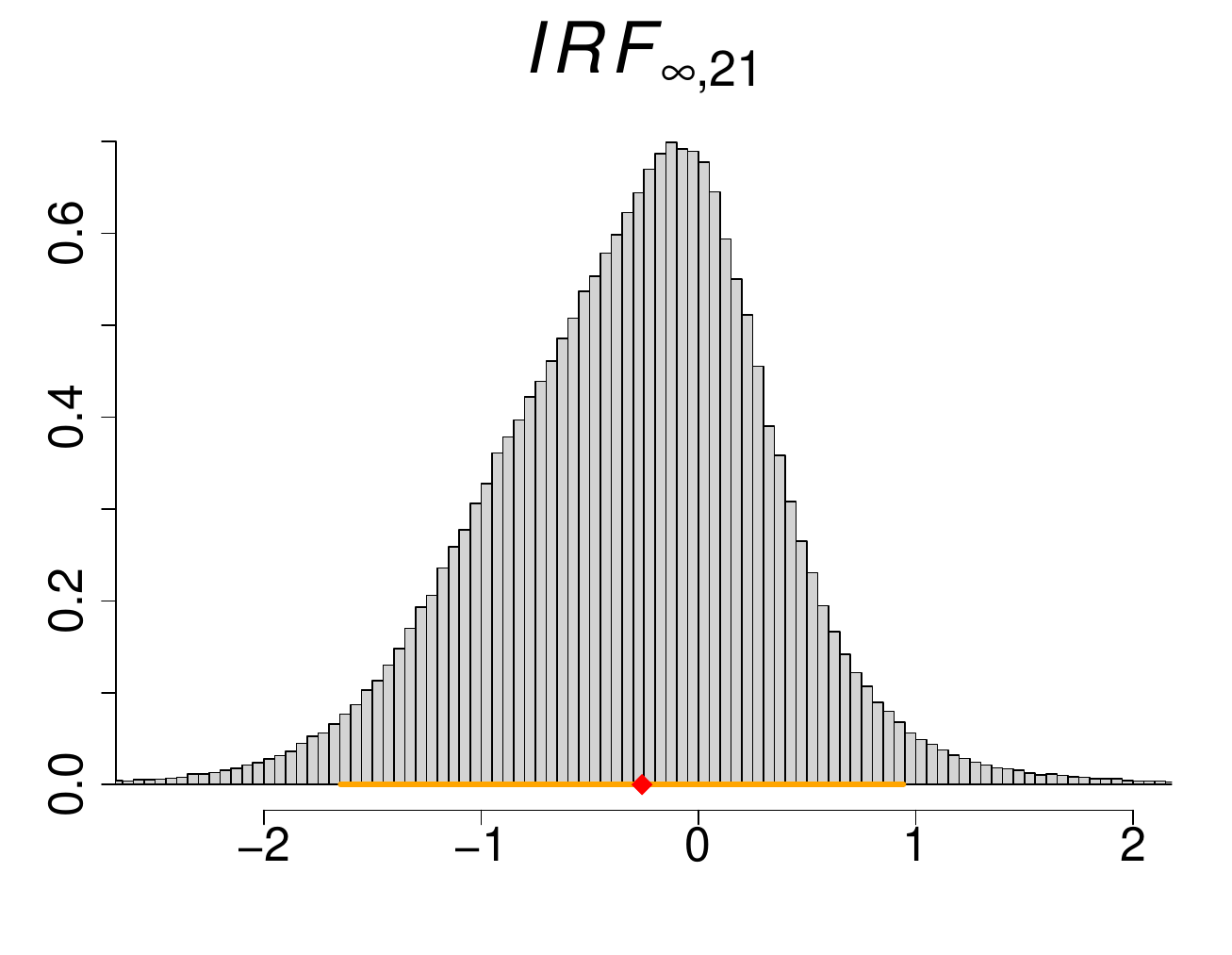} \tiny{$Me = -0.259,\ HPD = (-1.647, 0.944)$} &
			\includegraphics[width=0.33\textwidth]{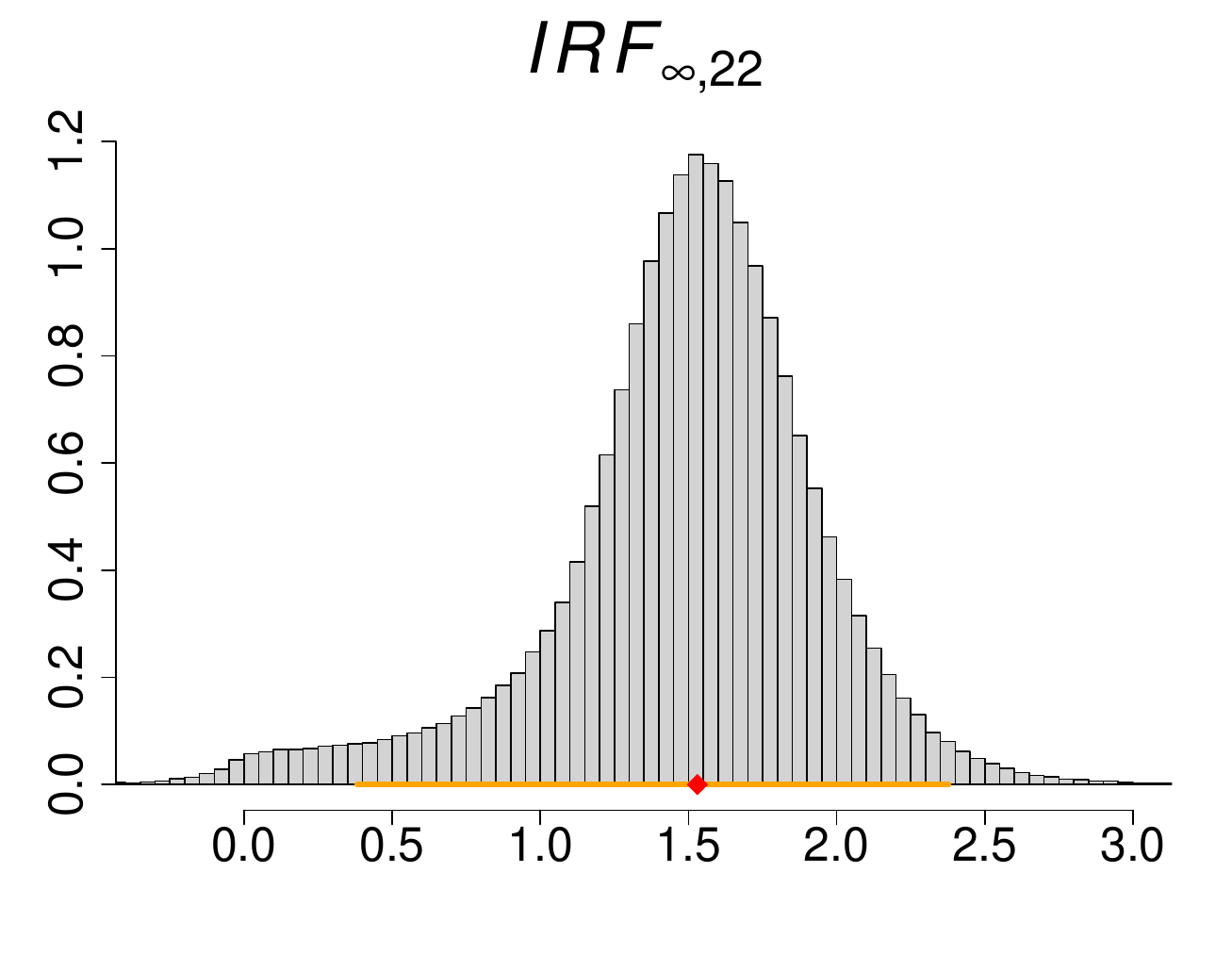} \tiny{$Me = 1.530,\ HPD = (0.381, 2.378)$}&
			\includegraphics[width=0.33\textwidth]{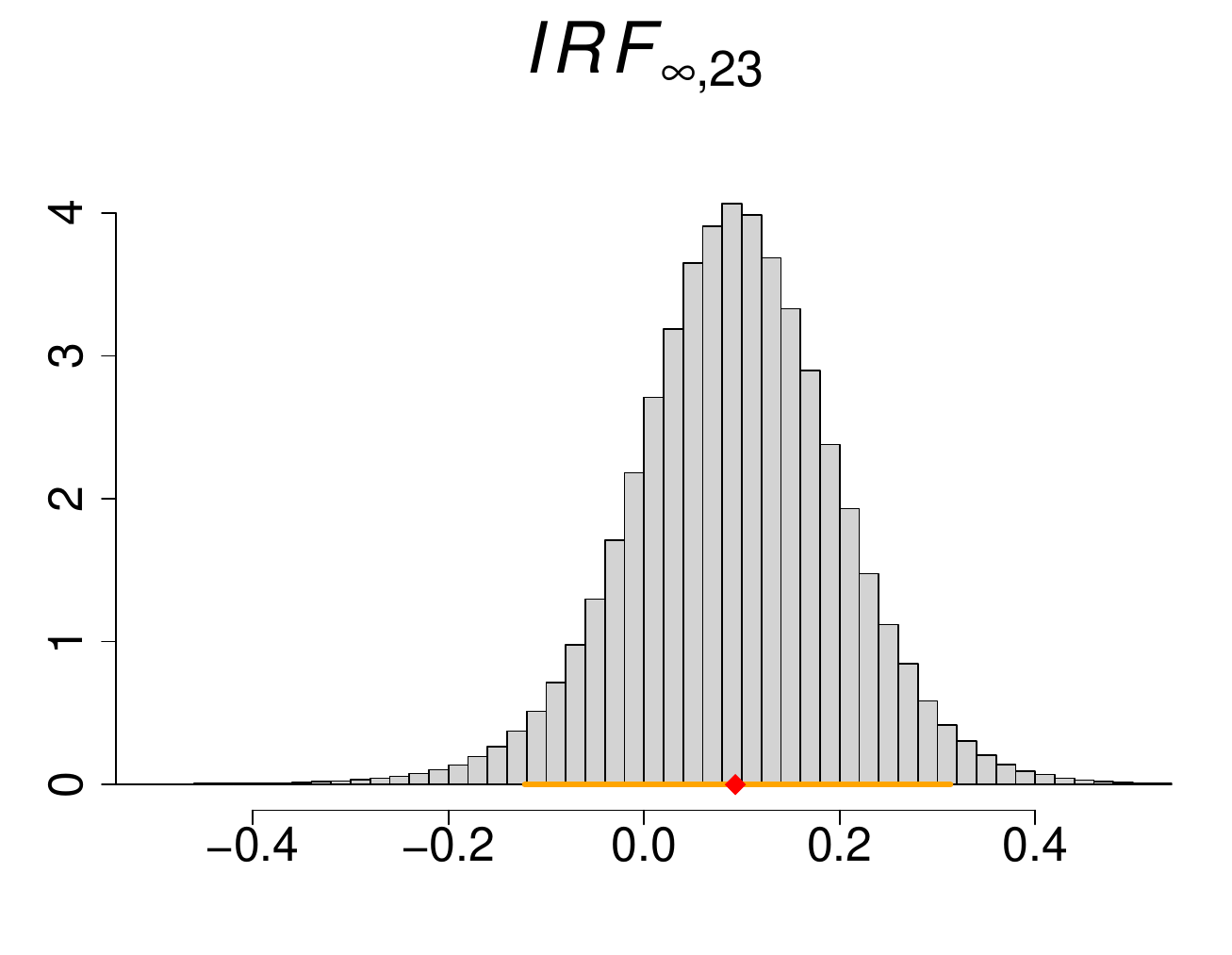} \tiny{$Me = 0.093,\ HPD = (-0.122, 0.313)$}\\
			\hline
			\includegraphics[width=0.33\textwidth]{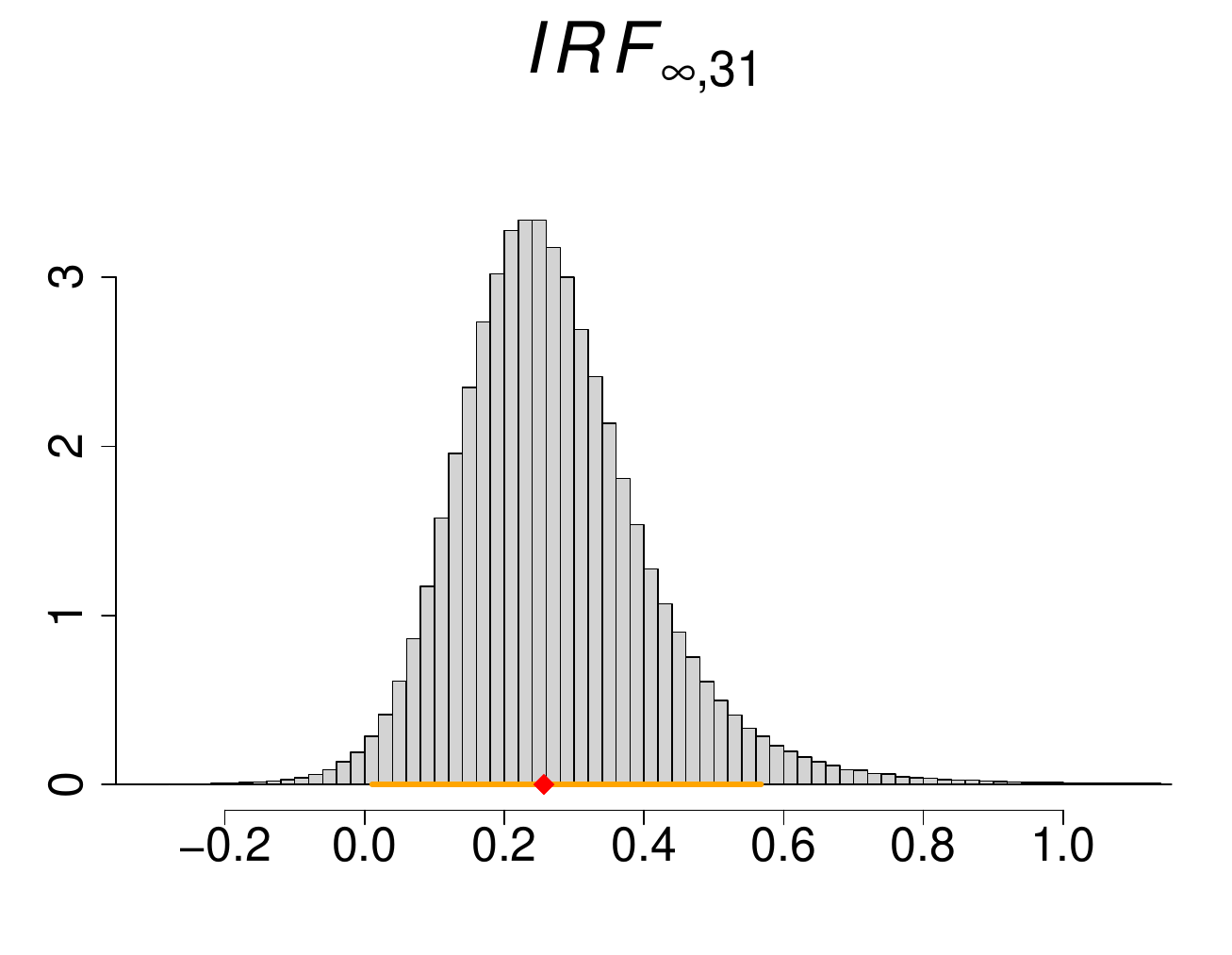} \tiny{$Me = 0.257,\ HPD = (0.011, 0.568)$} &
			\includegraphics[width=0.33\textwidth]{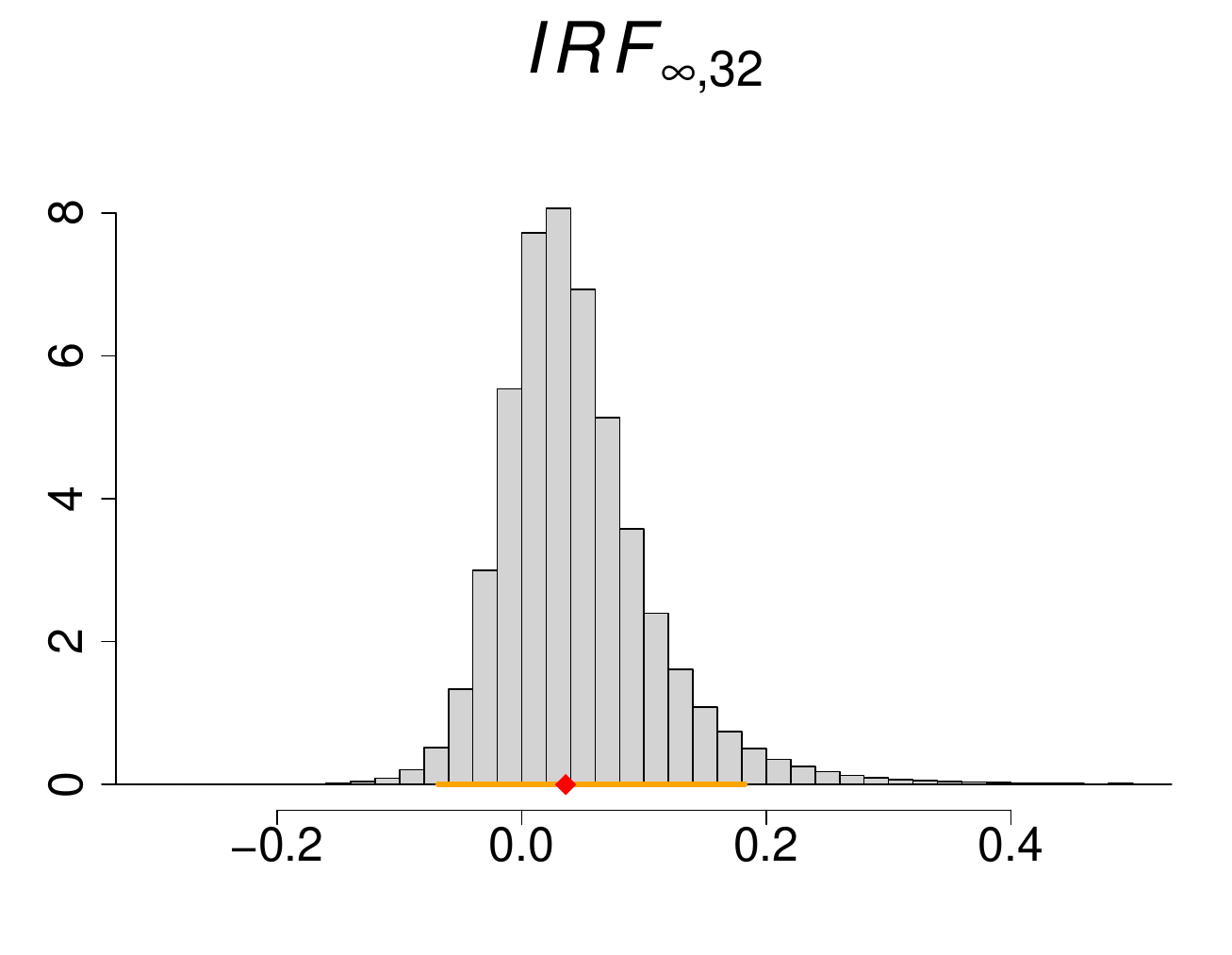} \tiny{$Me = 0.036,\ HPD = (-0.068, 0.183)$}&
			\includegraphics[width=0.33\textwidth]{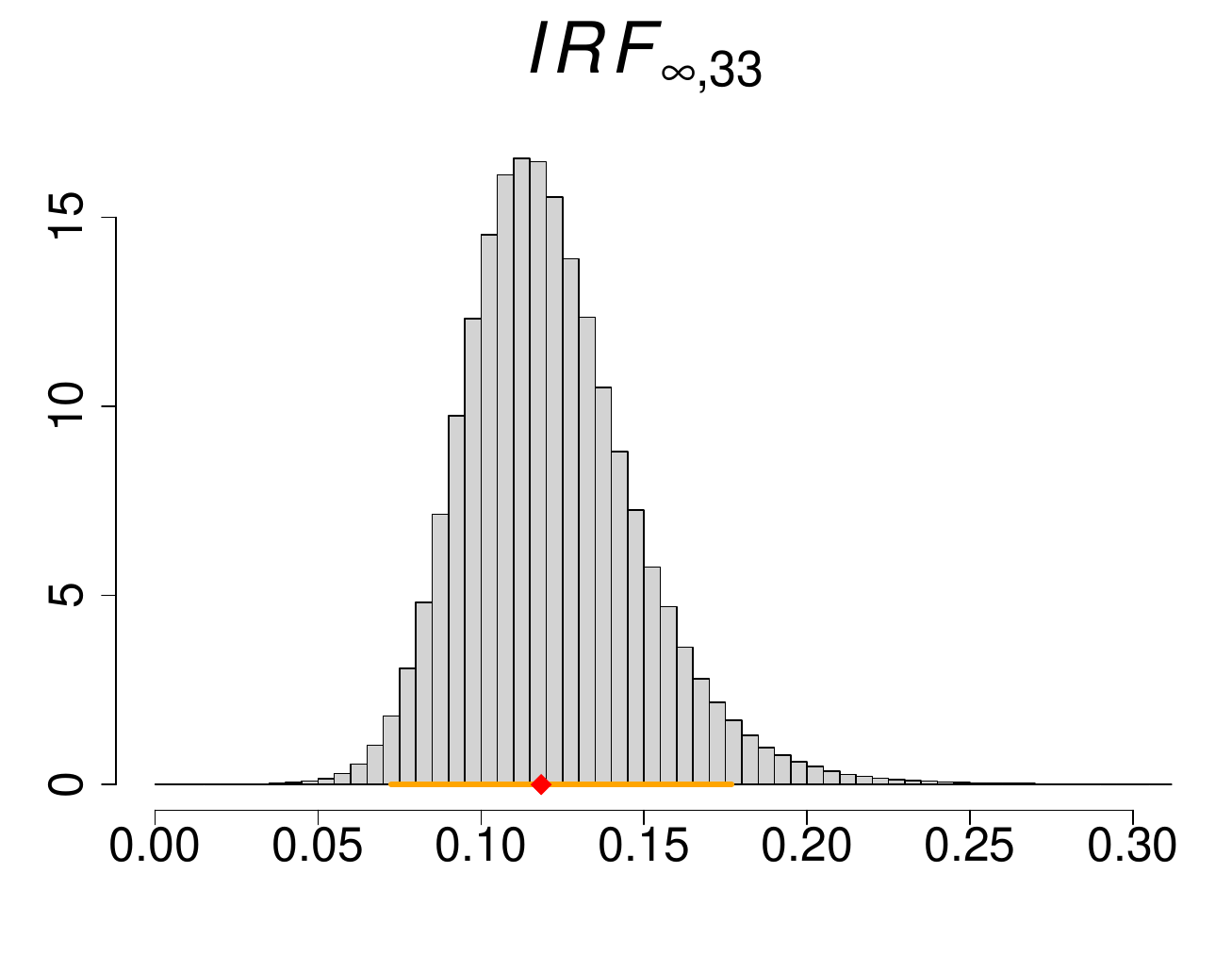} \tiny{$Me = 0.118,\ HPD = (0.072, 0.177)$}\\
			\hline
		\end{tabular}    
	\end{center}
	\caption{Posterior histograms for the elements of the long-run matrix in the empirical example in the first state, with the posterior medians (red markers) and 95\% HPD intervals (orange line).}
	\label{fig:IRF_inf_real_1}
\end{figure}

\begin{figure}
	\begin{center}
		\begin{tabular}{|>{\centering\arraybackslash}m{.33\linewidth} |>{\centering\arraybackslash}m{.34\linewidth} |>{\centering\arraybackslash}m{.33\linewidth}|}
			\hline
			\includegraphics[width=0.33\textwidth]{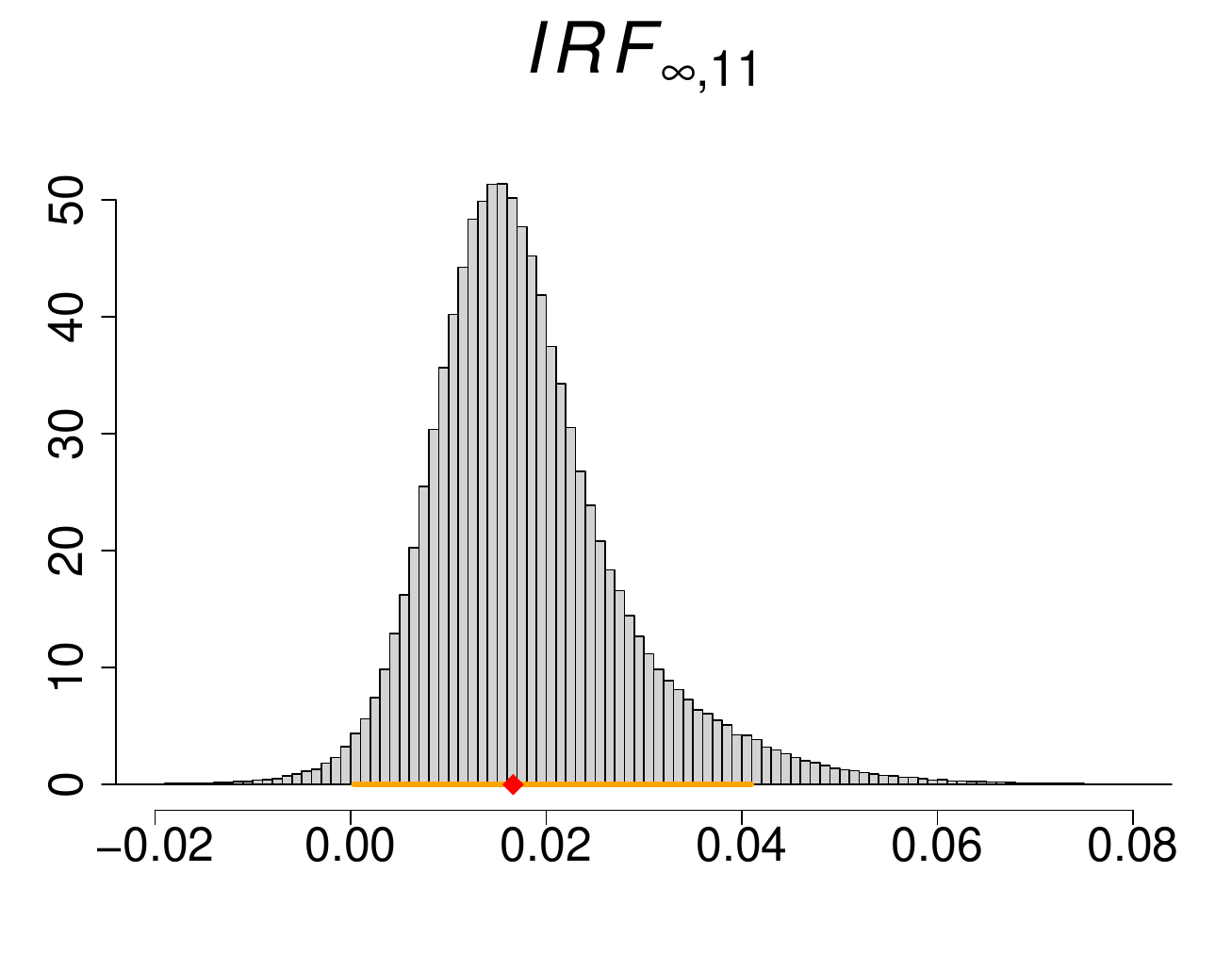} \tiny{$Me = 0.017,\ HPD = (0.000, 0.041)$} &
			\includegraphics[width=0.33\textwidth]{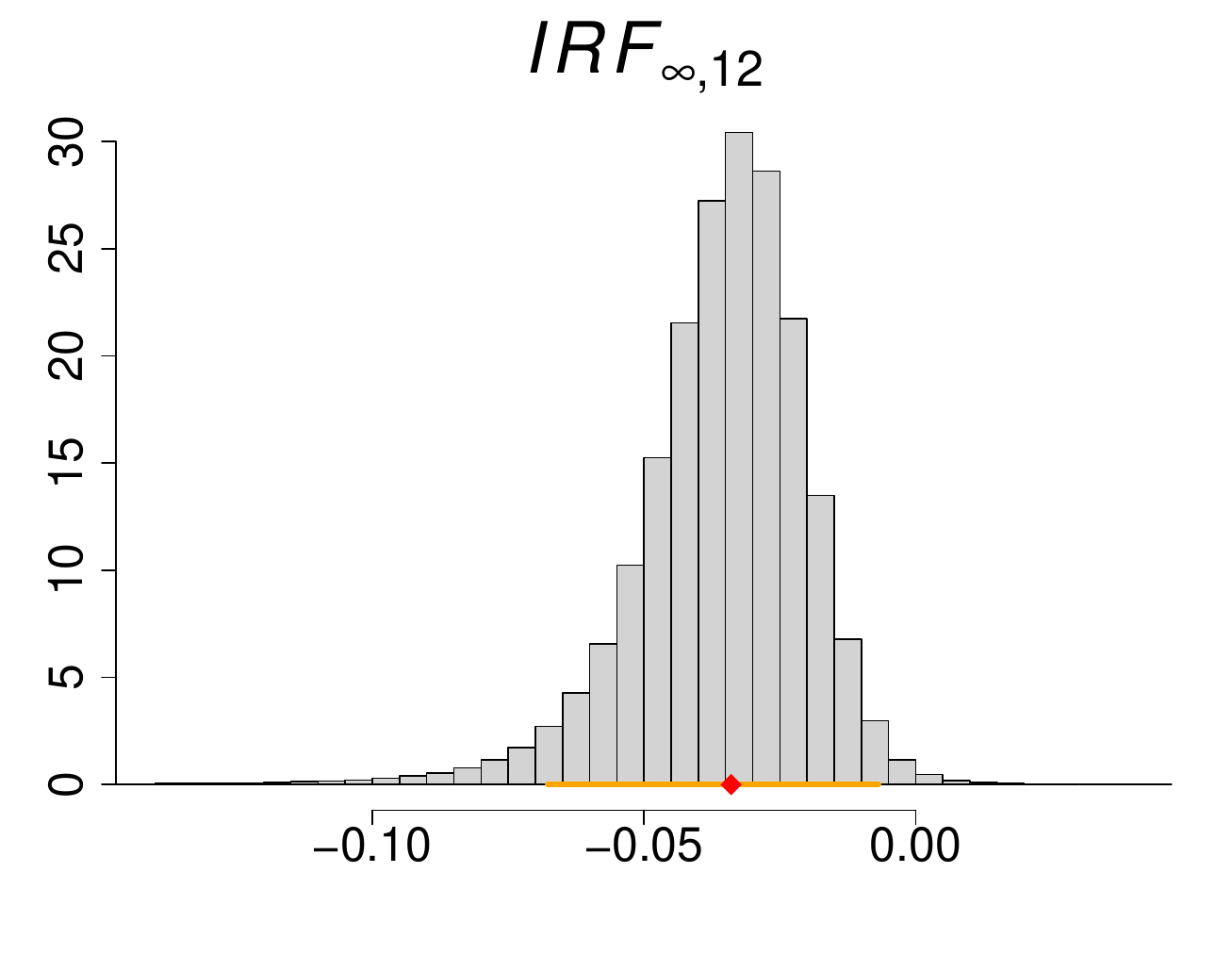} \tiny{$Me = -0.034,\ HPD = (-0.068, -0.007)$}&
			\includegraphics[width=0.33\textwidth]{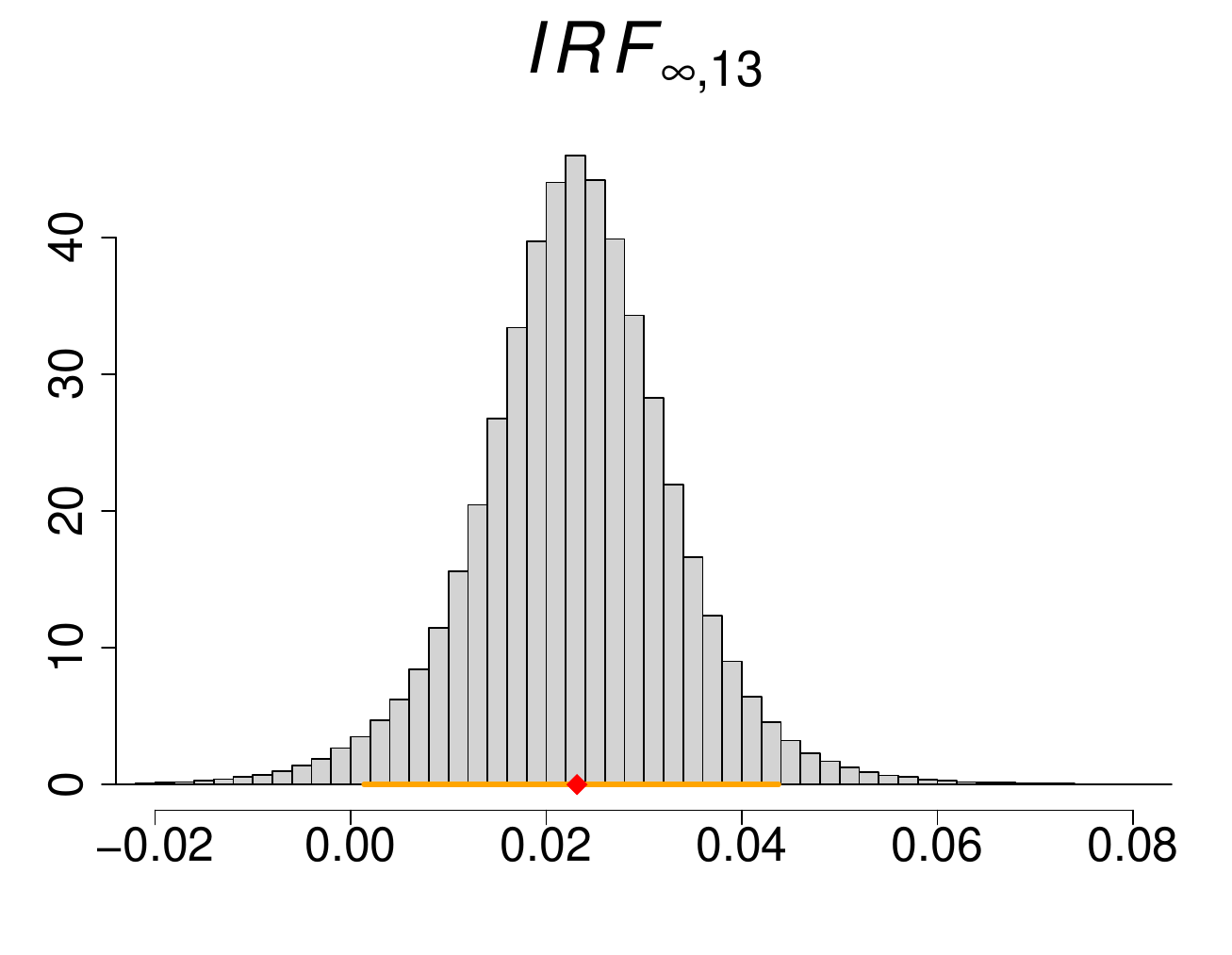} \tiny{$Me = 0.023,\ HPD = (0.001, 0.044)$}\\
			\hline
			\includegraphics[width=0.33\textwidth]{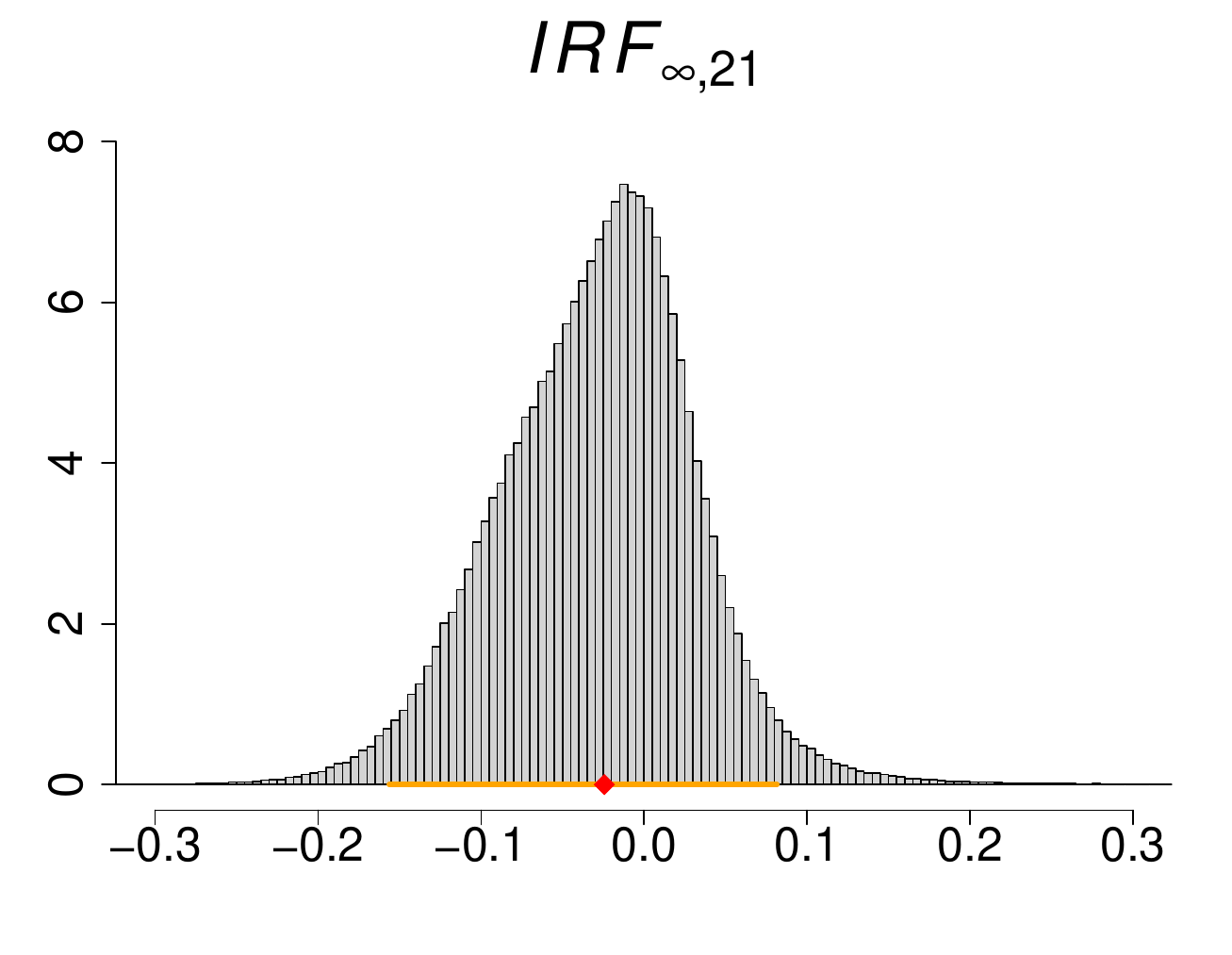} \tiny{$Me = -0.024,\ HPD = (-0.156, 0.082)$} &
			\includegraphics[width=0.33\textwidth]{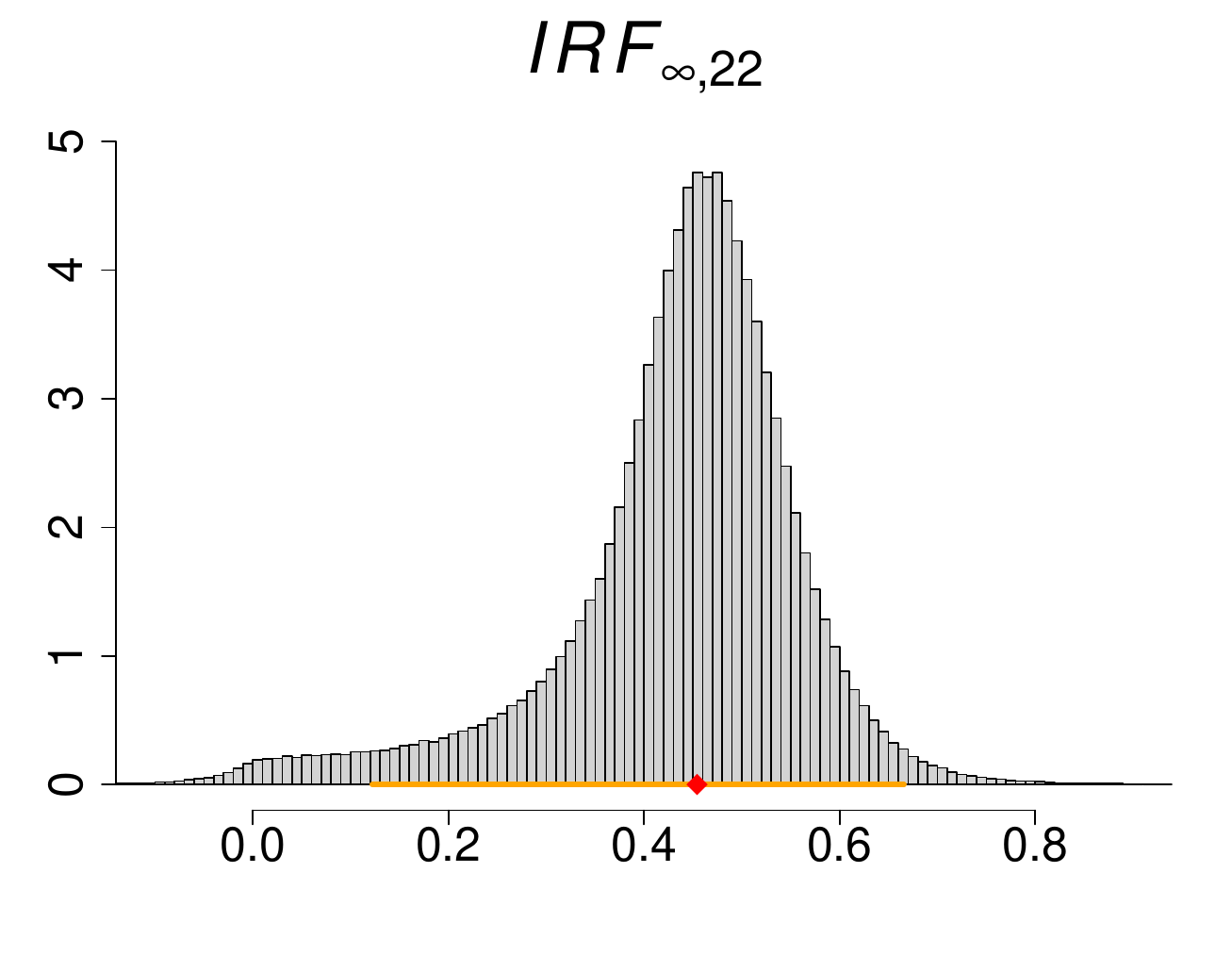} \tiny{$Me = 0.454,\ HPD = (0.122, 0.666)$}&
			\includegraphics[width=0.33\textwidth]{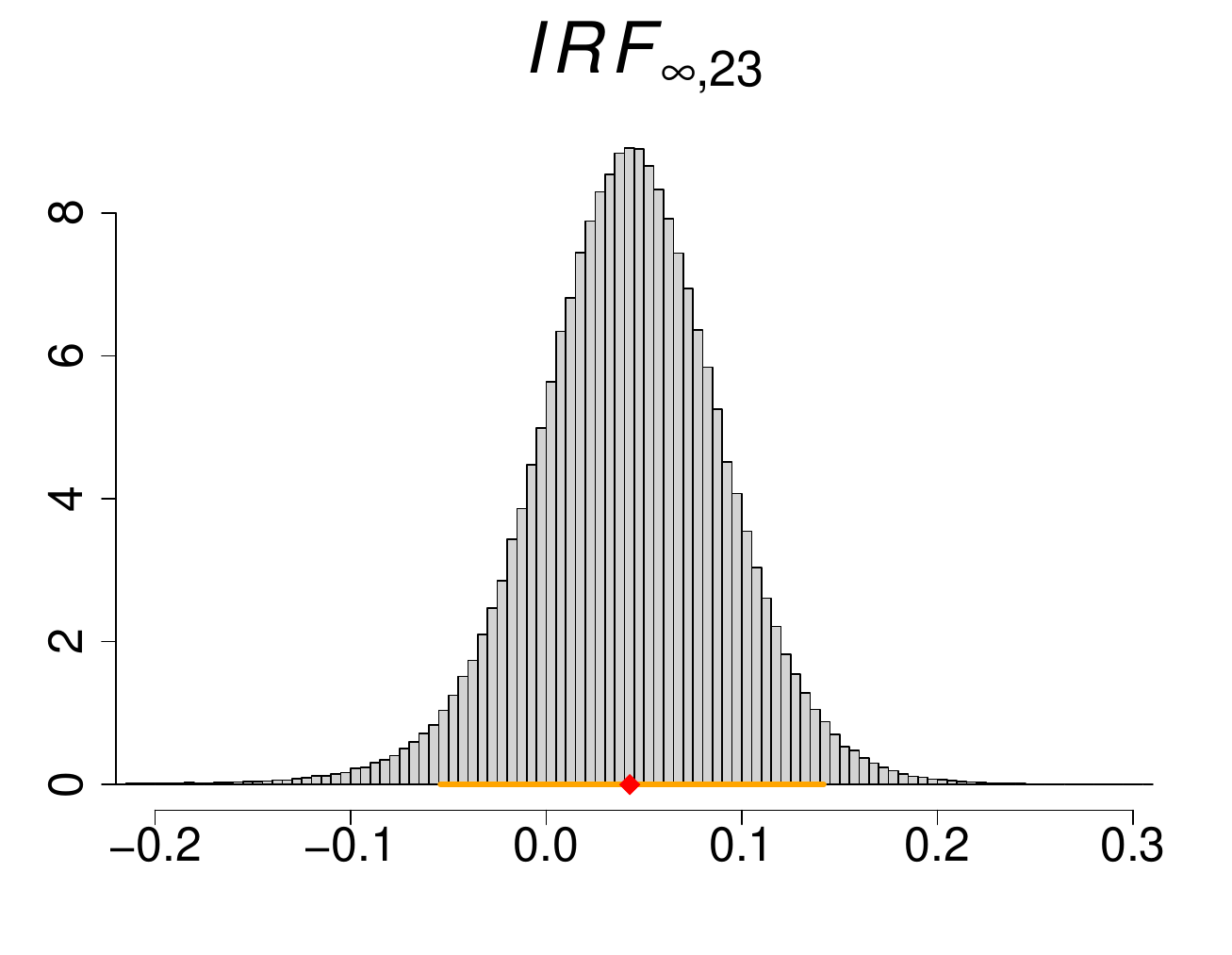} \tiny{$Me = 0.043,\ HPD = (-0.054, 0.142)$}\\
			\hline
			\includegraphics[width=0.33\textwidth]{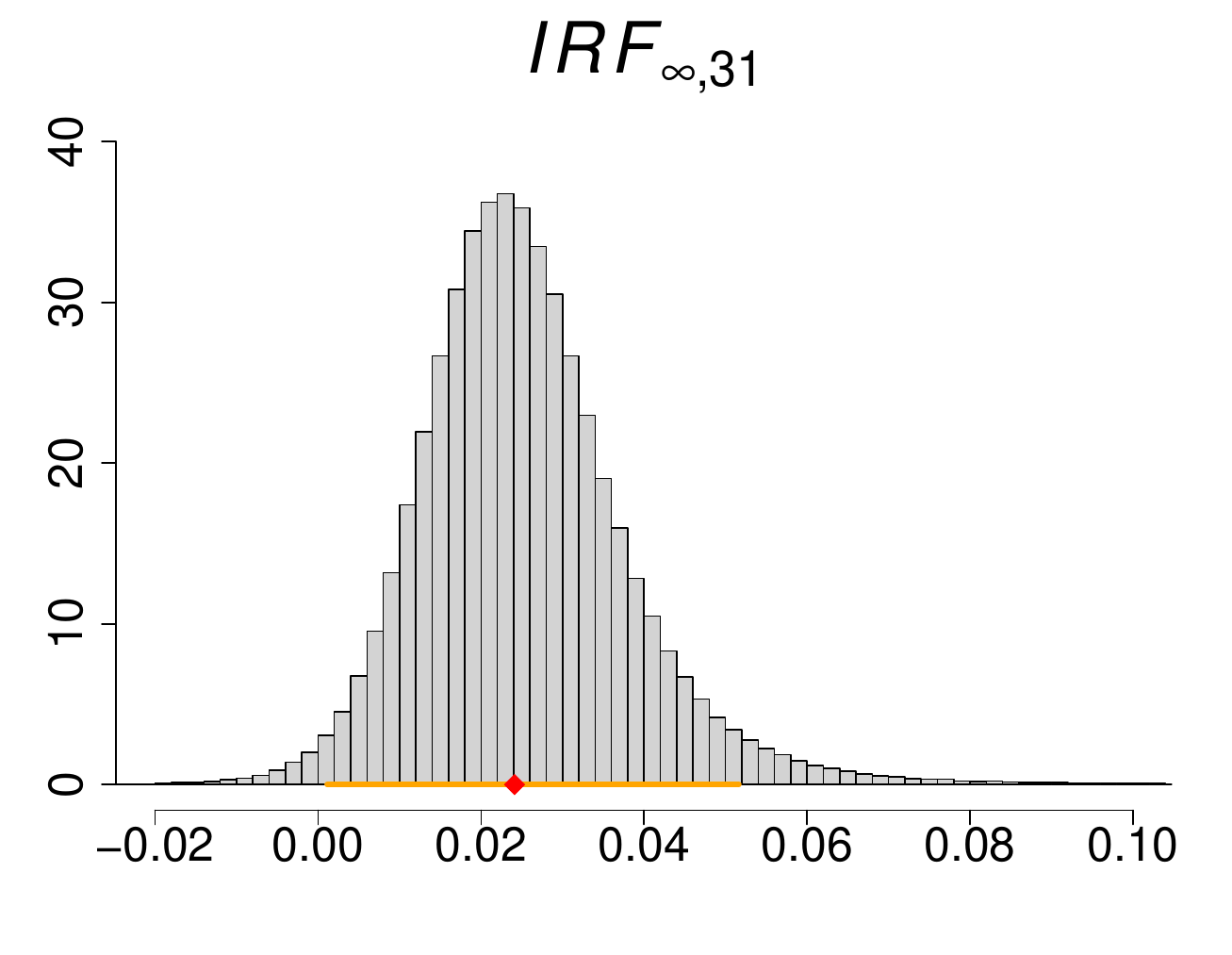} \tiny{$Me = 0.024,\ HPD = (0.001, 0.052)$} &
			\includegraphics[width=0.33\textwidth]{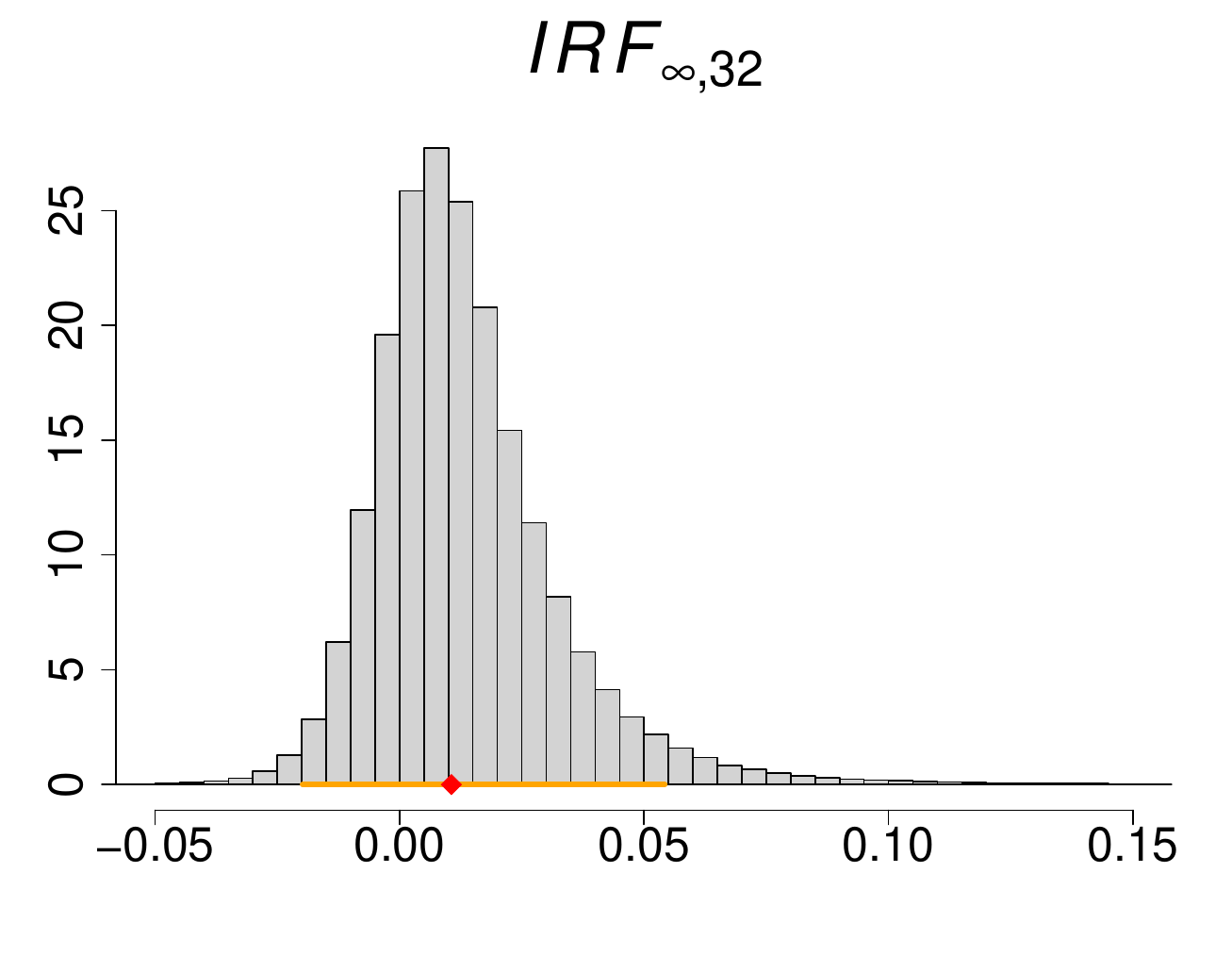} \tiny{$Me = 0.011,\ HPD = (-0.020, 0.054)$}&
			\includegraphics[width=0.33\textwidth]{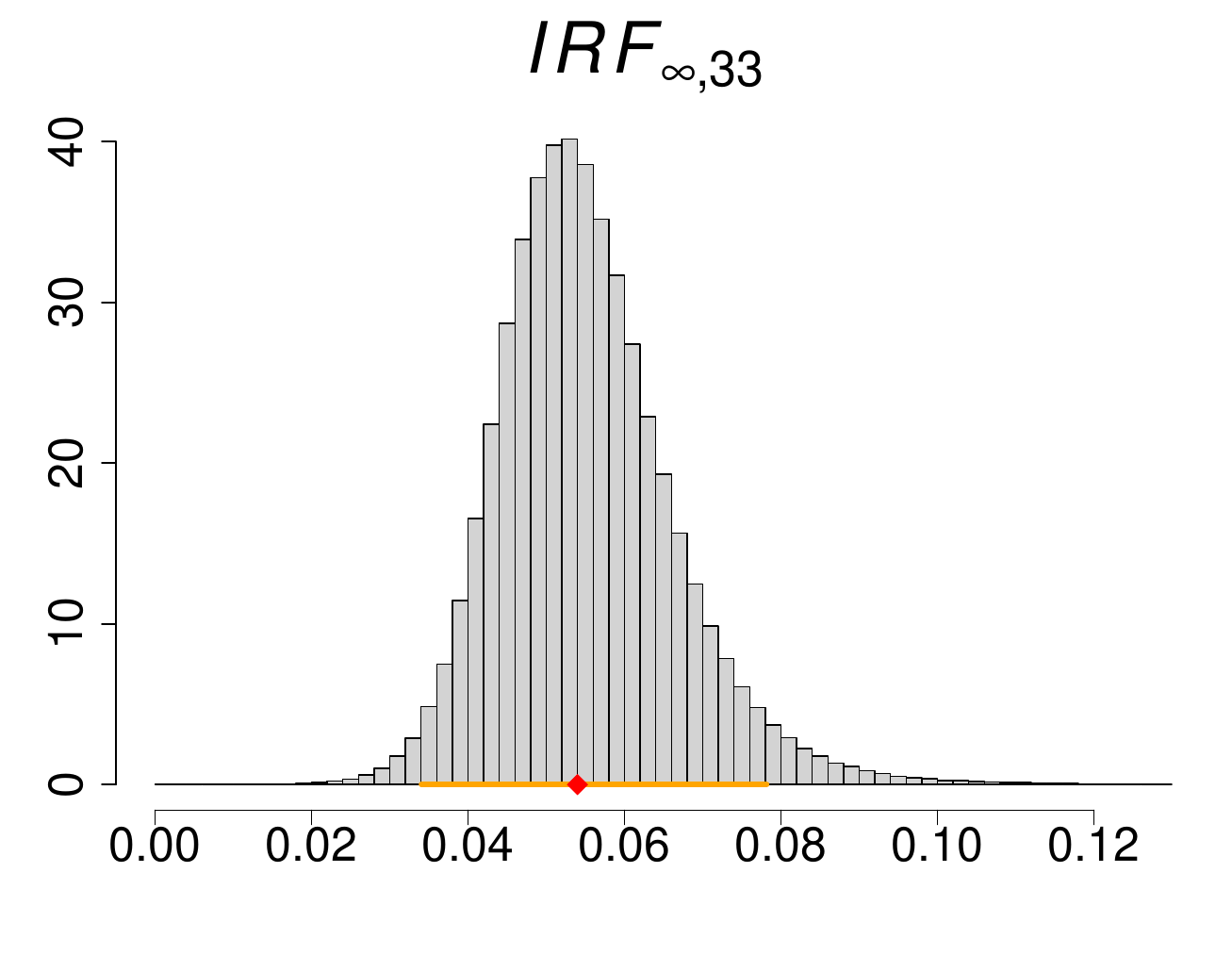} \tiny{$Me = 0.054,\ HPD = (0.034, 0.078)$}\\
			\hline
		\end{tabular}    
	\end{center}
	\caption{Posterior histograms for the elements of the long-run matrix in the empirical example in the second state, with the posterior medians (red markers) and 95\% HPD intervals (orange line).}
	\label{fig:IRF_inf_real_2}
\end{figure}

Figures~\ref{fig:B_real}-\ref{fig:IRF_inf_real_2} present the histograms of the instantaneous and long-run reaction of the variables to structural shocks.

\end{document}